# Conjectures on exact solution of three - dimensional (3D) simple orthorhombic Ising lattices

Zhi-dong Zhang

Shenyang National Laboratory for Materials Science, Institute of Metal Research and International Centre for Materials Physics, Chinese Academy of Sciences, 72 Wenhua Road, Shenyang, 110016, P.R. China

We report the conjectures on the three-dimensional (3D) Ising model on simple orthorhombic lattices, together with the details of calculations for a putative exact solution. Two conjectures, an additional rotation in the fourth curled-up dimension and the weight factors on the eigenvectors, are proposed to serve as a boundary condition to deal with the topologic problem of the 3D Ising model. The partition function of the 3D simple orthorhombic Ising model is evaluated by spinor analysis, by employing these conjectures. Based on the validity of the conjectures, the critical temperature of the simple orthorhombic Ising lattices could be determined by the relation of $KK^* = KK' + KK'' + K'K''$ or $\sinh 2K \cdot \sinh 2(K' + K'' + \frac{K'K''}{K}) = 1$. For a simple cubic Ising lattice, the critical point is putatively determined to locate exactly at the golden ratio $x_c = e^{-2K_c} = \frac{\sqrt{5}-1}{2}$, as derived from $K^* = 3K$ or $\sinh 2K \cdot \sinh 6K = 1$. If the conjectures would be true, the specific heat of the simple orthorhombic Ising system would show a logarithmic singularity at the critical point of the phase transition. The spontaneous magnetization of the simple orthorhombic Ising ferromagnet is derived explicitly by the perturbation procedure, following the conjectures. The spin correlation functions are discussed on the terms of the Pfaffians, by defining the effective skew-symmetric matrix $A_{eff}$. The true range $\kappa_x$ of the correlation, and the susceptibility of the simple orthorhombic Ising system are determined by the procedures similar to those used for the two-dimensional Ising system. The putative critical exponents derived explicitly for the simple orthorhombic

Ising lattices are $\alpha = 0$, $\beta = 3/8$, $\gamma = 5/4$, $\delta = 13/3$, $\eta = 1/8$ and $\nu = 2/3$, showing the universality behavior and satisfying the scaling laws. The cooperative phenomena near the critical point are studied and the results obtained based on the conjectures are compared with those of the approximation methods and the experimental findings. The putative solutions have been judged by several criterions. The deviations of the approximation results and the experimental data from the solutions are interpreted. Based on the solution, it is found that the 3D – to – 2D crossover phenomenon differs with the 2D – to – 1D crossover phenomenon and there is a gradual crossover of the exponents from the 3D values to the 2D ones. Special attentions are also paid on the extra energy caused by the introduction of the fourth curled-up dimension, the states at/near infinite temperature as revealed by the weight factors of the eigenvectors. The physics beyond the conjectures and the existence of the extra dimension are discussed. The present work may not only have significance in statistic physics and condensed matter physics, but also fill the gap between the fields of the quantum field theory, the cosmology theory, high-energy particle physics, graph theory and computer sciences.

Corresponding to: zdzhang@imr.ac.cn

# I. INTRODUCTION

The Ising model has been well – known to be a simple model providing profound physical significances, which is helpful for discovering principles in our physical world.[1] It has been not only conceived as a description of magnetism in crystalline materials, but also applied to various phenomena as diverse as the order-disorder transformation in alloys,[2-7] the transition of liquid helium to its suprafluid state, the freezing and evaporation of liquids, the behavior of glassy substances, and even the folding of protein molecules into their biologically active forms. In accordance with the Yang and Lee's theorems,[8,9] the problem of an Ising model in a magnetic field is mathematically equivalent to a lattice gas. The widespread interest focusing on the Ising model is primarily derived from the fact that it is one of the simplest examples describing a system of interacting particles (or atoms or spins). The Ising model forms an excellent test case for any new approximation method of investigating systems of interacting particles, specially, of understanding the cooperative phenomena and the critical behaviors at/near the critical point of a continuous phase transition. Furthermore, the 3D Ising model can serve as a testing model for the evolution of a system of interacting particles (or spins) from infinite temperature down to zero, as one could see the analogy of temperature in the thermodynamics to a variable of time in the dynamics. Therefore, the exact solutions obtained are quite helpful for understanding the evolution of an equilibrium infinite system not only for a magnet, but also even for our Universe. In addition, the formal theory of equilibrium phase transitions has found applications in problems such as contiuous quantum phase transitions,[10,11] constructing field and string theories of elementary particles, the transition to chaos in dynamical systems, the long – time behavior of systems out of equilbrium and dynamic critical phenomena.[12]

It is well understood that only the exact solution of a system of interacting particles can be used to reveal fully the cooperative phenomena and the critical behaviors at/near the critical point. The two-dimensional (2D) Ising model is among a few examples that have been solved explicitly.[13] The partition function for the 2D Ising model was evaluated exactly by Onsager,[13] using the approach introduced by Kramers and Wannier,[14,15] and Montroll.[16] Later on, the problem was solved exactly by a simple and elegant spinor analysis developed by Kaufman and Onsager.[17-19] The temperature dependence of magnetization of a square, rectangular or triangular Ising magnet was calculated by Yang,[20] Chang[21] and Potts,[22] respectively, using a perturbation method. Newell[23] showed the equalization between a cylindrical crystal studied by Onsager and Kaufman,[13,17-19] and a screw one studied by Kramers and Wannier.[14,15] The statistical mechanics of 2D Ising triangular, honeycomb and Kagomé nets was worked out by various authors.[24-38] A general lattice – statistical model, which included soluble 2D models of phase transitions, such as the ice model,[39,40] the hydrogen-bonded ferroelectrics and antiferroelectrics models,[41-51] was proposed.[52] Only a limited number of three-dimensional (3D) systems have been solved, including the four-spin interaction Ising model solved by Suzuki,[53] the Zamolodchikov model[54] solved by Barter[55] and its N – state extension by Bazhanov and Baxter,[56] the 3D dimer model solved by Huang, Popkov and Wu.[57] However, the Suzuki model turns out to be a 2D system in disguise,[53,57] while the Zamolodchikov model and its extension involve unphysical negative Boltzmann weights.[55-57] The Huang, Popkov and Wu's 3D dimer model,[57] consisting of layered honeycomb dimer lattices with a specific layer–layer interaction, has been the only solvable 3D lattice model with physical Boltzmann weights, but describes dimer configurations in which dimers are confined in planes. As a consequence the critical behavior of this 3D dimer model is essentially two - dimensional.[58]

The exact solution of the 3D Ising model presents difficulties of a very fundamental nature. The most reliable information on the behavior of the 3D Ising model has been provided by exact series expansions of the partition function at low and high temperatures,[59-125] and by renormalization group theory near the critical point,[126-191] and Monte Carlo simulations.[154-186] Although the region near the critical point has been explored by various approximation methods and its physical properties can be determined numerically with a high precision,[59-191] up to now, physicists fail to provide the exact mathematical solution for the 3D Ising model. It is clear that the 3D Ising model cannot be exactly solved within the framework of the procedure for solving the 2D Ising lattice. This is a bit disappointing for the 3D physicists who would like to understand 3D matter in our physical world.

The difficulty for solving completely the 3D Ising model is evident, since the problem becomes much more complicated than the 2D Ising model that is already very much complex. Attempts to apply the algebraic method used for solving the 2D model to the 3D problem are seriously hindered at an early stage, because the operators of interest generate a much large Lie algebra being so large that it would seem to be of little value.[59] It seems that all previous algebraic methods take advantage of very special properties of the operators, and it has not been possible to generalize them in any very interesting way to deal with the 3D Ising system successfully.[59] No spinors, Lie algebras, or other specialized algebraic techniques of the type used in the matrix method solution are required in the combinatorial method, developed by Kac and Ward.[60] However, the combinatorial method introduces some problems in topology that have not been rigorously solved. This combinatorial method of counting the closed graph cannot be generated in any obvious way to the 3D problem, since the peculiar topological property is that a polygon in three dimensions does not divide the space into an "inside and outside".[59] Realizing that it is hard to obtain any results owing to considerable mathematical difficulties, many authors have tried various methods to generate approximation results, such series expansions,[61-125] renormalization group and Monte Carlo techniques,[126-207] etc. However, any approximation method cannot prove exact information at/near the critical point, since whenever the thermodynamic functions have an essential singularity it is difficult to perform any computation by successive approximation because the convergence of approximation by analytic functions in such cases is notoriously slow.[13]

In this work, we shall try to derive a putative exact solution of the 3D Ising model on simple orthorhombic lattices. We try to do so because we believe that such exact solution must exist in the nature. Of course, what we are sure is that for this purpose, we must develop a completely new mathematical technique to overcome the difficulties hindering our path. This novel mathematical technique must be in a certain sense out of the bound of the previous methods, although we have to follow mainly the processes developed by Onsager, Kaufman and Yang, etc.[13,17-20] The partition function of the simple orthorhombic Ising model will be evaluated by the spinor analysis, by introducing two conjectures employing an additional rotation in the fourth curled-up dimension and the weight factors on the eigenvectors. These conjectures serve as a boundary condition to deal with the topologic problem of the 3D Ising model, so that its partition function could be evaluated successfully. The simple and beautiful solution comes out of the very complicated system, automatically and spontaneously, only by introducing few conjectures. The solutions will be compared with the results of various approximations and also experiments. The putative exact solutions have been judged by several criterions. The deviations of the approximation results and the experimental data from the putative solutions are interpreted. The physics beyond the conjectures and the existence of the extra dimension are discussed. Our simple and beautiful results in elegant forms suggest that we might hit our target exactly. Nevertheless, it should be emphasized that the validity of the putative solutions depends on that of the conjectures. In Sec. II, the simple orthorhombic Ising model will be described briefly and the matrix problem will be set up. In Sec. III, the partition function of the simple orthorhombic Ising model will be evaluated by the spinor analysis with help of two conjectures, and the specific heat of the simple orthorhombic Ising system will be studied. In Sec. IV, the spontaneous magnetization of the simple orthorhombic Ising magnet will be derived by the perturbation procedure, also based on the validity of the conjectures. In Sections V and VI, the correlation function and susceptibility will be investigated. In Sec. VII, the critical exponents at/near the critical point will be compared with the previous results of various approximations and experiments. Sections VIII and IX are for discussions and summary. The evaluation of the weight factors will be performed in Appendixes A and B, respectively, for the simple cubic lattice and the simple orthorhombic lattice. The purpose of the present article is just to present the calculation procedure and the final results of our solutions and to compare with other approximation methods and the experimental data, not to attempt to give a full and comprehension review of all advances in the Ising models, which is beyond the scope of this article. The readers, who are interested in the advances in the various approximation techniques, such as series expansions, the Monte Carlo simulations and the renormalization group techniques, etc., as well as in the experiments, refer to the existed comprehension reviews[59,103-107,141-144,149,152-160,192-202,208-225] and references therein.

## II. MODEL AND SETTING UP THE MATRIX PROBLEM

As most of real systems in 3D, the atoms occupy blocks on a 3D lattice, like a collection of stacked boxes, to establish a 3D Ising model. Our physical model is a simple orthorhombic lattice with m rows and n sites per row in one of l planes. Each site in the lattice could be indexed by (i, j, k) for its location in the coordinate system (rows, column, plane). These sites are to be occupied by two kinds of constituent atoms, each of which can have its magnetic poles pointing in one of two opposite orientations. In our 3D Ising model of spin 1/2, only are the interactions between the nearest neighboring atoms taken into account. Within one plane, the energy of interaction is +J between unlike neighbors in a row, and +J’ between unlike neighbors in a column; but –J, -J’ between like neighbors in a row, or column, respectively. The energy is +J’’ (or –J’’) for interaction between unlike (or like) neighbors connecting two neighboring planes. The unlike or like neighbors correspond to the anti-parallel or parallel arrangement of the neighboring spins. Clearly, our simple orthorhombic Ising model is an extension of the rectangular Ising model dealt with by Onsager[13] and Kaufman.[17] The Hamiltonian of the 3D Ising model on simple orthorhombic lattices is written as:

$$\hat{H} = -J\sum_{\tau=1}^{n}\sum_{\rho=1}^{m}\sum_{\delta=1}^{l} s_{\rho,\delta}^{(\tau)} s_{\rho,\delta}^{(\tau+1)} - J'\sum_{\tau=1}^{n}\sum_{\rho=1}^{m}\sum_{\delta=1}^{l} s_{\rho,\delta}^{(\tau)} s_{\rho+1,\delta}^{(\tau)} - J''\sum_{\tau=1}^{n}\sum_{\rho=1}^{m}\sum_{\delta=1}^{l} s_{\rho,\delta}^{(\tau)} s_{\rho,\delta+1}^{(\tau)} . \quad (2.1)$$

The probability of finding the simple orthorhombic Ising lattices in a given configuration, at the temperature T, is proportional to exp{$-E_c/k_BT$}, where $E_c$ is the total energy of the configuration and $k_B$ is the Boltzmann constant. The exponent appearing in the expression for the probability is always of the form:

$$(n_c \cdot J + n_c' \cdot J' + n_c'' \cdot J'')/k_B T$$

Here $n_c$, $n_c'$ and $n_c''$ are integers depending on the configuration of the lattice. Again, it is convenient to introduce the variables $K \equiv J/k_BT$, $K' \equiv J'/k_BT$ and $K'' \equiv J''/k_BT$ instead of J, J' and J''. Please notice that the notation K is the same as H in Onsager and Kaufman's papers,[13,17,18] and Yang's one.[20] Then the probability of a configuration reads as:

$$\frac{1}{Z}\exp\{n_c K + n_c' K' + n_c'' K''\}$$

where Z is the partition function for the lattice:

$$Z = \sum_{\substack{all \\ configurations}} e^{n_c K + n_c' K' + n_c'' K''} . \quad (2.2)$$

The thermodynamic functions for the simple orthorhombic Ising model can be found from knowledge of Z, but unfortunately, the problem becomes much more complicated than the case of a 2D Ising model, since the number of terms in Z is $2^{m \cdot n \cdot l}$. Following the procedure developed by Kaufman,[17] we also introduce the fiction of spin attributed to each atom. All atoms of one kind will be given the spin +1, while the other kind –1. So, the interaction between two neighboring atoms with spins μ, μ' is: $-\mu\,\mu' K$ (or $-\mu\,\mu' K'$ or $-\mu\,\mu' K''$) for row (or column or plane) neighbors. The configurations of the magnet can now be specified either by stating the value of μ at each site of the magnet or by considering the row configurations. The latter is more convenient. Since within one plane there are n atoms in a row and there are l planes, there are $2^{n\cdot l}$ possible configurations, $1 \leq \nu \leq 2^{n\cdot l}$. Then the configuration of the simple orthorhombic Ising model is given by the set $\{\nu_1, \nu_1, \ldots \nu_m\}$.

The energy due to interactions within the i-th row in all the planes is denoted by $E'(\nu_i)$; the energy due to interaction between two adjacent rows in all the planes by $E(\nu_i, \nu_{i+1})$; the energy due to interaction between two i-th rows in two adjacent planes by $E''(\nu_i)$. As a result, the energy of a configuration of the crystal is represented as:

$$E_c = \sum_{i=1}^{m} E'(\nu_i) + \sum_{i=1}^{m} E''(\nu_i) + \sum_{i=1}^{m} E(\nu_i, \nu_{i+1}) . \tag{2.3}$$

For purpose of the symmetry, it is assumed that the m-th row in each plane of the crystal interacts with the first row in that plane. To do so, we actually apply the cylindrical crystal model preferred by Onsager,[13] and Kaufman,[17] in which we wrap our crystal on cylinders. However, in the present 3D case, there are l coaxial cylinders corresponding to l planes, while in the 2D case there is only a cylinder. Making the abbreviations:

$$(V_1)_{\nu_i \nu_{i+1}} \equiv \exp\{-E(\nu_i, \nu_{i+1}) / k_B T\} , \tag{2.4a}$$

$$(V_2)_{\nu_i \nu_i} \equiv \exp\{-E'(\nu_i) / k_B T\} , \tag{2.4b}$$

$$(V_3)_{\nu_i \nu_i} \equiv \exp\{-E''(\nu_i) / k_B T\} , \tag{2.4c}$$

one finds that the probability of a configuration is proportional to

$$\begin{aligned} e^{-E_c / k_B T} &= (V_3)_{\nu_1 \nu_1} (V_2)_{\nu_1 \nu_1} (V_1)_{\nu_1 \nu_2} (V_3)_{\nu_2 \nu_2} (V_2)_{\nu_2 \nu_2} (V_1)_{\nu_2 \nu_3} \times \ldots \\ &\times (V_3)_{\nu_m \nu_m} (V_2)_{\nu_m \nu_m} (V_1)_{\nu_m \nu_1} \end{aligned} . \tag{2.5}$$

Therefore the partition function becomes:

$$Z = \sum_{\nu_1,\nu_2,\ldots\nu_m} (V_3)_{\nu_1\nu_1} (V_2)_{\nu_1\nu_1} (V_1)_{\nu_1\nu_2} \ldots (V_3)_{\nu_m\nu_m} (V_2)_{\nu_m\nu_m} (V_1)_{\nu_m\nu_1} \equiv trace(V_3 V_2 V_1)^m . \qquad (2.6)$$

Since for each i: $1 \le \nu_i \le 2^{n\cdot l}$, we find that $\mathbf{V_1}$, $\mathbf{V_2}$ and $\mathbf{V_3}$ are $2^{n\cdot l}$-dimensional matrices and, $\mathbf{V_2}$ and $\mathbf{V_3}$ are diagonal. $\mathbf{V_1}$, $\mathbf{V_2}$ and $\mathbf{V_3}$ can be given explicitly as:

$$V_3 = \exp\{K''\cdot \sum_{r=1}^{n}\sum_{s=1}^{l} s''_{r,s}\, s''_{r,s+1}\} \equiv \exp\{K''\cdot A''\}, \qquad (2.7a)$$

$$V_2 = \exp\{K'\cdot \sum_{s=1}^{l}\sum_{r=1}^{n} s'_{r,s}\, s'_{r+1,s}\} \equiv \exp\{K'\cdot A'\}, \qquad (2.7b)$$

$$V_1 = (2\sinh 2K)^{\frac{n\cdot l}{2}} \cdot \exp\{K^*\cdot \sum_{s=1}^{l}\sum_{r=1}^{n} C_{r,s}\}. \qquad (2.7c)$$

Here $\mathbf{s''_{r,s}}$, $\mathbf{s'_{r,s}}$ and $\mathbf{C_{r,s}}$ are $2^{n\cdot l}$-dimensional quaternion matrices:

$$s''_{r,s} \equiv 1\otimes 1\otimes \ldots \otimes 1\otimes s''\otimes 1\otimes \ldots \otimes 1, \qquad (2.8a)$$

$$s'_{r,s} \equiv 1\otimes 1\otimes \ldots \otimes 1\otimes s'\otimes 1\otimes \ldots \otimes 1, \qquad (2.8b)$$

$$C_{r,s} \equiv 1\otimes 1\otimes \ldots \otimes 1\otimes C\otimes 1\otimes \ldots \otimes 1, \qquad (2.8c)$$

there are n·l factors in each direct-product, with **s''**, **s'** and **C** appearing in the (r,s)-th position. **s''**, **s'** and **C** are generators of the Pauli spin matrices:

$$s'' \equiv \begin{bmatrix} 0 & -1 \\ 1 & 0 \end{bmatrix}, s' \equiv \begin{bmatrix} 1 & 0 \\ 0 & -1 \end{bmatrix}, C \equiv \begin{bmatrix} 0 & 1 \\ 1 & 0 \end{bmatrix}, 1 = \begin{pmatrix} 1 & 0 \\ 0 & 1 \end{pmatrix}. \qquad (2.9)$$

K* is defined by

$$e^{-2K} \equiv \tanh K^*. \qquad (2.10)$$

Here, to simplicity, at the beginning of the diagonalization procedure, we set up only the largest one among K, K' and K'' as the standard axis for definition of K*. This specialization will be discussed in details later.

We redefine $\mathbf{V_1}$ so as to remove the scalar coefficient from it:

$$V_1 \equiv \exp\{K^* \cdot \sum_{s=1}^{l} \sum_{r=1}^{n} C_{r,s}\} \equiv \exp\{K^* \cdot B\}. \tag{2.11}$$

Then the partition function is reduced to

$$Z = (2\sinh 2K)^{\frac{m \cdot n \cdot l}{2}} \cdot trace(V_3 V_2 V_1)^m \equiv (2\sinh 2K)^{\frac{m \cdot n \cdot l}{2}} \cdot \sum_{i=1}^{2^{n \cdot l}} \lambda_i^m. \tag{2.12}$$

where $\lambda_i$ are the eigenvalues of $\mathbf{V} \equiv \mathbf{V_3} \cdot \mathbf{V_2} \cdot \mathbf{V_1}$.

## III. PARTITION FUNCTION

In this section, we shall try to evaluate the partition function of the 3D simple orthorhombic Ising crystal, by the spinor analysis developed by Kaufman,[17] and by introducing two conjectures.

Before dealing with the problem, one would need to analyze what the root of the difficulty of the 3D Ising model is, and what the essential difference between 2D and 3D models is. Obviously, the 2D is flat, whereas the 3D is not because of the additional third dimension. But the essential difference between the 2D and 3D Ising models is even more than that, the key is the difference in the topology — the pattern of connections between the nearest-neighbor sites. After comparing the formulae in section II with those for the 2D Ising model,[17] one finds that in the 3D case many of the bonds would be nonplanar. These bonds in the 3D Ising model cross over one another with those in other planes, whereas a 2D square (even more complicated one, like triangular or hexagonal) lattice can always be drawn without crossings, except for a 2D lattice with next nearest neighbors plus nearest neighbors.[59] A 2D triangular net can be transformed from a square net with an additional interaction along one of the diagonals.[59] Unfortunately, it has not been possible to solve the case of the interactions along both diagonals for the 2D models. Indeed, none of the cases, which have been solved exactly, involve interactions with such crossings. This topological distinction seems to be at the root of the difficulty of the 3D Ising lattice where the topology of closed paths involve knots.[59]

### A. Spin matrices' representation of V

It is easily seen that matrices of the type exp(a''·$\mathbf{s''_{r,s}s''_{r,s+1}}$), exp(a'·$\mathbf{s'_{r,s}s'_{r+1,s}}$), exp(b·$\mathbf{C_{r,s}}$), and their products, form a $2^{n\cdot l}$-dimensional representation of the group of rotations in 2n·l-dimensions. Thus, the matrix **V** itself would be the representative of some such rotations.

We start out with a set of 2n·l-quantities $\mathbf{\Gamma_k}$:

$$\Gamma_{2r-1} \equiv C \times C \times \cdots \times s \times 1 \times 1 \times \cdots \equiv P_r ,$$

$$\Gamma_{2r} \equiv -C \times C \times \cdots \times isC \times 1 \times 1 \times \cdots \equiv Q_r , \qquad 1 \leq r \leq nl, \qquad (3.1)$$

where n·l factors appear in each product; **s** or i**sC** appear in the rth place. The $\mathbf{\Gamma_k}$ are $2^{n\cdot l}$-dimensional matrices, which obey the commutation rules

$$\Gamma_k^2 = 1 , \Gamma_k \Gamma_l = -\Gamma_l \Gamma_k ,(1 \leq k, l, \leq 2n\cdot l). \qquad (3.2)$$

All possible product of the $\mathbf{\Gamma_k}$ form a set of $2^{2n\cdot l}$ matrices so that any $2^{n\cdot l}$-dimensional matrix can be written as a linear combination of these base matrices. Following the work of Kaufman,[17] the matrices $\mathbf{V_1}$, $\mathbf{V_2}$ and $\mathbf{V_3}$ can be represented in term of the base matrices as:

$$V_3 = \prod_{r=1}^{n} \prod_{s=1}^{l-1} \exp\{-iK'' P_{r,s+1} Q_{r,s}\} \cdot \exp\{iK'' P_{r,1} Q_{r,l} U''_r\} ; \qquad (3.3a)$$

$$V_2 = \prod_{s=1}^{l} \prod_{r=1}^{n-1} \exp\{-iK' P_{r+1,s} Q_{r,s}\} \cdot \exp\{iK' P_{1,s} Q_{n,s} U'_s\}; \quad (3.3b)$$

$$V_1 = \prod_{s=1}^{l} \prod_{r=1}^{n} \exp\{iK^* \cdot P_{r,s} Q_{r,s}\}. \quad (3.3c)$$

since

$$C_{r,s} = iP_{r,s} Q_{r,s} = 1 \times 1 \times \cdots \times C \times 1 \times \cdots, \quad (3.4a)$$

$$s'_{r,s} = C_{1,s} C_{2,s} \cdots C_{r-1,s} P_{r,s} = 1 \times 1 \times \cdots \times s \times 1 \times \cdots, \quad (3.4b)$$

$$s''_{r,s} = C_{r,1} C_{r,2} \cdots C_{r,s-1} P_{r,s} = 1 \times 1 \times \cdots \times isC \times 1 \times \cdots. \quad (3.4c)$$

The end factors in Eqs. (3.3a) and (3.3b) differ from others, which originate from the boundary conditions. As mentioned above, it is clear from these boundary conditions that in the 3D Ising model the many of the bonds are nonplanar and that these bonds cross over one another with those in other planes.

We have: [17] If the set $\mathbf{\Gamma_k}$ is a matrix realization of the commutation rules of (3.2), all the sets $\mathbf{S\Gamma_k S^{-1}}$ will be also realizations of (3.2). If both sets of matrices, $\mathbf{\Gamma_k}$ and $\mathbf{\Gamma_k^*}$, obey the commutation rules of (3.2), a transformation $\mathbf{S}$ can be found such that $\mathbf{\Gamma_k^*} = \mathbf{S\Gamma_k S^{-1}}$. The immediate consequences are: Two relations between two sets of matrices, $\mathbf{\Gamma_k}$ and $\mathbf{\Gamma_k^*}$, both obey the commutation rules of (3.2), can be referred as a rotation in 2n·l-space, and its spin representation in $2^{n \cdot l}$-space of the rotation in 2n·l-space. If there is a rotation in 2n·l-space, one can always find a spin representation in $2^{n \cdot l}$-space for the rotation in 2n·l-space.

Compared with the 2D Ising model, our 3D model is much more complex, because of the topologic problem, the knots of interactions between the sites. In what follows, we shall introduce a novel transformation to remove the crossover of the nonplanar bonds to overcome the topologic difficulty of the 3D Ising model, which is the key to solve the problem we are dealing with. What we need is to find a transformation representing rotations, which must remove simultaneously the crossings of the connections for the interactions J' and J'' between the neighbors in the lattice, while rearranging the elements in the matrix **V**. It is necessary to introduce a conjecture as follow:

***Conjecture 1:*** The topologic problem of a 3D Ising system can be solved by introducing an additional rotation in a four-dimensional (4D) space, since the knots in a 3D space can be opened by a rotation in a 4D space. One can find a spin representation in $2^{n \cdot l \cdot o}$-space for this additional rotation in 2n·l·o-space with o = $(n \cdot l)^{1/2}$. Meanwhile, the matrices $\mathbf{V_1}$, $\mathbf{V_2}$ and $\mathbf{V_3}$ have to be represented and rearranged, also in the 2n·l·o-space.

This additional rotation in the 2n·l·o-space appears in **V** as an additional matrix $\mathbf{V'_4}$:

$$V'_4 = \prod_{t=1}^{n \cdot l \cdot o - 1} \exp\{-iK''' P_{t+1} Q_t\} \cdot \exp\{iK''' P_1 Q_{n \cdot l \cdot o} U\} . \tag{3.5}$$

with:

$$K''' = \frac{K'K''}{K} \quad \text{for } K \neq 0, \tag{3.6}$$

considering the symmetry of the system and the topologic problem we deal with. The form of $\frac{K'K''}{K}$ stands for the crossings and/or knots in the 3D Ising model. The introduction of the additional dimension can be treated as the introduction of a boundary condition in order to deal with the topologic problem and the non-local behavior in the 3D physical system. This procedure is in a certain sense similar to the introduction of the well – known Born-Kármán periodic boundary condition for dealing with the energy band of infinite crystal with free boundary in the solid - state physics. In the thermodynamic limit (N → ∞), that periodic boundary condition is equalized mathematically to the free boundary. In our case, the special boundary condition attached by introducing the additional rotation would be equalized mathematically to the free boundary of the 3D model in the thermodynamic limit. In mathematic, there are many similar techniques, such as, first to rent something and then to pay it back. The 3D Ising model have to be dealt with within a (3 + 1) – dimensional frame because of the well – known topologic problem of the 3D model. The introduction of an additional rotation in an additional dimension serves to take into account properly an important hidden intrinsic property, i.e., the topologic knots of interacting spins and thus their non-local behaviors, of the 3D model.

It is important to ensure that K''' must be not larger than K' or K'', since the additional rotation is performed in a curled-up dimension. It would be unreasonable if the strength of the rotation or interaction in the curled-up dimension were larger than that in the normal dimensions. This verifies that one would have to take the crystallographic axis with largest exchange interaction as the standard axis for initiating the procedure. Namely, the conditions of K ≥ K' and K ≥ K'' should be valid, as we start our procedure by defining K* by $e^{-2K} \equiv \tanh K^*$. Meanwhile, the matrices $\mathbf{V_1}$, $\mathbf{V_2}$ and $\mathbf{V_3}$ could be represented and rearranged also in the 2n·l·o-space as:

$$V'_3 = \prod_{t=1}^{n\cdot l\cdot o-1} \exp\{-iK'' P_{t+1} Q_t\} \cdot \exp\{iK'' P_1 Q_{n\cdot l\cdot o} U\}; \tag{3.7a}$$

$$V'_2 = \prod_{t=1}^{n\cdot l\cdot o-1} \exp\{-iK' P_{t+1} Q_t\} \cdot \exp\{iK' P_1 Q_{n\cdot l\cdot o} U\}; \tag{3.7b}$$

$$V'_1 = \prod_{t=1}^{n\cdot l\cdot o} \exp\{iK^* \cdot P_t Q_t\}. \tag{3.7c}$$

It is clear that the additional rotation in the 2n·l·o-space (i.e., $2(n\cdot l)^{3/2}$ because of o = $(n\cdot l)^{1/2}$) with the spin representation in the $2^{n\cdot l\cdot o}$ (i.e., $2^{(n\cdot l)^{3/2}}$)-space extends the original rotations in the 2n·l-space with the spin representations in the $2^{n \cdot l}$-space to be ones in the 2n·l·o-space with the spin representations in the $2^{n \cdot l \cdot o}$-space. The operators of the 3D Ising lattice generate a very large Lie algebra, so large in fact that it cannot be dealt with in the frame of the spin representations in the $2^{n \cdot l}$-space, but only in the $2^{n \cdot l \cdot o}$-space. The three groups of $2^{n \cdot l} \times 2^{n \cdot l}$ matrices $\mathbf{C}_{\alpha\beta}$, $\mathbf{s'}_{\alpha\beta}$, $\mathbf{s''}_{\alpha\beta}$ ($\alpha$ = 1,2,…n; $\beta$ = 1,2,…l) in Eq. (2.8) are extended to be:

$$C_t = iP_t Q_t = 1 \times 1 \times \cdots \times C \times 1 \times \cdots, \tag{3.8a}$$

$$s'_t = C_1 C_2 \cdots C_{t-1} P_t = 1 \times 1 \times \cdots \times s \times 1 \times \cdots, \tag{3.8b}$$

$$s''_t = C_1 C_2 \cdots C_{t-1} P_t = 1 \times 1 \times \cdots \times isC \times 1 \times \cdots, \tag{3.8c}$$

Here **C**, **s** and **isC** are located on the position of the $\eta$-th factor ($\eta$ = 1, 2, …n·l·o) of the $2^{n \cdot l \cdot o} \times 2^{n \cdot l \cdot o}$ (i.e., $2^{(n \cdot l)^{3/2}} \times 2^{(n \cdot l)^{3/2}}$ because of o = $(n \cdot l)^{1/2}$) spin matrices. By doing so, we have performed already the transformations of n → $n^{3/2}$ and l → $l^{3/2}$ on the lattice.

The problem at hand is to evaluate the eigenvalues of the new matrices $V' \equiv V'_4 \cdot V'_3 \cdot V'_2 \cdot V'_1$. The new matrices **V'** are much larger than the original one **V**, with an additional energy of interaction $J''' = \frac{J' J''}{J}$ along an additional curled-up dimension. It is noticed that as J''' → 0 (i.e., one of J' and J'' approaches zero) the model turns automatically back to the 2D one. The appearance of $\mathbf{V'_4}$ in the **V'** and the extension of the space for the spin representations of $\mathbf{V_1}$, $\mathbf{V_2}$ and $\mathbf{V_3}$ should not change the values for the maximal eigenvalues of the matrix **V**. However, it indeed over-accounts the total free energy of the system. In order to compensate such over-accounting, one needs to introduce another new conjecture, *Conjecture 2* (as will be proposed below), of the weight factors on the eigenvectors. Furthermore, when it is necessary, one could perform the transformations of n → $n^{2/3}$ and l → $l^{2/3}$ to transform the $2^{n\cdot l\cdot o} \times 2^{n\cdot l\cdot o}$ (i.e., $2^{(n\cdot l)^{3/2}} \times 2^{(n\cdot l)^{3/2}}$) spin matrices back to the $2^{n\cdot l} \times 2^{n\cdot l}$ ones. The addition of this new matrix **V'₄** is very important to overcome the difficulty of dealing with the 3D Ising models, because only can its corresponding rotations in a larger dimensional space remove the topological problem of the 3D Ising system and take into account the non – local property in the Hamiltonian. It is understood that the matrix $\mathbf{V'_4}$ is actually acting as a bridge connecting the roads toward solving the 3D Ising problem. This $\mathbf{V'_4}$ must vanish if one reduces the dimension of the system to be two, since no such topological difficulty occurs in that case. However, the detailed action of the matrix $\mathbf{V_4'}$ would lead to a 3D to 2D crossover phenomenon. The additional matrix $\mathbf{V_4'}$ with K''' is attached directly on the matrix **V** to arrange it in higher dimensional spaces. The eigenvalues before/after such attachment should be equalized, because the 3D Ising model has to be set up within the (3 + 1) – dimensional framework (as a boundary condition), which might be also due to that we are actually living in the four non-compact dimensions. If one of K, K' and K'' vanished, both the models would immediately return to the 2D one. If two of K, K' and K'' vanished, both the models would immediately turn back to the 1D one. In the next sub-section, we shall try to find the eigenvalues and eigenvectors of the new matrix **V'**.

### B. Eigenvalues and eigenvectors of the matrix V'

Following the finding in Kaufman's paper,[17] one could treat the last "boundary " factor and select the eigenvalues in two subspaces similarly. The complete partition function for the lattice could be written down as:

$$
\begin{aligned}
Z &= (2\sinh 2K)^{mnl/2} \cdot \sum_{i=1}^{2^{nl}} \lambda_i^m \\
&= (2\sinh 2K)^{mnl/2} \cdot \{\sum \exp[\frac{m}{2}(\pm\gamma_2 \pm \gamma_4 \pm \cdots)] + \sum \exp[\frac{m}{2}(\pm\gamma_1 \pm \gamma_3 \pm \cdots)]\}
\end{aligned}
\tag{3.9}
$$

For eigenvalues and eigenvectors of the matrix V⁻, we could have:

$$V_0^- \equiv \prod_{t=1}^{n\cdot l\cdot o} \exp(\frac{i}{2} K^* P_t Q_t) \cdot \prod_{t=1}^{n\cdot l\cdot o} \exp(-iK' P_{t+1} Q_t) \cdot \prod_{t=1}^{n\cdot l\cdot o} \exp(-iK'' P_{t+1} Q_t)$$
$$\cdot \prod_{t=1}^{n\cdot l\cdot o} \exp(-iK''' P_{t+1} Q_t) \cdot \prod_{t=1}^{n\cdot l\cdot o} \exp(\frac{i}{2} K^* P_t Q_t) \equiv S(R_0^-) \quad . \qquad (3.10)$$

The first (and last) product represents the rotation:

$$\begin{bmatrix} \cosh K^* & i\sinh K^* & & & & & \\ -i\sinh K^* & \cosh K^* & & & & & \\ & & \cosh K^* & i\sinh K^* & & & \\ & & -i\sinh K^* & \cosh K^* & & & \\ & & & & \cdot & & \\ & & & & & \cdot & \\ & & & & & & \cdot \end{bmatrix} . \qquad (3.11)$$

The middle three products have the same form:

$$\begin{bmatrix} \cosh 2K_i & & & & & & & & -i\sinh 2K_i \\ & \cosh 2K_i & i\sinh 2K_i & & & & & & \\ & -i\sinh 2K_i & \cosh 2K_i & & & & & & \\ & & & \cosh 2K_i & i\sinh 2K_i & & & & \\ & & & -i\sinh 2K_i & \cosh 2K_i & & & & \\ & & & & & \cdot & & & \\ & & & & & & \cdot & & \\ & & & & & & & \cdot & \\ i\sinh 2K_i & & & & & & & & \cosh 2K_i \end{bmatrix},$$

(3.12)

but with different quantities $K_i$ (i = 1, 2, 3, for K', K'' and K'''). Compared with those in Kaufman's procedure,[17] only are differences in our procedure the appearances of two additional middle products with K'' (because of the third dimension) and K''' (because of the introduction of the fourth curled-up dimension). However, the dimension of each matrix in the present case becomes 2n·l·o, instead of 2n in the 2D case.

$R_0^-$ could be written schematically as:

$$R_0^- = \begin{bmatrix} a & b & 0 & 0 & \cdot & \cdot & \cdot & 0 & 0 & b^* \\ b^* & a & b & 0 & 0 & \cdot & \cdot & 0 & 0 & 0 \\ 0 & b^* & a & b & 0 & \cdot & \cdot & \cdot & \cdot & \cdot \\ \cdot & \cdot & \cdot & \cdot & \cdot & \cdot & \cdot & \cdot & \cdot & \cdot \\ b & 0 & \cdot & \cdot & \cdot & \cdot & \cdot & 0 & b^* & a \end{bmatrix}, \tag{3.13}$$

where

$$a = \begin{pmatrix} \cosh 2(K'+K''+K''')\cdot\cosh 2K^* & -i\cosh 2(K'+K''+K''')\cdot\sinh 2K^* \\ i\cosh 2(K'+K''+K''')\cdot\sinh 2K^* & \cosh 2(K'+K''+K''')\cdot\cosh 2K^* \end{pmatrix}, \tag{3.14a}$$

$$b = \begin{pmatrix} -\frac{1}{2}\sinh 2(K'+K''+K''')\cdot\sinh 2K^* & i\sinh 2(K'+K''+K''')\cdot\sinh^2 K^* \\ -i\sinh 2(K'+K''+K''')\cdot\cosh^2 K^* & -\frac{1}{2}\sinh 2(K'+K''+K''')\cdot\sinh 2K^* \end{pmatrix}, \tag{3.14b}$$

$$b^* = \begin{pmatrix} -\frac{1}{2}\sinh 2(K'+K''+K''')\cdot\sinh 2K^* & i\sinh 2(K'+K''+K''')\cdot\cosh^2 K^* \\ -i\sinh 2(K'+K''+K''')\cdot\sinh^2 K^* & -\frac{1}{2}\sinh 2(K'+K''+K''')\cdot\sinh 2K^* \end{pmatrix}. \tag{3.14c}$$

The eigenvectors are:

$$\frac{1}{(2n\cdot l\cdot o)^{\frac{1}{2}}} \begin{bmatrix} \varepsilon^{2t}\cdot W_{2t} \\ \varepsilon^{4t}\cdot W_{2t} \\ \cdot \\ \cdot \\ \cdot \\ \cdot \\ \varepsilon^{2(n\cdot l\cdot o)t}\cdot W_{2t} \end{bmatrix}$$

where $W_{2t}$ is an eigenvector of the 2-dimensional matrix $\alpha_{2t}$. Therefore, the 2n·l·o-eigenvalues of $\mathbf{R_0}^-$ are the eigenvalues of the n·l·o 2-dimensional matrices:

$$\alpha_{2t} = a + \varepsilon^{2t} \cdot b + \varepsilon^{2(nlo-1)t} \cdot b^* = a + \varepsilon^{2t} \cdot b + \varepsilon^{-2t} \cdot b^* . \tag{3.15}$$

where

$$\varepsilon^{2t} = w_x e^{i\frac{\pi}{n}} + w_y e^{i\frac{\pi}{l}} + w_z e^{i\frac{\pi}{o}} . \tag{3.16a}$$

and

$$\varepsilon^{2(t_{x,y,z})t} = w_x e^{i\frac{t_x \pi}{n}} + w_y e^{i\frac{t_y \pi}{l}} + w_z e^{i\frac{t_z \pi}{o}} . \tag{3.16b}$$

At this step, we need to introduce our second conjecture.

***Conjecture 2:*** The weight factors $w_x$, $w_y$ and $w_z$, varying in range of [-1, 1], on the eigenvectors represent the contribution of $e^{i\frac{t_x \pi}{n}}$, $e^{i\frac{t_y \pi}{l}}$ and $e^{i\frac{t_z \pi}{o}}$ in the 4D space to the energy spectrum of the system.

By introducing the *Conjecture 1,* the 3D physical system is embedded in the (3 + 1) – dimensional space, with the maximal eigenvalues the same as those of the original 3D model. By introducing the *Conjecture 2,* the over-accounting of the total free energy of the (3 + 1) – dimensional model is compensated by the weight factors, in order to unchange the total free energy.

The determinant of this matrix is +1. Its eigenvalues could be written as exp(± $\gamma_{2t}$), and $\gamma_{2t}$ could be determined by:

$$\begin{aligned} &\frac{1}{2} trace(\alpha_{2t}) = \frac{1}{2}(e^{\gamma_{2t}} + e^{-\gamma_{2t}}) = \cosh \gamma_{2t} = \cosh \gamma_{2t_x, 2t_y, 2t_z} \\ &= \cosh 2K^* \cdot \cosh 2(K' + K'' + K''') - \sinh 2K^* \cdot \sinh 2(K' + K'' + K''') \\ &\times (w_x \cos(2t_x \pi / n) + w_y \cos(2t_y \pi / l) + w_z \cos(2t_z \pi / o)) \end{aligned} \tag{3.17}$$

Here t stands for variation of $t_x$, $t_y$ and $t_z$. Similar to the 2D Ising case, $\gamma_{2t}$ is geometrically the third side of a hyperbolic triangle, whose other two sides, 2(K' + K'' + K''') and 2K*. The angle between the two sides 2(K' + K'' + K''') and 2K* is determined by the combinational effects of three angles $\omega_{2t_x} = \frac{2t_x\pi}{n}$, $\omega_{2t_y} = \frac{2t_y\pi}{l}$, and $\omega_{2t_z} = \frac{2t_z\pi}{o}$. The situation here is similar to the band structure of a three-dimensional material, which is determined by three wave-vectors $k_x$, $k_y$ and $k_z$ along three crystallographic axes. The effects of the three wave-vectors altogether contribute to the band structure of a 3D crystal. In the present case, all the three angles $\omega_{2t_x}$, $\omega_{2t_y}$, and $\omega_{2t_z}$ contribute the angle between the two sides 2(K' + K'' + K''') and 2K*, however, with different weights $w_x$, $w_y$ and $w_z$. For the 2D Ising model, only one angle $\omega_{2t}$ exists, as constructed by the two-dimensional coordinates. One could find easily the analogy of only one angle $\omega_{2t}$ in the 2D Ising system to the energy band of a 1D spin chain. For the 3D Ising model, the three angles $\omega_{2t_x}$, $\omega_{2t_y}$, and $\omega_{2t_z}$ exist as constructed by the three-dimensional coordinates and the extra-dimensional coordinate. The analogy of the three angles $\omega_{2t_x}$, $\omega_{2t_y}$, and $\omega_{2t_z}$ in the 3D Ising system to the energy band of a 3D crystal is realized only by introducing the extra-dimensional coordinate. One has to notice that the unit of the dimensions changes, due to such introduction of the extra-dimensional coordinate. The effects of $e^{i\frac{t_x\pi}{l}}$, $e^{i\frac{t_y\pi}{l}}$ and $e^{i\frac{t_z\pi}{o}}$ with their weights $w_x$, $w_y$ and $w_z$ will be discussed in details in Appendixes.

Introducing the angle $\delta'_{2t}$ between 2K* and $\gamma_{2t}$ simplifies the matrix $\alpha_{2t}$. Since the procedure is similar to the 2D Ising case, we quote other relations:

$$\begin{aligned} \sinh\gamma_{2t}\cdot\cos\delta_{2t}' &= \sinh 2K^*\cdot\cosh 2(K'+K''+K''') \\ &-\cosh 2K^*\cdot\sinh 2(K'+K''+K''')\times(w_x\cos\omega_{2t_x}+w_y\cos\omega_{2t_y}+w_z\cos\omega_{2t_z}) \end{aligned} \tag{3.18}$$

$$\sinh\gamma_{2t}\cdot\sin\delta_{2t}'=\sinh 2(K'+K''+K''')\cdot(w_x\sin\omega_{2t_x}+w_y\sin\omega_{2t_y}+w_z\sin\omega_{2t_z}). \quad (3.19)$$

Therefore, the matrix $\boldsymbol{\alpha_{2t}}$ could be reduced to:

$$\begin{aligned}\alpha_{2t}&=\cosh\gamma_{2t}\cdot\begin{pmatrix}1&0\\0&1\end{pmatrix}+\sinh\gamma_{2t}\cdot\begin{pmatrix}0&\sin\delta_{2t}'-i\cos\delta_{2t}'\\\sin\delta_{2t}'+i\cos\delta_{2t}'&0\end{pmatrix}\\&=\cosh\gamma_{2t}-i\sinh\gamma_{2t}\cdot\begin{pmatrix}0&e^{i\delta_{2t}'}\\-e^{-i\delta_{2t}'}&0\end{pmatrix}\end{aligned}. \quad (3.20)$$

The normalized eigenvectors of $\boldsymbol{\alpha_{2t}}$ are:

$$\frac{1}{\sqrt{2}}\begin{pmatrix}e^{\frac{i}{2}\delta_{2t}'}\\ie^{-\frac{i}{2}\delta_{2t}'}\end{pmatrix},\ \frac{1}{\sqrt{2}}\begin{pmatrix}ie^{\frac{i}{2}\delta_{2t}'}\\e^{-\frac{i}{2}\delta_{2t}'}\end{pmatrix},$$

corresponding to the eigenvalues $\exp(\gamma_{2t})$, $\exp(-\gamma_{2t})$, respectively. The 2n·l·o-normalized eigenvectors of $\mathbf{R_0}^-$ would behave as:

$$u_{2t}\equiv\frac{1}{(2n\cdot l\cdot o)^{\frac{1}{2}}}\begin{bmatrix}(w_xe^{i\omega_{2tx}}+w_ye^{i\omega_{2ty}}+w_ze^{i\omega_{2tz}})\cdot e^{\frac{i}{2}\delta_{2t}'}\\i(w_xe^{i\omega_{2tx}}+w_ye^{i\omega_{2ty}}+w_ze^{i\omega_{2tz}})\cdot e^{-\frac{i}{2}\delta_{2t}'}\\(w_xe^{i\omega_{4tx}}+w_ye^{i\omega_{2ty}}+w_ze^{i\omega_{2tz}})\cdot e^{\frac{i}{2}\delta_{2t}'}\\i(w_xe^{i\omega_{4tx}}+w_ye^{i\omega_{2ty}}+w_ze^{i\omega_{2tz}})\cdot e^{-\frac{i}{2}\delta_{2t}'}\\\cdot\\\cdot\\\cdot\\i(w_xe^{i\omega_{2ntx}}+w_ye^{i\omega_{2lty}}+w_ze^{i\omega_{2otz}})\cdot e^{-\frac{i}{2}\delta_{2t}'}\end{bmatrix}, \quad (3.21a)$$

and

$$v_{2t} \equiv \frac{1}{(2n \cdot l \cdot o)^{\frac{1}{2}}} \begin{bmatrix} i(w_x e^{i\omega_{2t_x}} + w_y e^{i\omega_{2t_y}} + w_z e^{i\omega_{2t_z}}) \cdot e^{\frac{i}{2}\delta_{2t}'} \\ (w_x e^{i\omega_{2t_x}} + w_y e^{i\omega_{2t_y}} + w_z e^{i\omega_{2t_z}}) \cdot e^{-\frac{i}{2}\delta_{2t}'} \\ i(w_x e^{i\omega_{4t_x}} + w_y e^{i\omega_{2t_y}} + w_z e^{i\omega_{2t_z}}) \cdot e^{\frac{i}{2}\delta_{2t}'} \\ (w_x e^{i\omega_{4t_x}} + w_y e^{i\omega_{2t_y}} + w_z e^{i\omega_{2t_z}}) \cdot e^{-\frac{i}{2}\delta_{2t}'} \\ \cdot \\ \cdot \\ \cdot \\ (w_x e^{i\omega_{2nt_x}} + w_y e^{i\omega_{2lt_y}} + w_z e^{i\omega_{2ot_z}}) \cdot e^{-\frac{i}{2}\delta_{2t}'} \end{bmatrix}. \qquad (3.21b)$$

The formation of the 2n·l·o-normalized eigenvectors above is in a sense analogous to the construction of a quaternion. As all physicists know, a complex number of the form z = x + yi, where $i = \sqrt{-1}$, can be represented by the point (x, y) on a Cartesian plane. Conversely, any point on the plane can be represented by a complex number. A quaternion is a 4D complex number in the form of q = *w* + *x*i + *y*j + *z*k, where i, j and k are all different square roots of –1. The quaternion can be regarded as an object composed of a scalar part, a real number *w*, and a 3D vector part, *x*i + *y*j + *z*k. For the procedure of solving the Ising models, as shown in Sec. II as example, wrapping our crystal on cylinder simplifies greatly the calculations. In the 2D Ising case,[17] this process led to the eigenvectors in the form of a 1D vector, as the wrapped dimension was treated as the form of a scalar. In the 3D Ising case, as shown in Eqs. (2.5) and (2.6), wrapping our crystal on cylinder again treats one of the three original dimensions as the form of a scalar, which does not contribute to the eigenvectors. Only when the additional fourth curled – up dimension is introduced and taken into account, the eigenvectors in the form of a 3D vector can be constructed successfully.

Similar to what Kaufman did,[17] let the matrix of these eigenvectors be denoted by **t** = **t₁** · **t₂** ·**t₃**, so that

$$t \cdot R^{-} \cdot t^{-1} = \lambda^{-1}, \qquad (3.22)$$

where $\lambda^{-}$ is the diagonal form of $\mathbf{R_0}^{-}$. Neither $\lambda^{-}$ nor **t** are orthogonal, which cannot be represented in the spin space. Thus, one needs to apply the transformation **I** to both sides of (3.22):

$$(It)\cdot R^{-}\cdot(t^{-1}I^{-1})\equiv T^{1}\cdot R^{-}\cdot T^{-1}=I\cdot\lambda^{-1}\cdot I^{-1}\equiv \mathrm{K}\,. \tag{3.23}$$

where **I** is so chosen that it brings $\lambda^{-}$ into its canonical form, and meanwhile, makes **T** = **I** · **t** orthogonal. The spin-representative of the canonical form **K** could be given by:

$$S(\mathrm{K})=\prod_{t=1}^{n\cdot l\cdot o}S(\mathrm{K}_t)=\prod_{t=1}^{n\cdot l\cdot o}\exp\{\frac{\theta_t}{2}\cdot\Gamma_{t1}\Gamma_{t2}\}\,, \tag{3.24}$$

Since **S(T)** is a complicated matrix we do not know the spin-representative of **T** explicitly. However, we must ensure that **T** is orthogonal, and does possess a spin-representative. The transformation **T** could be given by:

**T**:

$$P_{a,b,c}\rightarrow\sum_{t_x=1}^{n}\sum_{t_y=1}^{l}\sum_{t_z=1}^{o}\sigma_{t_x a,t_y b,t_z c}P_{t_x,t_y,t_z}+\sum_{t_x=1}^{n}\sum_{t_y=1}^{l}\sum_{t_z=1}^{o}\tau_{t_x a,t_y b,t_z c}Q_{t_x,t_y,t_z}\,, \tag{3.25}$$

$$Q_{a,b,c}\rightarrow\sum_{t_x=1}^{n}\sum_{t_y=1}^{l}\sum_{t_z=1}^{o}\sigma_{t_x a,t_y b,t_z c}'P_{t_x,t_y,t_z}+\sum_{t_x=1}^{n}\sum_{t_y=1}^{l}\sum_{t_z=1}^{o}\tau_{t_x a,t_y b,t_z c}'Q_{t_x,t_y,t_z}\,, \tag{3.26}$$

where

$$\sigma_{t_x a,t_y b,t_z c}=\frac{1}{(nlo)^{\frac{1}{2}}}\times \left\{w_x\cos\left[\frac{2t_x a\pi}{n}+\frac{\delta_{2t}'}{2}\right]+w_y\cos\left[\frac{2t_y b\pi}{l}+\frac{\delta_{2t}'}{2}\right]+w_z\cos\left[\frac{2t_z c\pi}{o}+\frac{\delta_{2t}'}{2}\right]\right\} \tag{3.27a}$$

$$\tau_{t_x a, t_y b, t_z c} = \frac{-1}{(nlo)^{\frac{1}{2}}} \times$$
$$\left\{ w_x \sin\left[\frac{2t_x a\pi}{n} + \frac{\delta_{2t}'}{2}\right] + w_y \sin\left[\frac{2t_y b\pi}{l} + \frac{\delta_{2t}'}{2}\right] + w_z \sin\left[\frac{2t_z c\pi}{o} + \frac{\delta_{2t}'}{2}\right]\right\} \tag{3.27b}$$

$$\sigma_{t_x a, t_y b, t_z c}' = \frac{1}{(nlo)^{\frac{1}{2}}} \times$$
$$\left\{ w_x \sin\left[\frac{2t_x a\pi}{n} - \frac{\delta_{2t}'}{2}\right] + w_y \sin\left[\frac{2t_y b\pi}{l} - \frac{\delta_{2t}'}{2}\right] + w_z \sin\left[\frac{2t_z c\pi}{o} - \frac{\delta_{2t}'}{2}\right]\right\} \tag{3.27c}$$

$$\tau_{t_x a, t_y b, t_z c}' = \frac{-1}{(nlo)^{\frac{1}{2}}} \times$$
$$\left\{ w_x \cos\left[\frac{2t_x a\pi}{n} - \frac{\delta_{2t}'}{2}\right] + w_y \cos\left[\frac{2t_y b\pi}{l} - \frac{\delta_{2t}'}{2}\right] + w_z \cos\left[\frac{2t_z c\pi}{o} - \frac{\delta_{2t}'}{2}\right]\right\} \tag{3.27d}$$

**T** is now orthogonal and possesses a spin-representative, so that the equation:

$$T \cdot (R^{-}) \cdot T^{-1} = \mathrm{K}, \tag{3.28}$$

yields:

$$S(T) \cdot (V_0^{-}) \cdot S(T)^{-1} = S(\mathrm{K}) = \prod_{t=1}^{n \cdot l \cdot o} \exp[\frac{i}{2} \gamma_t P_t Q_t]. \tag{3.29}$$

But **S(K)** is still not diagonal, because our coordinate system $\mathbf{iP_tQ_t} = \mathbf{C_t}$. In order to diagonalize **S(K),** we would have to use the transformation

$$g = 2^{nlo/2} \cdot (C+s) \times (C+s) \times \cdots \times (C+s) = g^{-1}, \tag{3.30}$$

$$gC_t g = s_t, \tag{3.31}$$

which is not the spin-representative of any rotation. Then we would find:

$$g \cdot S(T) \cdot (V_0^-) \cdot S(T)^{-1} \cdot g = g \cdot S(\mathrm{K}) \cdot g = \Lambda^- . \tag{3.32}$$

Furthermore, we would have:

$$V_0 = V^{-\frac{1}{2}} \cdot V^- \cdot V^{\frac{1}{2}} \equiv S(H) \cdot (V^-) \cdot S(H)^{-1} . \tag{3.33}$$

Therefore,

$$g \cdot S(TH) \cdot (V^-) \cdot S(TH)^{-1} \cdot g \equiv \Psi_- \cdot (V^-) \cdot \Psi_-^{-1} = \Lambda^- , \tag{3.34}$$

with

$$\Psi_- = g \cdot S(TH) . \tag{3.35}$$

Here **H** stands for the rotation represented by $\mathbf{V_1}^{-1/2}$, i.e., the reciprocal of the rotation in (3.11):

**H**:

$$P_t \rightarrow \cosh K^* \cdot P_t - i \sinh K^* \cdot Q_t , \tag{3.36a}$$

$$Q_t \rightarrow i \sinh K^* \cdot P_t + \cosh K^* \cdot Q_t . \tag{3.36b}$$

Again, similar to the 2D case,[17] it is not feasible to write down explicitly the components of **Ψ-** because of the complexity of **S(T)**. On the other hand, similarly, one could easily evaluate the eigenvalues and eigenvectors of $V^+ = S(R^+)$.[17]

The partition function of the simple orthorhombic lattices could be expressed as

$$\begin{aligned} N^{-1} \ln Z = \ln 2 + \frac{1}{2(2\pi)^4} \int_{-\pi}^{\pi}\int_{-\pi}^{\pi}\int_{-\pi}^{\pi}\int_{-\pi}^{\pi} \ln\big[\cosh 2K \cosh 2(K'+K''+K''') - \sinh 2K \cos \omega' \\ - \sinh 2(K'+K''+K''')(w_x \cos \omega_x + w_y \cos \omega_y + w_z \cos \omega_z)\big] d\omega' d\omega_x d\omega_y d\omega_z \end{aligned} .$$

(3.37)

In accordance with details of the weights $w_y$ and $w_z$ revealed in Appendixes, the putative exact solution for the partition function of the 3D simple orthorhombic (and simple cubic) Ising lattices could be fitted to the high temperature series expansion at/near infinite temperature.[80,93,107] Eq. (3. 37) contains yet – to – be determined coefficients, i.e., three weights, which are in form of series. However, as shown in Appendixes, all of the series can be represented or curled inside the square form, with very regular laws, such as all high - order terms are regularly negative for i ≥ 1.

Because $w_y$ and $w_z$ become zero for finite temperatures, one could immediately obtain:

$$\cosh \gamma_0 = \cosh 2(K^* - K' - K'' - K''') , \quad (3.38)$$

from which the critical point is determined by $\gamma_0 = 0$, i.e.,

$$K^* = K' + K'' + K''' . \quad (3.39)$$

Namely, from Eq. (3.6), one would have:

$$KK^* = KK' + KK'' + K'K'' . \quad (3.40)$$

The following relations could also derive the critical point of the simple orthorhombic lattice Ising system:

$$\sinh 2K \cdot \sinh 2(K' + K'' + K''') = 1 . \quad (3.41)$$

or

$$\tanh^{-1} e^{-2K} = K' + K'' + K''' . \quad (3.42)$$

These formulae would be the same as those of the 2D rectangular Ising lattices if one of K' and K'' equaled to zero, the 1D Ising one if both K' and K'' equaled to zero, or the simple cubic Ising one if K = K' = K''.

The partition function (3.37) yields directly the free energy F of the crystal, from which the internal energy U and the specific heat C are derived by differentiation with regard to the temperature T. For a crystal of N atoms, we would have the expressions:

$$\begin{aligned} F &= U - TS = -Nk_B T \log\lambda, \\ U &= F - T\frac{dF}{dT} = Nk_B T^2 \frac{d(\log\lambda)}{dT}, \\ C &= \frac{dU}{dT}. \end{aligned} \tag{3.43}$$

For the present 3D system, it is convenient to evaluate the internal energy U and the specific heat C by adopting notations $K = J/k_B T$ and $\widetilde{K} = (J' + J'' + J'J''/J)/k_B T$:

$$\begin{aligned} U &= -NJ\frac{\partial \log\lambda}{\partial K} - N(J'+J''+\frac{J'J''}{J})\frac{\partial \log\lambda}{\partial \widetilde{K}} \\ &= -Nk_B T\left[H\frac{\partial \log\lambda}{\partial K} + \widetilde{K}\frac{\partial \log\lambda}{\partial \widetilde{K}}\right] \end{aligned}, \tag{3.44}$$

$$C = Nk_B\left[K^2\frac{\partial^2 \log\lambda}{\partial K^2} + 2K\widetilde{K}\frac{\partial^2 \log\lambda}{\partial K\partial \widetilde{K}} + \widetilde{K}^2\frac{\partial^2 \log\lambda}{\partial \widetilde{K}^2}\right]. \tag{3.45}$$

The discussion carried by Onsager[13] for the energy and the specific heat of the rectangular Ising lattice could be easily extended to the simple orthorhombic Ising lattices. However, in the present case, two terms of (3.44) are not separated since both K and $\widetilde{K}$ are related with J. Nevertheless, we could discuss the problem, using $\widetilde{K} = (J' + J'' + J'J''/J)/k_B T$ in the 3D instead of $K' = J'/k_B T$ in the 2D. The following formulas could be derived:

$$\frac{dK^*}{dK} = -\sinh 2K^* = -\frac{1}{\sinh 2(K'+K''+K''')}, \tag{3.46}$$

$$\begin{aligned}
\frac{\partial \gamma}{\partial \widetilde{K}} &= 2\cos\delta^*, \\
\frac{\partial \gamma}{\partial K^*} &= 2\cos\delta', \\
\frac{\partial^2 \gamma}{\partial \widetilde{K}^2} &= 4\sin^2\delta^*\coth\gamma, \\
\frac{\partial^2 \gamma}{\partial K^{*2}} &= 4\sin^2\delta'\coth\gamma, \\
\frac{\partial^2 \gamma}{\partial \widetilde{K}\partial K^*} &= -\frac{4\sin\delta^*\sin\delta'}{\sinh\gamma}.
\end{aligned} \tag{3.47}$$

We would have:

$$\begin{aligned}
\frac{\partial \log\lambda}{\partial \widetilde{K}} &= \int_0^\pi \cos\delta^* d\omega/\pi, \\
\frac{\partial \log\lambda}{\partial K} &= \cosh 2K^* - \sinh 2K^* \int_0^\pi \cos\delta' d\omega/\pi,
\end{aligned} \tag{3.48}$$

and

$$\begin{aligned}
\frac{\partial^2 \log\lambda}{\partial \widetilde{K}^2} &= \frac{2}{\pi}\int_0^\pi \sin^2\delta^*\coth\gamma d\omega, \\
\frac{\partial^2 \log\lambda}{\partial K\partial \widetilde{K}} &= 2\sinh 2K^* \int_0^\pi \frac{\sin\delta^*\sin\delta'}{\pi\sinh\gamma} d\omega, \\
\frac{\partial^2 \log\lambda}{\partial K^2} &= 2\sinh^2 2K^* \left(-1 + \coth 2K^* \int_0^\pi \cos\delta' d\omega/\pi + \int_0^\pi \sin^2\delta'\coth\gamma d\omega/\pi\right)
\end{aligned} \tag{3.49}$$

The integrals (3.48) are continuous functions of $\widetilde{K}$ and K (or K*) for all values of these parameters, even for $\widetilde{K}$ = K* (critical point), whereas the three integrals (3.49) are infinite at the critical point, otherwise finite. Fig. 1 shows the temperature dependence of the specific heat C for the 3D simple orthorhombic Ising lattices with K' = K'' = K, 0.5 K, 0.1 K and 0.0001 K. The critical point decreases with decreasing K' and K'', until the singularity disappears as no ordering occurs in the 1D system.

Clearly, if the conjectures were valid, the analytic nature of the singularity of the specific heat for the 3D simple orthorhombic Ising lattices would be the same as in the 2D Ising lattices.[13]

For a simple cubic Ising lattice, K' = K'' = K, resulting in K''' = K. The eigenvalues of the matrix **V** could be represented as:

$$\begin{aligned}\cosh\gamma_{2t} &= \cosh 2K^* \cdot \cosh 6K - \sinh 2K^* \cdot \sinh 6K \cdot \\ &\cdot (w_x \cos(2t_x\pi/n) + w_y \cos(2t_y\pi/l) + w_z \cos(2t_x\pi/o)) \\ &= \cosh 2K^* \cdot \cosh 6K - \sinh 2K^* \cdot \sinh 6K \cdot (w_x \cos\omega_{2t_x} + w_y \cos\omega_{2t_y} + w_z \cos\omega_{2t_z})\end{aligned} \quad (3.50)$$

and we would have:

$$\begin{aligned}&\sinh\gamma_{2t} \cos\delta_{2t}' = \sinh 2K^* \cosh 6K \\ &- \cosh 2K^* \sinh 6K (w_x \cos\omega_{2t_x} + w_y \cos\omega_{2t_y} + w_z \cos\omega_{2t_z})\end{aligned} \quad (3.51)$$

$$\sinh\gamma_{2t} \sin\delta_{2t}' = \sinh 6K (w_x \sin\omega_{2t_x} + w_y \sin\omega_{2t_y} + w_z \sin\omega_{2t_z}). \quad (3.52)$$

Fig. 2 gives the plots of $\gamma \sim K$ for different values of $\omega_{2t_x} = \pi$, $3\pi/4$, $\pi/2$, $\pi/4$ and 0, neglecting the effects of $\omega_{2t_y}$ and $\omega_{2t_z}$. The minimum of the $\gamma \sim K$ curve shifts toward small K range (i.e., high temperature range). The minimum of the $\gamma \sim K$ curve for $\omega_{2t_x} = \pi$ is located at $x_d = e^{-2K_d} = \frac{\sqrt{10}-1}{3} = 0.72075922.......$ However, the behavior of $\gamma_0$ for $\omega_{2t_x} = 0$ dominates most sensitively the behaviors of the physical quantities at the critical point $K_c$ of the phase transition, where $\gamma_0 = 0$. For finite temperatures, it is easy to reduce Eq. (3.50) into the following expression:

$$\cosh\gamma_0 = \cosh 2(K^* - 3K), \quad (3.53)$$

from which we could determine the critical point $x_c = e^{-2K_c} = \frac{\sqrt{5}-1}{2} = 0.61803398874989484820458683436563811......$ by $\gamma_0 = 0$, i.e., K* = 3K. The following formulas are held also for the critical point:

$$\sinh 2K_c = \frac{1}{2}, \tag{3.54}$$

$$\cosh 2K_c = \frac{\sqrt{5}}{2}, \tag{3.55}$$

$$K_c = 0.24060591...... \tag{3.56}$$

$$1/K_c = 4.15617384...... \tag{3.57}$$

The putative critical point of the 3D simple cubic Ising system is located at $x_c = e^{-2K_c} = \frac{\sqrt{5}-1}{2}$, exactly one of the golden solutions of the equation $x^2 + x - 1 = 0$. One could compare it with the critical point of the 2D square Ising system, which is located at $x_c = e^{-2K_c} = \sqrt{2}-1$, exactly one of the silver solutions of the equation $x^2 + 2x - 1 = 0$. One could compare also with the formulas of $\sinh 2K_c = 1$ and $\cosh 2K_c = \sqrt{2}$ for the critical point of the 2D square Ising system. The similarity between the exact solution for the critical points of the simple cubic and the square Ising lattices is seen more clearly when the golden and silver solutions are expressed as following continued fractions:

$$\frac{\sqrt{5}-1}{2} = \cfrac{1}{1+\cfrac{1}{1+\cfrac{1}{1+\cfrac{1}{1+\cdots}}}}, \tag{3.58a}$$

$$\frac{\sqrt{5}+1}{2}=1+\cfrac{1}{1+\cfrac{1}{1+\cfrac{1}{1+\cfrac{1}{1+\cdots}}}}, \tag{3.58b}$$

$$\sqrt{2}-1=\cfrac{1}{2+\cfrac{1}{2+\cfrac{1}{2+\cfrac{1}{2+\cdots}}}}, \tag{3.58c}$$

$$\sqrt{2}+1=2+\cfrac{1}{2+\cfrac{1}{2+\cfrac{1}{2+\cfrac{1}{2+\cdots}}}}. \tag{3.58d}$$

In addition, the golden and silver solutions can be also expressed in the infinite series of square roots:

$$\frac{\sqrt{5}\pm 1}{2}=\sqrt{1\pm\sqrt{1\pm\sqrt{1\pm\sqrt{1\pm\sqrt{1\pm\ldots}}}}}, \tag{3.59a}$$

$$\sqrt{2}\pm 1=\sqrt{1\pm 2\sqrt{1\pm 2\sqrt{1\pm 2\sqrt{1\pm 2\sqrt{1\pm\ldots}}}}}, \tag{3.59b}$$

These formulae in Eqs. (3.58) and (3.59) could be related with the conceptions of self-similarity and fractals.

The putative critical point of the simple cubic Ising system could be derived by the following relations:

$$\sinh 2K_c \cdot \sinh 6K_c = 1. \tag{3.60}$$

or

$$\tanh^{-1}\left(e^{-2K_c}\right) = 3K_c. \tag{3.61}$$

One may notice that these formulas are the same as those for the 2D asymmetric Ising lattice with K' = 3K. Although the solution of the golden ratio does exist also in this 2D Ising system, it can be eliminated by setting the larger one between K and K' as the starting standard axis (as discussed below in Sec. VIII).

The partition function of the simple cubic Ising model reads as

$$\begin{aligned} N^{-1}\ln Z = \ln 2 + \frac{1}{2(2\pi)^3}\int_{-\pi}^{\pi}\int_{-\pi}^{\pi}\int_{-\pi}^{\pi}\int_{-\pi}^{\pi} \ln[\cosh 2K \cosh 6K - \sinh 2K \cos\omega' \\ - \sinh 6K(w_x \cos\omega_x + w_y \cos\omega_y + w_z \cos\omega_z)] d\omega' d\omega_x d\omega_y d\omega_z \end{aligned} \tag{3.62}$$

Again, as revealed in Appendix A, at/near infinite temperature, the partition function (3.62) of the 3D simple cubic Ising lattices equals to the high - temperature series expansion.[80,93,107] This actually forms a closed – form solution, as long as *Ansatz 1* in Appendix A is true. This *Ansatz* is an uncertain section of the whole approach, but from regular tendency of the parameters $b_1$ - $b_{10}$, it should be true for all of high order terms $b_i$ (i > 11). It is believed that the *Ansatz* is true, although it has been not proved rigorously.

### C. Critial point

It is important to compare the putative exact solution for the critical point with the results of the previous approximation methods. We shall compare first with the data obtained in last six decades and then with those obtained most recently. It is understood that the exact value for $1/K_c$ should be lower than the values obtained by various approximation methods. The mean field theory yields $1/K_c = z$ (where z is the coordination number), which is correct only for $d \geq 4$.[212,226,227] For the 3D Ising model, the mean field value of $1/K_c = 6$ is the highest one among the approximation values, which is not quantitatively correct, since the mean field theory overestimates the critical point in every case of $d < 4$. Oguchi concluded that the range of the existence of the Curie point for the 3D ferromagnet is in $0.21 < K_c < 0.24$,[62-64] correspondingly, $4.7619 > 1/K_c > 4.16667$. Our putative exact solution $K_c = 0.24060591……$ (i.e., $1/K_c = 4.15617384……$) is exactly located at the upper border of $K_c$ (or the lower border of $1/K_c$) of Oguchi's estimation,[62-64] within the error of ~ 0.25 %. The value $K_c$ of the Bethe's first approximation is equal to 0.202 (i.e., $1/K_c = 4.939$).[74,228] The putative exact solution for $1/K_c$ is much smaller than that of the Bethe's first approximation.[74,228] The putative exact solution is also lower than Kikuchi's estimation $4.2221 < \tau_c < 4.6097$ and $\tau_t \approx 4.5810$ (where $\tau_c$ or $\tau_t$ is our $1/K_c$).[85] Actually, this solution is very close to the low limit of $1/K_c$ of Kikuchi's estimation,[85] within the error of 1.6%. Meanwhile, the solution $x_c = e^{-2K_c} = \frac{\sqrt{5}-1}{2} = 0.618033988......$ is much lower than the values obtained by various approximation methods, such as Wakefield's method, 0.641 (i.e., $1/K_c = 4.497$);[73,74] Bethe's first approximation, 0.667 (i.e., $1/K_c = 4.939$);[74,228] Bethe's second approximation, 0.656 (i.e., $1/K_c = 4.744$);[74,228] Kirkwood's method, 0.658 (i.e., $1/K_c = 4.778$); [74,229] and Burley's best known value, 0.642 (i.e., $1/K_c = 4.513$).[209] It has been well - known that on the one hand, the method with higher order approximations has a lower value for the critical temperature and the exact solution must have the lowest one and, on the other hand, the corrections of the higher order terms are much lower than those of the lower order terms, specially, the first leading term. The mean field theory can be treated as the zero order approximation. The correction of the Bethe's first approximation on the mean field value is evaluated by $\delta_1\left(\frac{1}{K_c}\right) = \frac{1}{K_c^{MF}} - \frac{1}{K_c^{Bethe-1}}$, which is about 1.061. The correction of the Bethe's second approximation on the value of the Bethe's first approximation is evaluated by $\delta_2\left(\frac{1}{K_c}\right) = \frac{1}{K_c^{Bethe-1}} - \frac{1}{K_c^{Bethe-2}}$, which is about 0.195. The value of $\delta_2\left(\frac{1}{K_c}\right)$ is about 18.38% of $\delta_1\left(\frac{1}{K_c}\right)$. It is reasonable to believe that this tendency is approximately held by all the high order terms, namely, all of $\delta_{i+1}\left(\frac{1}{K_c}\right)$ is about one order less of $\delta_i\left(\frac{1}{K_c}\right)$. Suppose the ratios between every neighboring terms are same in the Bethe's approximations, we obtain the value of $1/K_c$ = 4.700 for the upper limit of the critical point of the 3D simple cubic Ising model. However, the ratios may not be the same, but vary in a certain range. The evaluation of the lower limit gives us a criterion that the exact solution $1/K_c$ for the critical point of the 3D simple cubic Ising model must be not smaller than 3.878 (= 6 - $2\delta_1\left(\frac{1}{K_c}\right)$), because the sum of all the high order terms of the corrections must be not larger than the first correction, i.e., $\sum_{i=2}^{\infty}\delta_i\left(\frac{1}{K_c}\right) < \delta_1\left(\frac{1}{K_c}\right)$.

In 1985, Rosengren conjectured that the critical point of the symmetric, simple cubic Ising model is given by $v_c \equiv \tanh\left(\frac{J}{k_B T_c}\right) = \left(\sqrt{5}-2\right)\cos\left(\frac{\pi}{8}\right) = 0.218098372......$, i.e., $K_c$ = 0.22165863…, in consideration of a certain circumstance for the 2D case and possible generalizations of the combinatorial solution to three dimensions.[230] One could easily check from $x_c = e^{-2K_c} = \frac{\sqrt{5}-1}{2}$ that our putative solution gives $v_c \equiv \tanh K_c = \sqrt{5}-2$. Surprisingly, what we obtained for the critical point of the simple cubic Ising model is exactly the same as the first factor $\left(\sqrt{5}-2\right)$ in the Rosengren's conjecture.[230] As commented by Fisher,[231] it is not unfair to say that the basis for the Rosengren's guess remains somewhat obscure: Rosengren did sketch an argument suggesting that a relevant class of weighted lattice walks with no backsteps would yield the factor $\left(\sqrt{5}-2\right)$; but the second factor in the Rosengren's conjecture was then selected to match various critical point estimates based on series and Monte Carlo studies published in 1981-1984.[120,121,173, 232-234] The factor cos(π/8) introduced many confusion, which certainly misled the Fisher's efforts on the critical polynomial, who finally claimed that the critical points of the truly 3D models may not be the root of any polynomial.[231] Although the Rosengren's conjecture of $K_c$ = 0.22165863… is still in fair good agreement with the most recent estimates of the high-temperature series extrapolation,[113,116,118,121,122,213] Monte Carlo and renormalization group techniques,[161-163,168,170,172-175,177,180,181,186,213,235] there are strong theoretical arguments again it to be as the exact solution.[231] This, on the other hand, implies that the most recent estimates of the Monte Carlo and renormalization group techniques are not very close to the exact one, although they have been determined with a high precision.

Various new approximation methods have been developed to apply for studying the critical phenomena in different systems,[106,107,141-144,149,152-160,192-202,208-225] since the discovery of the renormalization group theory in 1971. In the last decades, further estimates of the critical point and the critical exponents have become available,[107-225] many being more precise, but it is hard to say more accurate. Two review articles, given respectively by Pelissetto and Vicari[154] and Binder and Luijten,[213] summarized recent results from the renormalization group theory and Monte Carlo simulations and other methods. As summarized in Table 2 of Binder and Luijten's review,[213] well established critical point,[113,116,118,121,122,161-163,168,170,172-175,177,180,181,186,235] occurs at $K_c$ = 0.221655(5), i.e., $1/K_c$ = 4.511505(5), in good agreements also with the results in various references.[119,164,167,191] This value of $1/K_c$ is slightly higher than our solution $1/K_c$ = 4.15617384…… as the approximation should be. It is well – known that in all the approximation methods, systematic errors are difficult to assess with confidence,[106,107,141-144,149,152-160,192-202,208-225] which might be the origin of such deviation. The reasons for the existences of systematic errors of the approximation methods will be discussed in details in section VIII.

In the following several paragraphs, we would like to compare the results of the critical points of the renormalization group theory and Monte Carlo simulations with the exact solutions of the 2D and 3D Ising models and even some possibly existing analytical solutions.

Before such comparison, it is interesting to have a look on the mathematical characters of the exact solutions of the 2D square and the 3D simple cubic Ising models. The exact solution of the 2D square Ising model is located exactly at $x_c = e^{-2K_c} = \sqrt{2} - 1$, $\sinh 2K_c = 1$ and $\cosh 2K_c = \sqrt{2}$, yielding $K_c$ = 0.44068679……, i.e., $1/K_c$ = 2.26918531……. The putative exact solution of the 3D simple cubic Ising model is located exactly at $x_c = e^{-2K_c} = \frac{\sqrt{5}-1}{2} = 0.618033988......$, $\sinh 2K_c = \frac{1}{2}$ and $\cosh 2K_c = \frac{\sqrt{5}}{2}$, yielding $K_c = 0.24060591......$, i.e., $1/K_c$ = 4.15617384…. We also found the minimum of the γ ~ K curve for $\omega_{2t_x} = \pi$ is located at $x_d = e^{-2K_d} = \frac{\sqrt{10}-1}{3} = 0.72075922......$, $\sinh 2K_d = \frac{1}{3}$ and $\cosh 2K_d = \frac{\sqrt{10}}{3}$, $K_d = 0.16372507......$and $1/K_d$ = 6.10779991…… Although the minimum of the γ ~ K curve for $\omega_{2t_x} = \pi$ does not correspond to the critical point or any phase transition. The nature shows the hidden intrinsic relationship between the 2D square and the 3D simple cubic Ising lattices, as revealed by the values of $\sinh 2K = 1$, $\frac{1}{2}$ and $\frac{1}{3}$. It is also interesting to compare the critical points of the 2D square and the 3D simple cubic Ising lattices together with that of the 2D triangular Ising model: for the 2D square lattice, $\frac{1}{\tanh K_c}$ = 1 + $\sqrt{2}$ = 2.414213562…; for the 2D triangular lattice, $\frac{1}{\tanh K_c}$ = 2 + $\sqrt{3}$ = 3.732050808…;[24-31,93] for the simple cubic lattice, $\frac{1}{\tanh K_c}$ = $2 + \sqrt{5}$ = 4.236067977…. It is worth noting that these values are simply related with the smallest three irregular numbers subsequently, again showing some hidden intrinsic relations between the three lattices. The value of $\frac{1}{\tanh K_c}$ for each of these three lattices equals to one of the two smallest integers plus one of the smallest three irregular numbers.

It has been understood that the competition between the interaction energy and the thermal activity balances at the critical temperature. The values of the critical points could be used for the evaluation of the contribution of the interactions on the ordering of the systems. For the 1D Ising model, there is no order, i.e., $1/K_c = 0$, and the value of $1/K_c$ per J equals to zero for the existence of one interaction per unit cell. For the 2D square Ising model, the critical point of $1/K_c = 2.26918531$...... and the existence of two interactions J per unit cell lead to the fact that the value of $1/K_c$ per J equals to 1.13459265...... For the 3D simple cubic Ising model, the critical point of $1/K_c = 4.15617384$...... and the existence of three interactions J per unit cell result in the value of $1/K_c$ per J being equal to 1.38539128...... For the models with their dimensions $d \geq 4$, the mean field theory yields $1/K_c = z$ (where z is the coordination number).[212,226,227] Namely, the value of $1/K_c$ per J is equal to 2 for the $d \geq 4$ models, in consideration of the existence of z/2 interactions J per unit cell. It is reasonable to see that the value of $1/K_c$ per J increases monotonously from 0 via 1.13459265...... and 1.38539128...... to 2 when the dimension of the Ising models alters from 1 via 2 and 3 to 4 and above. Namely, the value of $1/K_c$ per J varies smoothly with dimensionality. This is because the correlations between the spins are strengthened with increasing the dimension of the system, which contribute more action to the ordering of the system.

The putative exact critical point of $1/K_c = 4.15617384$...... for the 3D simple cubic Ising model is derived by the introduction of the extra dimension, in accordance with the topologic problem of the three dimensions. One might think that the procedure put the extra energy to the final result. Now let us treat the third interaction of the 3D simple cubic Ising model to be the same as the role of the interactions in the 2D Ising model. Simply, taking the sums of the two interactions into the expressions of the eigenvalues as well as the partition function, one would derive the values of $x_c = e^{-2K_c} = \sqrt[3]{\frac{17}{27} + \sqrt{\frac{11}{27}}} + \sqrt[3]{\frac{17}{27} - \sqrt{\frac{11}{27}}} - \frac{1}{3} = 0.543689012......$ , and $K_c$ = 0.30468893......, i.e., $1/K_c = 3.28203585$...... by $\cosh\gamma_0 = \cosh 2(K^* - 2K)$, or $\sinh 2K \cdot \sinh 4K = 1$. However, this value of $1/K_c = 3.28203585$...... is evidently lower than the real one, since it does not take the real effects of the three dimensions into account. Actually, this value is even smaller than the exact solution of the 2D triangular Ising model, which is located exactly at $x_c = e^{-2K_c} = \frac{1}{\sqrt{3}} = 0.577350269......$ i.e., $1/K_c$ = 3.6409569……[24-31,93] It is known that the 2D triangular Ising model is equalized to the 2D square Ising model with only one next nearest neighboring interaction.[31,83,93] The critical point of these two models must be lower than the 2D Ising model with two next nearest neighboring interactions and the 3D simple cubic Ising model, because the latter two models have the topologic problems with the crosses/knots. It is a criterion that the critical point of the 3D simple cubic Ising model must be much higher than that of the 2D triangular Ising model. This criterion can be verified by the following consideration: It is a fact that the mean field theory shows none of the sensitive dependence on the lattice geometry which we might expect and predicts better results in higher dimensions. The mean field theory predicts the same critical point for the simple cubic lattice in three dimensions and the triangular lattice in two dimensions, since both have the coordination number z = 6. The difference between the exact value and the mean field one in the simple cubic lattice should be smaller than that in the triangular lattice, indicating clearly that the critical point of the former must be higher than that of the latter. The solution of $1/K_c$ = 4.15617384…… we found satisfies this criterion for the 3D simple cubic Ising model. It is thought that the difference between the critical points of the simple cubic lattice and the triangular lattice can be treated as the pure contribution of the 3D lattice, of course, which originates from not only the third dimension and but also the curled-up fourth dimension.

Finally, we would like to compare the putative exact critical points in more details with the results of the mean field theory. The mean field theory predicts that the critical point should depend on the geometry of the model only through the coordination number z, namely, $1/K_c$ is equal to the coordination number, and it is non-zero for all $z \neq 0$. It has been well – known that the mean field theory gives the correct predictions only for d ≥ 4 and it overestimates the critical point in every case of d < 4. Specially, the mean field theory is obviously wrong for the 1D Ising model, because it predicts $1/K_c = 2$, in contradiction with the fact that the exact calculation proves no order exists at finite temperatures. For the 2D square Ising model, the mean field theory suggests $1/K_c = 4$, which is much higher than the Onsager's exact solution of $1/K_c = 2.26918531$...... For the 3D simple cubic Ising model, the mean field theory gives $1/K_c = 6$, which is also higher than our exact solution of $1/K_c = 4.15617384$...... The feature of the mean field theory is that it identifies the order parameter of the system and tries to describe it as simply as possible.[212] It assumes that one needs only to take account of configurations in which the order parameter is uniform, and therefore that every spin, bond or whatever behaves in an average manner, regardless of what its neighbors are doing. This means that it neglects all fluctuations in the order parameter in which nearby parts of the system, while remaining correlated with each other, do something different from the average. That is all Fourier components with $q \neq 0$ are suppressed.[236] This neglect is responsible for the consistent overestimation of the critical point in 2D and 3D, and the wrong prediction of the existence of the order in 1D. It is evident that such neglect is more serious in lower dimensions, because the correlations affect more dramatically the physical properties. In order to compare the effects of this neglect in models of different dimensions, we define here a parameter, $\Delta\left(\frac{1}{K_c}\right) = \frac{\frac{1}{K_c^{MF}} - \frac{1}{K_c^{Exact}}}{\frac{1}{K_c^{MF}}}$ to evaluate the difference between the critical points of the mean field theory and the exact solution. Immediately, we obtain that $\Delta\left(\frac{1}{K_c}\right)$ equals to 100%, ~ 43.27%, ~ 30.73%, and 0 for 1D, 2D, 3D and 4D, respectively. Clearly, the error due to this neglect decreases monotonously with increasing the dimension of the system. It is relevant that the mean field theory predicts better results in higher dimensions.

There are something more on the facts concerning on the critical point: a) The critical point predicted by high – temperature expansions is even higher than the exact value obtained by introducing an additional dimension, rotation, and thus energy. b) The approximation value obtained by series expansions is usually higher than the exact value, whereas that obtained by removing one of interactions should be lower than the exact value. If the golden ratio were not for the 3D model (but, say, for a (3 + 1) – dimensional model), then the value obtained by high – temperature expansions would correspond to a model with even higher dimensionality since the higher dimensionality, the higher critical point. The question is how to construct the function of the free energy to obtain such over high value predicted by high – temperature expansions? The reasonable situation might be that the golden ratio is exactly for the 3D model, while the value of the critical point obtained by the high – temperature expansions might be inexact, but as high as an approximate should be.

What we uncovered for the 3D Ising model have been judged by several criterions: 1) At/near infinite temperature, the putative exact solution for the partition function of the 3D simple orthorhombic (and simple cubic) Ising lattices equals to the high temperature series expansion;[80,93,107] 2) The formulae for the eigenvalues, the eigenvectors, the partition function and the critical point of the 3D simple orthorhombic can return to those of the 2D rectangular Ising lattices if one of K' and K'' vanishes, the 1D Ising one if both K' and K'' vanish, and the simple cubic Ising one if K = K' = K''. 3) Our putative exact solution coincides with the first factor of the Rosengren's conjecture for the critical point of the 3D simple cubic Ising model,[230] while certainly the second factor of the Rosengren's conjecture has to be omitted;[231] 4) The putative exact solution of the 3D simple cubic Ising model is lower than the approximation values obtained by various series expansion methods, such as Kikuchi's estimation (please notice: the exact solution is very close to the low limit of the Kikuchi's estimation, within the error of 1.6%),[85] Wakefield's method,[73,74] Bethe's first and second approximations,[74,228] Kirkwood's method,[74,229] etc.; 5) The putative exact solution of the 3D simple cubic Ising model is in good agreement with the range of $4.16667 < 1/K_c < 4.7619$, as Oguchi estimated, and actually, it is exactly located at the lowest boundary of the Oguchi' estimations within the error of ~ 0.25 %;[62-64] 6) The exact solution for the critical point $1/K_c$ of the 3D simple cubic Ising model must be not smaller than 3.878, because the corrections of all the terms of the Bethe high order approximations on the mean field theory must be not larger than twice the correction of the Bethe first approximation; 7) The critical point $1/K_c$ of the 3D simple cubic Ising model must be much higher than that (3.6409569……) of the 2D triangular Ising model; 8) The putative exact solution is close to and lower than the value of $1/K_c$ = 4.511505(5) in the Binder and Luijten's review,[213] which was well established from the results of high-temperature series extrapolation, Monte Carlo renormalization group, Monte Carlo and finite – size scaling in the most recent years; 9) The value of $1/K_c$ per J increases monotonously with increasing the dimension of the Ising model because the correlations between spins are strengthened, contributing more efficiently to the ordering of the system in higher dimensions; 10) The parameter $\Delta\left(\frac{1}{K_c}\right)$, as the evaluation of the difference between the exact solution and the mean field one, decreases monotonously with increasing the dimension of the system, as a consequence of that the mean field theory predicts better results in higher dimensions because the errors due to the neglects of fluctuations in the order parameter become smaller; 11) The exact solutions of the 3D simple cubic and 2D square Ising models are intimately correlated with similar mathematical structures, such as the golden and the silver ratios as the solutions of the equations $x^2 + x - 1 = 0$ and $x^2 + 2x - 1 = 0$, the simplest two continued fractions, the critical point formulas of $\sinh 2K_c = \frac{1}{2}$ and $\sinh 2K_c = 1$, etc.; 12) The exact solution satisfies the principles of simple, symmetry and beauty with aesthetic appeal, which are the most important among the principles for judging the validness and correctness of a theory in case that no body knows the answer. It is well – known that the principles of simple, symmetry and beauty have been employed widely for establishing the elegance theories, like Einstein's general relativity, Dirac's equation, Feynman's path integrals, and Onsager's solution, etc.

# IV. SPONTANEOUS MAGNETIZATION

## A. Perturbation procedure

The spontaneous magnetization of the square Ising magnet was calculated exactly by Yang, using a perturbation procedure.[20] Chang[21] and Potts[22] derived the spontaneous magnetization of the rectangular Ising lattice, and Potts also dealt with the triangular Ising lattice. Schultz, Mattis and Lieb investigated the two-spin correlation function in an infinite 2D lattice in terms of many fermions and reconciled the different approaches of previous authors, discussed the definitions of the spontaneous magnetization.[237] Although the definition of the spontaneous magnetization was argued by Schultz, Mattis and Lieb,[237] in this section, we shall follow Yang's method to calculate the spontaneous magnetization of the 3D Ising model, based on the two conjectures introduced. We shall focus our interest first on the spontaneous magnetization of the simple orthorhombic lattices and then reduce to the simple cubic lattice.

As a weak magnetic field $\aleph$ is introduced, the partition function of the 3D simple orthorhombic Ising magnet could become:

$$Z_{\mathrm{K}} = (2\sinh 2K)^{mnl/2}\, trace(V_5 V_4 V_3 V_2 V_1)^m, \tag{4.1}$$

where

$$V_5 = \exp\{\aleph \sum_{1}^{n \cdot l} s_t\}. \tag{4.2}$$

For a large crystal, as discussed above, only the eigenvector of $\mathbf{V} = \mathbf{V_5 V_4 V_3 V_2 V_1}$ with the largest eigenvalue is important. The limiting form of this eigenvector as $\aleph \to 0$ is our interest. The largest eigenvalue of $\mathbf{V_4 V_3 V_2 V_1}$ is doubly degenerate below the critical temperature. This is evidently also true of the symmetrical matrix $\mathbf{V_1^{1/2}V_4V_3V_2V_1^{1/2}}$. Let $\psi_+$ and $\psi_-$ be the even and odd eigenvectors corresponding to the largest eigenvalue $\lambda$. Introducing the operator

$$U = C_1 C_2 \cdots C_{nl} \quad , \tag{4.3}$$

that reverses the spins of all atoms, we would have

$$U\psi_+ = \psi_+ , \qquad U\psi_- = -\psi_- , \tag{4.4}$$

for the even and odd eigenvectors respectively.

When the magnetic field $\aleph$ is applied, the degeneracy is removed. Analogous consideration to Yang's,[20] we shall perform a perturbation calculation, since we are only interested in the limit of $\aleph \to 0$. We shall consider only the largest eigenvalue of

$$V_1^{\frac{1}{2}} V V_1^{-\frac{1}{2}} = V_1^{\frac{1}{2}} V_5 V_4 V_3 V_2 V_1^{\frac{1}{2}} = V_1^{\frac{1}{2}} V_4 V_3 V_2 V_1^{\frac{1}{2}} + \aleph V_1^{\frac{1}{2}} (\sum_1^{nl} s_t) V_4 V_3 V_2 V_1^{\frac{1}{2}} . \tag{4.5}$$

The last term is a real symmetrical matrix anticommuting with **U**, which has no diagonal matrix element with respect to either $\psi_+$ or $\psi_-$. Ordinary perturbation theory shows that as $\aleph \to 0$ the eigenvector of (4.5) with the largest eigenvalue approaches

$$\psi_{\max} = \frac{1}{\sqrt{2}} (\psi_+ + \psi_-) , \tag{4.6}$$

if the phase of $\psi_+$ and $\psi_-$ are so chosen that they are real and that:

$$\psi_+ ' V_1^{\frac{1}{2}} (\sum_1^{nl} s_t) V_4 V_3 V_2 V_1^{\frac{1}{2}} \psi_- \geq 0 . \tag{4.7}$$

From the general definition of the matrix method, the average magnetization per atom reads as:

$$
\begin{aligned}
I &= \frac{1}{mnl}\frac{m\cdot trace(V_5V_4V_3V_2V_1)^m\sum_1^{nl}s_t}{trace(V_5V_4V_3V_2V_1)^m} \\
&= \frac{1}{nl}\frac{trace(V_1^{\frac{1}{2}}V_5V_4V_3V_2V_1^{\frac{1}{2}})^m(V_1^{\frac{1}{2}}\sum_1^{nl}s_tV_1^{-\frac{1}{2}})}{trace(V_1^{\frac{1}{2}}V_5V_4V_3V_2V_1^{\frac{1}{2}})^m}, \\
&= \frac{1}{nl}\psi_{\max}'V_1^{\frac{1}{2}}\sum_1^{nl}s_tV_1^{-\frac{1}{2}}\psi_{\max}
\end{aligned}
\tag{4.8}
$$

As $\aleph \to 0$, (4.8) becomes by (4.6)

$$
I = \frac{1}{2nl}(\psi_+'+\psi_-')V_1^{\frac{1}{2}}\sum_1^{nl}s_tV_1^{-\frac{1}{2}}(\psi_+ + \psi_-)\,. \tag{4.9}
$$

Similar to the discussion in Yang's paper,[20] the spontaneous magnetization is

$$
I = \frac{1}{nl}\psi_-'V_1^{\frac{1}{2}}\sum_1^{nl}s_tV_1^{-\frac{1}{2}}\psi_+\,. \tag{4.10}
$$

By replacing the summation $\Sigma s_t$, it can be written as:

$$
I = \psi_-'V_1^{\frac{1}{2}}s_1V_1^{-\frac{1}{2}}\psi_+\,. \tag{4.11a}
$$

The relation would be expressed as:

$$
I^{\frac{1}{3}} = \varphi_-'V_1^{\frac{1}{2}}s_1V_1^{-\frac{1}{2}}\varphi_+\,. \tag{4.11b}
$$

where $\varphi_-$‘ and $\varphi_+$ are the reduced normalized eigenvectors in consideration of weights at finite temperature. The 2n-reduced normalized eigenvectors is represented as:

$$u_{2t} \equiv \frac{1}{(2n)^{\frac{1}{2}}} \begin{bmatrix} e^{i\omega_{2t_x}} \cdot e^{\frac{i}{2}\delta_{2t}'} \\ ie^{i\omega_{2t_x}} \cdot e^{-\frac{i}{2}\delta_{2t}'} \\ e^{i\omega_{4t_x}} \cdot e^{\frac{i}{2}\delta_{2t}'} \\ ie^{i\omega_{4t_x}} \cdot e^{-\frac{i}{2}\delta_{2t}'} \\ \cdot \\ \cdot \\ \cdot \\ ie^{i\omega_{2nt_x}} \cdot e^{-\frac{i}{2}\delta_{2t}'} \end{bmatrix},$$

and

$$v_{2t} \equiv \frac{1}{(2n)^{\frac{1}{2}}} \begin{bmatrix} ie^{i\omega_{2t_x}} \cdot e^{\frac{i}{2}\delta_{2t}'} \\ e^{i\omega_{2t_x}} \cdot e^{-\frac{i}{2}\delta_{2t}'} \\ ie^{i\omega_{4t_x}} \cdot e^{\frac{i}{2}\delta_{2t}'} \\ e^{i\omega_{4t_x}} \cdot e^{-\frac{i}{2}\delta_{2t}'} \\ \cdot \\ \cdot \\ \cdot \\ e^{i\omega_{2nt_x}} \cdot e^{-\frac{i}{2}\delta_{2t}'} \end{bmatrix}.$$

The power 1/3 for the spontaneous magnetization I comes automatically from the dimensional unit of the 3D system as one uses the reduced eigenvectors. The physical significance of I is the same as defined for the 2D Ising magnet. Following Yang's work,[20] we introduce an artificial limiting process and reduce the problem to an eigenvalue problem of an n × n matrix. One could arrive after a little algebra at:

$$I^{\frac{1}{3}} = \underset{a \to i\infty}{Lim} (2\cos a)^{-n} \, trace V_1^{\frac{1}{2}} s_1 V_1^{-\frac{1}{2}} S(T_+^{-1} M T_-). \qquad (4.12)$$

The next step is to follow the procedure of subsections B, C, D and E of section II of Yang's paper.[20]

The procedure for an infinite crystal could be simplified greatly. With the consideration of the weights $w_x \equiv 1$, $w_y = w_z = 0$ for finite temperatures, the relationship between δ' and ω shown in Eq. (3.17) could be reduced explicitly, in term of $z = e^{i\omega_x}$ ($\omega_x = t_x\pi/n$, $t_x = 1, 2, \cdots n$), to:

$$e^{2i\delta'} = \frac{\tanh^2 K^*(z - \coth(K'+K''+K''')\coth K^*)(z - \tanh(K'+K''+K''')\coth K^*)}{(z - \coth(K'+K''+K''')\tanh K^*)(z - \tanh(K'+K''+K''')\tanh K^*)}.$$

(4.13)

Then $e^{i\delta'}$ behaves as

$$\Theta(z) = e^{i\delta'} = \frac{1}{(AB)^{\frac{1}{2}}}\left[\frac{(z-A)(z-B)}{(z-A^{-1})(z-B^{-1})}\right]^{\frac{1}{2}}, \tag{4.14}$$

where

$$A = \coth(K'+K''+K''')\coth K^* = \frac{1 + x_2x_3x_4}{x_1(1 - x_2x_3x_4)}, \tag{4.15a}$$

$$B = \tanh(K'+K''+K''')\coth K^* = \frac{1 - x_2x_3x_4}{x_1(1 + x_2x_3x_4)}, \tag{4.15b}$$

with

$$x_1 = e^{-2K};\ \ x_2 = e^{-2K'};\ \ x_3 = e^{-2K''};\ \ x_4 = e^{-2K'''}.$$

For $T < T_c$, $A > B > 1$. $\Theta(z)$ is analytic everywhere except at the points $z = A$, $B$, $1/A$, or $1/B$ where it has branch points. The square root in Eq. (4.14) is defined to be that branch of the function that takes the value –1 at $z = 1$.[20] Similar to Yang's procedure,[20] one has:

$$F^{-2} = \frac{4}{\pi}\frac{1}{A-B}k_{-1}^{\frac{1}{2}}K(k_{-1}), \quad (4.16)$$

where

$$k_{-1} = \left[\frac{(A^2-1)^{\frac{1}{2}} - (B^2-1)^{\frac{1}{2}}}{A(B^2-1)^{\frac{1}{2}} + B(A^2-1)^{\frac{1}{2}}}\right]^2, \quad (4.17)$$

and $K(k_{-1})$ is the complete elliptic integral of the first kind. It is convenient to change the modulus:

$$k = \frac{2k_{-1}^{\frac{1}{2}}}{1+k_{-1}} = \frac{4x_1x_2x_3x_4}{(1-x_1^2)(1-x_2^2x_3^2x_4^2)} = \sinh^{-1} 2K \sinh^{-1} 2(K'+K''+K'''). \quad (4.18)$$

Then

$$F^{-2} = \frac{2kK(k)}{\pi(A-B)}. \quad (4.19)$$

and

$$I^{\frac{4}{3}} = [\prod_{2}^{nl}(l_\alpha^2/4)]F^4A^{-2}B^{-2}\cosh^4 K^*, \quad (4.20)$$

which can be further simplified to

$$I^{\frac{4}{3}} = \left(\prod_{2}^{nl}\frac{l_\alpha^2}{4}\right)\frac{\pi^2}{4}\left[\frac{1}{K(k)}\right]^2. \quad (4.21)$$

Similar to Chang,[21] the elliptic transformation (81) in Yang's paper[20] is replaced by:

$$z = -\frac{(cnu - i(1+k_1)^{\frac{1}{2}}snu)(dnu - i(k_1+k_1k_2)^{\frac{1}{2}}snu)}{1+k_1sn^2u}, \quad (4.22)$$

where the modulus is given by Eq. (4.18). It is easy to verify that:

$$\frac{1}{z}\frac{dz}{du} = -i\frac{1-k^2}{(1+k_1)^{\frac{1}{2}}}\frac{1}{dnu - \left[\frac{k_1(1+k_2)}{1+k_1}\right]^{\frac{1}{2}} cnu}. \tag{4.23}$$

where $k_1 = \sinh^{-2} 2K$ and $k_2 = \sinh^{-2} 2(K' + K'' + K''')$. The essential properties of the variable z as a function of u remain the same as in the 2D lattices. There are still two singularities per unit cell (4K × 4iK') (please notice: only here the denotions K and K' are the same as Yang's K and K' in u-plane[20,21]), although their positions are changed. Finally, we quote:

$$\prod_{2}^{\infty}\frac{l_\alpha^2}{4} = \frac{4}{\pi^2}[K(k)]^2(1-k^2)^{\frac{1}{2}} = \frac{4}{\pi^2}K^2\frac{1}{(1-x_1^2)(1-x_2^2x_3^2x_4^2)}\times$$
$$\left[(1-x_1^2+4x_1x_2x_3x_4-x_2^2x_3^2x_4^2+x_1^2x_2^2x_3^2x_4^2)(1-x_1^2-4x_1x_2x_3x_4-x_2^2x_3^2x_4^2+x_1^2x_2^2x_3^2x_4^2)\right]^{\frac{1}{2}}$$

(4.24)

The spontaneous magnetization I for the simple orthorhombic lattices is obtained from Eqs. (4.21) and (4.24) as:

$$I = \left\{\frac{\left[(1-x_1^2+4x_1x_2x_3x_4-x_2^2x_3^2x_4^2+x_1^2x_2^2x_3^2x_4^2)(1-x_1^2-4x_1x_2x_3x_4-x_2^2x_3^2x_4^2+x_1^2x_2^2x_3^2x_4^2)\right]^{\frac{1}{2}}}{(1-x_1^2)(1-x_2^2x_3^2x_4^2)}\right\}^{\frac{3}{4}}$$

(4.25)

The temperature dependence of the spontaneous magnetization I for several simple orthorhombic lattices with K' = K'' = K, 0.5 K, 0.1 K and 0.0001 K is represented in Fig. 3(a). The spontaneous magnetization decreases with increasing temperature to be zero at the critical point. The critical point decreases with decreasing K' and K'', until disappearing as K' = K'' = 0 as a 1D system. For a simple cubic lattice, because $x_1 = x_2 = x_3 = x_4 = x$, k is reduced to:

$$k = \frac{2k_{-1}^{\frac{1}{2}}}{1+k_{-1}} = \frac{4x^4}{(1-x^2)(1-x^6)} = \sinh^{-1} 2K \sinh^{-1} 6K\,. \tag{4.26}$$

Then the spontaneous magnetization I for the simple cubic lattices could be:

$$I = \left\{ \frac{1}{(1-x^2)(1-x^6)} \left[ (1-x^2+4x^4-x^6+x^8)(1-x^2-4x^4-x^6+x^8) \right]^{\frac{1}{2}} \right\}^{\frac{3}{4}}. \tag{4.27}$$

or

$$I = \left[ 1 - \frac{16x^8}{(1-x^2)^2(1-x^6)^2} \right]^{\frac{3}{8}}. \tag{4.28}$$

At low temperatures, this gives the expansion in power of x as:

$$\begin{aligned} I &= 1 - 6x^8 - 12x^{10} - 18x^{12} - 36x^{14} - 84x^{16} - 192x^{18} - 408x^{20} - 864x^{22} - 1970x^{24} \\ &- 4680x^{26} - 10980x^{28} - 25480x^{30} - 59970x^{32} - 143940x^{34} - 347730x^{36} \\ &- 838956x^{38} - 2028870x^{40} - 7790088x^{42} \cdots \end{aligned} \tag{4.29}$$

This series is convergent all the way up to the critical point, where $x = x_c = \frac{\sqrt{5}-1}{2}$.

Near the critical point, I has a branch point:

$$I \cong \left[ \left(\frac{5\sqrt{5}(\sqrt{5}+1)}{2}\right)(x_c - x) \right]^{3/8}. \tag{4.30}$$

In Fig. 3 (b), the spontaneous magnetization I of the simple cubic Ising lattices is plotted again the temperature, in comparison with the spontaneous magnetization obtained by Yang for the square Ising model,[20] and the result of the series expansion[111,238] of the simple cubic Ising model. It is interesting to note that two golden solutions, $\frac{\sqrt{5}-1}{2}$ and $\frac{\sqrt{5}+1}{2}$ of the equation $x^2 + x - 1 = 0$ appear as $x_c$ and the constant in the formula for the critical behavior of the spontaneous magnetization of the 3D simple cubic Ising model, while two silver solutions of the equation $x^2 + 2x - 1 = 0$ show up in the Yang's formula for the 2D square Ising model.[20] Being plotted as a function of $T/T_c$, the different critical behaviors of the spontaneous magnetization of the 3D and 2D Ising lattices are clearly seen, which originate from their different powers of 3/8 and 1/8. We believe that the low temperature series expansion is not exact and the appearance of plus sign is clearly incorrect,[59] which is to compensate the incorrectness of $x^6$ term (this issue will be discussed in details later on in Sec. VIII). Nevertheless, the series expansion of the simple cubic Ising model numerically fits well with the putative exact one, and oscillates around it, up to $T \approx 0.9\ T_c$, and then deviates from it. The spontaneous magnetization obtained by the low – temperature series expansion depends sensitively on how many terms are taken into account. It is seen from Fig. 3 (b) that the curve with terms up to the 52nd order one (with a positive coefficient) goes up above ~ 0.9 $T_c$,[111,238] while the curve with terms with the 54th order (with a negative coefficient) as its last term drops down monotonously.[238] This indicates clearly that the low – temperature series diverges. Moreover, the curve with more term is closer to the putative exact solution. It is expected that if more terms were taken into account, the curves of the low – temperature series would numerically fit better with the putative solution in wider temperature range (however, the curves of the low – temperature series are still divergent, depending sensitively on the sign of the last term). This implies that our putative exact solution might be correct.

### B. 3D – to – 2D crossover phenomenon

It is interesting to see how the exponent $\beta = 3/8$ for 3D becomes the famous $\beta = 1/8$ for 2D and whether the 3D – to – 2D crossover phenomenon is similar to the 2D – to – 1D crossover phenomenon. There was no evident indication in Yang's formula[20] to show how the famous β = 1/8 disappears when one of $x_1$ and $x_2$ equals to 1, as it was not represented directly as a function of $x_1$ and $x_2$. Of course, one can understand the 2D – to – 1D crossover phenomenon embodied in the expression in the brackets of their formula, certainly, not directly from the power 1/8 itself. One can understand the 2D – to – 1D crossover phenomenon more easily by Chang's general formula for a rectangular lattice,[21] compared with Yang's for a square lattice.[20] As Chang discovered,[21] the exponent 1/8 does not change with varying ratios of the vertical and horizontal interactions. One was tempted to conclude that the exponent is dependent only on the dimensionality of the lattice and not on the number of nearest neighbors.[21]

For the 3D – to – 2D crossover phenomenon, one might expect the similar situation happens. From the first feeling, as $x_3 = x_4 \equiv 1$ (i.e., K'' = K''' ≡ 0), the spontaneous magnetization (Eq. (4.25)) for the simple orthorhombic lattices should automatically return the Onsager's original 2D Ising model and hence the critical exponent β automatically becomes 1/8. This could be realized, since one does not need the additional rotation in the 2D limit. However, a careful inspect uncovered another mechanism on the 3D – to – 2D crossover phenomenon based on the validity of our solution. Namely, there should be a gradual crossover between the 3D and 2D behaviors when $x_3 = x_4 \to 1$ as K'' = K''' → 0. The criterion for illustrating the existence of such crossover is described as following: At the same temperature, the spontaneous magnetization of a 3D Ising system with $x_3 = x_4 \neq 1$ must be always higher than that of a 2D Ising system with the same values of $x_1$ and $x_2$.

This criterion is not based on a hypothesis, but a fact with physical significances: The spontaneous magnetization (i.e., the order parameter) of the system depends on the competition between the order energy (controlled by the Hamiltonian) and the disorder energy (i.e., the thermal activity). How the order parameter decreases with increasing temperature reveals in details this competition. To compare the spontaneous magnetizations of different systems at a same temperature actually remains to compare only their order energy, because with the fixed temperature the thermal activity is kept to be the same for these systems. From the Hamiltonian, it is clear that the order energy of a 3D Ising system with K" ≠ 0 is always larger than that of a 2D Ising system with the same values of K and K', no matter how small the K" is. That is, this criterion can apply also in the regime where one of the coupling interactions of the 3D ferromagnet approaches to zero, namely, even when $T_{c,3D}$ to $T_{c,2D}$. The 3D Ising system can be constructed by connecting the l planes of the 2D Ising systems by the third interaction K". With the help of the third interaction K" to bid defiance to the thermal activity, certainly, the order parameter of such 3D Ising system is always higher than that of the 2D Ising system at the same temperature. Clearly, this criterion is always true, no matter whether one or both of the systems are in the critical regions.

The exponent β = 3/8 gives a curve lower than that of the exponent β = 1/8 for 2D if plotting the spontaneous magnetization as a function of $T/T_c$. When K'' and K''' are large enough to have a Curie point for 3D high enough to keep its spontaneous magnetization is always higher than 2D plots, the system behaves as a real 3D one with β = 3/8. Else, the system behaves as a crossover with an exponent β in range of 1/8 – 3/8, though the small values K'' and K''' do not vanish yet. The range of such crossover could be determined numerically to be from K'' = K''' ≈ 0.195 K (in case of K = K') to zero. One could derive the district for the 3D – to - 2D crossover in the parametric diagram for the whole system (see Fig. 4). The dashed curve of $\frac{K'}{K}+\frac{K''}{K}+\frac{K'K''}{K^2}=1$ in Fig. 4 corresponds to the points with the critical temperature of the silver solution. The 3D to 2D crossover phenomenon appears in the district between the dashed curve and the dash dot one of $\frac{K'}{K}+\frac{K''}{K}+\frac{K'K''}{K^2}\approx 1.39$. All the points with the critical temperature below the silver solution would have the 2D critical exponent, while all the points in the area above the dash dot curve would behave as a real 3D system. Of course, it is hard to represent in the exponent of the formula mathematically to illustrate in details how the exponent β = 3/8 for 3D changes to be the famous β = 1/8 for 2D in this crossover.

In the following, the occurrence of such crossover can be proved briefly: As shown in Fig. 3(b), being plotted as a function of $T/T_c$, the spontaneous magnetization of the 3D Ising model is always lower than that of the 2D Ising model at every normalized temperature. This is always true, no matter whether we choose the exponent of 3/8 or 5/16 (as high – temperature series expansions suggested). Supposed we have a function $m_{3D}(x_1, x_2, x_3)$ for the expression in the brackets for the spontaneous magnetization of the 3D Ising model, while a function $m_{2D}(x_1, x_2)$ for 2D. From the two functions, one can determine directly the critical point, without considering how big the exponent is. With a value of the third interaction being small enough (say, K'' $\rightarrow 0^+$ in case of K = K'), the difference between the critical points of the 3D and 2D Ising models can be also small enough ($T_{c,3D} \rightarrow T_{c,2D}{}^+$). Then the large difference between the exponents of the two systems results in that most of the spontaneous magnetization of the 3D Ising model at temperatures below $T_{c,2D}$ is lower than that of the 2D system, while the spontaneous magnetization of the 3D Ising model at temperatures above $T_{c,2D}$ is higher than that of the 2D system. There is a cross point between the two curves for the spontaneous magnetization of the two systems. In principle, this situation must happen, no matter how the functions $m_{3D}$ and $m_{2D}$ look like and whether the exponent β for the 3D Ising model is 3/8 or 5/16. On the other hand, it is unreasonable to believe that the spontaneous magnetization of the 3D Ising model is smaller than that of the 2D Ising model, because the existence of the third interaction should increase the energy, and hence enhance the spontaneous magnetization. A reasonable mechanism is that the exponent β in this limit is not 3/8 (or 5/16), but it can be just slightly larger than that of the 2D Ising model (i.e., $\beta \rightarrow (1/8)^+$). The similar analysis reveals that as K'' is decreased to be less than about 0.195 K, the exponent β should be slightly less than 3/8 (i.e., $\beta \rightarrow (3/8)^-$). This indicates clearly that there is a crossover of the exponent β from the 3D value of 3/8 to the 2D value of 1/8, as K'' decreases down to zero. This is a fact, which does not depend on the detail expression of the solution. Such kind of the 3D – to – 2D crossover phenomenon implies that the action of the additional rotation becomes weaker gradually as K'' decreases to zero. A similar prove can be performed also for the cases with both K’ and K’’ decreasing. One easily derives the condition of $\frac{K'}{K}+\frac{K''}{K}+\frac{K'K''}{K^2}\approx 1.39$ for the border between the districts for the real 3D behaviors and the 3D – to – 2D crossover. The present work shows that the 3D – to – 2D crossover phenomenon differs with the 2D – to – 1D crossover phenomenon. This is because from 3D to 2D, one undergoes the crossover between two different ordering systems, whereas from 2D to 1D, one undergoes the crossover between ordering and disordering ones. This is also because from 3D to 2D, one undergoes the crossover between two systems with and without the topologic problem respectively, whereas from 2D to 1D, one undergoes the crossover between two systems both without the topologic problem.

It has been accepted by the community (mainly based on the numerical calculation) that the 3D system always shows the 3D critical behavior, no matter how the relative ratios between the strengths of the interactions along three crystallographic axes are. It is known from the numerical results that even with a small enough interlayer interaction, the system shows the 3D critical behavior for a narrow range near the critical temperature. This range becomes narrower when the interlayer interaction becomes smaller. However, to fit the 3D critical exponent well in a narrower range near the critical temperature means that the 3D critical behaviors are becoming weaker, while other terms (like, the subleading order in expansions) of different critical behaviors become comparatively stronger. If one insisted to fit with the 3D critical exponents as the interlayer interaction becomes extremely smaller, the critical region would be extremely narrower (even to be zero or infinitesimal to be no physical meaning) and then one would meet a problem: how does the 3D system with a very narrow critical region and the 3D critical exponents jump suddenly to the 2D system with a much wider critical region and the 2D critical exponents as the infinitesimal interaction K” vanishes? Such sudden jump should not occur, because the 3D system with the infinitesimal interaction K” should have the 2D-like behavior because the infinitesimal interaction K” make the l planes of the 2D Ising systems to be almost independent each other. The 3D system in this case is close to many 2D

Ising systems separated from each other. The critical behavior of the 3D system with the infinitesimal interaction K'' should be near to the 2D critical behavior. It is emphasized here that any numerical results obtained by fitting the data points of the calculations cannot serve as a standard for discussion on the present topic, due to the limitation of the accuracy of the numerical calculation (though it might be in a high precision), such as their systematic errors originated from the disadvantage of the approximations and the computer powers (dealing with the cooperative phenomena of infinite spins in the system).

To discuss the situation in details, one could assume that the 3D magnetization were of the typical form in the expansion: $M_{3D}(T) \sim A(T_c) \mid T - T_c \mid^{\beta} + B(T_c) \mid T - T_c \mid^{\beta'}$ + higher order terms, ($\beta \neq \beta'$). Then, one could argue that to satisfy the criterion above, there is no need of such crossover, specifically, by taking into account the next leading order term in the 3D magnetization expansion close to $T_c$. $\beta < \beta'$ is not the case here, since it may correspond to the 3D to 4D crossover. For $\beta > \beta'$ ($\beta \sim 3/8$ (or 5/16) and $\beta' = 1/8$ for the 3D to 2D crossover), one could believe that it might happen that in the limit $T_{c,3D}$ to $T_{c,2D}$, the subleading order could take over with a divergent amplitude B such that: $B(T_c) \mid T - T_c|^{\beta'}$ to $C(T_c) \mid T - T_c|^{1/8}$. However, the action of the subleading order depends sensitively on the relative ratio $r_{AB}$ (= A/B) between the amplitudes A and B. In the limit case of a pure 3D system, the subleading order is negligible since the amplitude A is dominant ($r_{AB} \to \infty$), whereas in the pure 2D case, the subleading order takes over since the amplitude B becomes dominant ($r_{AB} \to 0$). In consideration of the continuity, there should exist a region in the parametric plane where the amplitudes A and B are comparable ($r_{AB} \sim 1$). In this region, the contributions from the leading and the subleading orders in the expansion above are comparably in the same order and one cannot neglect the effects of both the terms. In this region, the 3D magnetization expansion above fails in an attempt to derive a unique critical exponent and the actual critical exponent of the system is neither $\beta$ nor $\beta'$. The only possibility is to describe the critical behaviors by $D(T_c) \mid T - T_c \mid^{\beta''}$ with $\beta > \beta'' > \beta'$ for $r_{AB} \sim 1$. Finally, it is noticed that $\beta'$ should be not smaller than 1/8;

else, it would be another difficulty to interpret how the critical behavior of the system changes from β' to the 2D value 1/8.

The fact is very clear and simple: During the 3D to 2D crossover, as one of the coupling interactions of the 3D ferromagnet approaches to zero (e.g. $K_3 \to 0$, $x_3 \to 1$), the difference between the functions $m_{3D}(x_1, x_2, x_3)$ and $m_{2D}(x_1, x_2)$ in the brackets of the expression for the spontaneous magnetization of the 3D and 2D Ising models can be negligible, while if the system still had a 3D critical exponent, the existence of the large difference between the 3D and 2D critical exponents would certainly violate the criterion that at the same temperature, the spontaneous magnetization of a 3D Ising system with K" ≠ 0 (i.e., $x_3 \neq 1$) must be always higher than that of a 2D Ising system with the same values of $x_1$ and $x_2$. The only possibility to satisfy this criterion is that the critical exponent of the system approaches the 2D one during such crossover.

## V. SPIN CORRELATION FUNCTION

The spin-spin correlations in the 2D Ising model were studied first by Kaufman and Onsager, [18] then by various authors.[239-251] The combinatorial method was used by Potts and Ward to calculate the partition function of a finite rectangular Ising lattice and the correlation functions of an infinite lattice.[60,239] The proofs necessary to make this solution rigorous were supplied by Sherman[240,241] and Burgoyne.[242] Kadanoff phrased the Onsager solution of the 2D Ising model in the language of thermodynamic Green's functions and discussed the spin correlation functions for temperatures just smaller/greater than the critical point.[243] The Pfaffians was first introduced by Hurst and Green to derive the solution of the Ising problem for a rectangular lattice.[244] The number of ways in which a finite quadratic lattice (with edges or periodic boundary conditions) can be fully covered with given numbers of "horizontal" and "vertical" dimers was rigorously calculated by a combinatorial method involving Pfaffians.[245] The Ising problem was shown to be equivalent to a generalized dimer problem and the Onsager's expression for the Ising partition function of a rectangular lattice graph was derived on the basis of this equivalence.[246-248] As revealed by Kasteleyn,[245,246] neither (C) nor (D) theorem in his paper is true if a nonplanar graph is represented in a plane (with intersecting lines). The Onsager – Kaufman formulas for the correlations and the Onsager formula for the spontaneous magnetization of the rectangular 2D Ising lattice were re-droved by Montroll, Potts and Ward.[249] The Pfaffian approach was used to derive the correlation in terms of Pfaffians, and for the correlations in a row, a single Toepltiz determinant was obtained which was proved equivalent to the Onsager – Kaufman result. The Pfaffian representation of the partition function of the 2D triangular lattice was also applied to derive the expressions for various two, four, and six spin correlations in terms of Pfaffians.[250] It is clear that one cannot simply apply the technique of Pfaffians to deal with the problem of the 3D Ising model. In addition, the short-range order parameters were evaluated for the triangular and honeycomb Ising nets in ferro- and antiferromagnetic cases by the method of Kaufman and Onsager.[251]

In this section, we shall investigate the spin correlation function for the 3D simple orthorhombic Ising lattices, also based on the introduction of the two conjectures. We shall first give the general formula for the spin correlation function of the simple orthorhombic Ising system. Then we shall discuss the spin correlation function, separately in its different features. The first is to follow Montroll, Potts and Ward's procedure to evaluate the spin correlation function,[249] which is related to the spontaneous magnetization in its long-range order. The second is to follow Wu's procedure to study the short-range order.[252] The third is to follow Fisher's procedure to discuss the true range of the spin correlation function,[247] which is related to the correlation length. The fourth is to extend the discussion of Kaufman and Onsager[18] on the short-range order to the three-dimensional binary Ising lattice.

### A. General formula

Each site in the lattice could be indexed by (i, j, k) for its location in the coordinate system (rows, column, plane). In general, the spin in the jth site in the ith row of the kth plane, when the crystal is in configuration {ν1, ν1, …νm}, is $(s_{jk})_{\nu_i \nu_i}$. The average value of this spin is: [18]

$$\begin{aligned}\bar{s}_{jk} &= \frac{1}{Z}\sum_{\nu_1,\nu_2,\ldots\nu_m}(s_{jk})_{\nu_k\nu_k}V_{\nu_1\nu_2}V_{\nu_2\nu_3}\ldots V_{\nu_{i-1}\nu_i}V_{\nu_i\nu_{i+1}}\ldots V_{\nu_m\nu_1} \\ &= \frac{1}{Z}\sum_{\nu_1,\nu_2,\ldots\nu_m}V_{\nu_1\nu_2}V_{\nu_2\nu_3}\ldots V_{\nu_{i-1}\nu_i}(s_{jk})_{\nu_i\nu_i}V_{\nu_i\nu_{i+1}}\ldots V_{\nu_m\nu_1} \quad . \\ &= 1/Z\cdot trace\{V^{i-1}s_{ij}V^{m-i+1}\} = (1/Z)trace(s_{jk}V^{m})\end{aligned} \tag{5.1}$$

This result is independent of i, and $\bar{s}_{jk}$ vanishes identically for every (j, k).

The correlation function between the spins of the site j in row i and the site b in row a within plane k, i.e., (i, j, k) and (a, b, k), is written as:

$$\begin{aligned}&< s_{ijk}s_{abk} >_{Av} = \frac{1}{Z}\sum_{\nu_1,\nu_2,\ldots\nu_m}V_{\nu_1\nu_2}\ldots V_{\nu_{i-1}\nu_i}(s_{jk})_{\nu_i\nu_i}V_{\nu_i\nu_{i+1}}\ldots V_{\nu_{a-1}\nu_a}(s_{bk})_{\nu_a\nu_a}V_{\nu_a\nu_{a+1}}\ldots \\ &= 1/Z\cdot trace\{V^{i-1}s_{jk}V^{a-i}s_{bk}V^{m-a+1}\} = (1/Z)trace(s_{jk}V^{a-i}s_{bk}V^{m-a+i})\end{aligned} \tag{5.2}$$

For the correlation function between the spins of the site located in plane k and row i and the one located in plane χ and row α within column j, i.e., (i, j, k) and (α, j, χ), we have:

$$\begin{aligned}&< s_{ijk}s_{\alpha j\chi} >_{Av} = \frac{1}{Z}\sum_{\nu_1,\nu_2,\ldots\nu_m}V_{\nu_1\nu_2}\ldots V_{\nu_{i-1}\nu_i}(s_{jk})_{\nu_i\nu_i}V_{\nu_i\nu_{i+1}}\ldots V_{\nu_{\alpha-1}\nu_\alpha}(s_{j\chi})_{\nu_\alpha\nu_\alpha}V_{\nu_\alpha\nu_{\alpha+1}}\ldots \\ &= 1/Z\cdot trace\{V^{i-1}s_{jk}V^{\alpha-i}s_{j\chi}V^{m-\alpha+1}\} = (1/Z)trace(s_{jk}V^{\alpha-i}s_{j\chi}V^{m-\alpha+i})\end{aligned} \tag{5.3}$$

The correlation function between the spins of site b in row a in plane k and site j in row α in plane χ, i.e., (a, b, k) and (α, j, χ), could be calculated by multiplication of the two correlation functions (5.2) and (5.3):

$$\begin{aligned}&< s_{abk}s_{\alpha j\chi} >_{Av} = < s_{abk}s_{ijk} >_{Av}\cdot < s_{ijk}s_{\alpha j\chi} >_{Av} = \\ &(1/Z)trace(s_{jk}V^{a-i}s_{bk}V^{m-a+i})\cdot(1/Z)trace(s_{jk}V^{\alpha-i}s_{j\chi}V^{m-\alpha+i})\end{aligned} \tag{5.4}$$

At zero temperature, all spins are aligned, and as a result, we have:

$$< s_{abk} s_{ijk} >_{Av} = +1 , \quad (5.5a)$$

$$< s_{ijk} s_{\alpha j\chi} >_{Av} = +1 , \quad (5.5b)$$

and

$$< s_{abk} s_{\alpha j\chi} >_{Av} = +1 . \quad (5.5c)$$

for all pairs of sites. At higher temperatures, the correlation functions decrease, and tend to zero for very high temperatures. It is clear that we will know the correlation function $<s_{abk}s_{\alpha j\chi}>_{Av}$ between the spins of two sites in different planes in the lattice, as soon as we know the correlation function $<s_{abk}s_{ijk}>_{Av}$ between spins of two sites within one of planes and the correlation function $<s_{ijk}s_{\alpha j\chi}>_{Av}$ between spins of two sites within one of columns. The procedures for evaluating the correlation functions $<s_{abk}s_{ijk}>_{Av}$ and $<s_{ijk}s_{\alpha j\chi}>_{Av}$ are the same, because of the symmetry of the lattice.

### B. Correlation functions

The partition function for the 3D simple orthorhombic Ising lattices could be written as:

$$\begin{aligned} Z &= \sum_{\sigma=\pm 1 n.n.n} \prod \exp\left(K\sigma_{\alpha,\beta,\gamma}\sigma_{\alpha,\beta+1,\gamma} + K'\sigma_{\alpha,\beta,\gamma}\sigma_{\alpha+1,\beta,\gamma} + K''\sigma_{\alpha,\beta,\gamma}\sigma_{\alpha,\beta,\gamma+1}\right) \\ &= (\cosh K \cosh K' \cosh K'')^N \times \\ &\sum_{\sigma=\pm 1 n.n.n} \prod (1 + z_1\sigma_{\alpha,\beta,\gamma}\sigma_{\alpha,\beta+1,\gamma})(1 + z_2\sigma_{\alpha,\beta,\gamma}\sigma_{\alpha+1,\beta,\gamma})(1 + z_3\sigma_{\alpha,\beta,\gamma}\sigma_{\alpha,\beta,\gamma+1}) \end{aligned} \quad (5.6)$$

where $z_1$ = tanh K, $z_2$ = tanh K' and $z_3$ = tanh K''. The square of the Pfaffian is the determinant of a skew-symmetric matrix A so that:

$$Z^2 = (2\cosh K \cosh K' \cosh K'')^{2N} |A| . \quad (5.7)$$

The correlation between two spins at the sites (1, 1, 1) and (1+m, 1+n, 1+l) could be defined as:

$$< \sigma_{1,1,1}\sigma_{1+m,1+n,1+l} >= \frac{1}{Z}(\cosh K \cosh K' \cosh K'')^N \sum_{\sigma=\pm1} \sigma_{1,1,1}\sigma_{1+m,1+n,1+l} \times \prod_{n.n.n} (1 + z_1\sigma_{\alpha,\beta,\gamma}\sigma_{\alpha,\beta+1,\gamma})(1 + z_2\sigma_{\alpha,\beta,\gamma}\sigma_{\alpha+1,\beta,\gamma})(1 + z_3\sigma_{\alpha,\beta,\gamma}\sigma_{\alpha,\beta,\gamma+1}) . \quad (5.8)$$

Similar to the 2D Ising case, the determinant of a skew-symmetric matrix A is equal to the square of the Pfaffian. However, the skew-symmetric matrix A for the 3D simple orthorhombic Ising lattices is actually a three-dimensional matrix and nobody knows the laws of its operations. It is difficulty to evaluate the spin correlation function of the 3D Ising lattices, simply following the Montroll, Potts and Ward's method.[249] Fortunately, we have found the putative solution (3.64) for the 3D simple orthorhombic Ising lattices for finite temperatures, based on the two conjectures. We could immediately find out the similarity between the formulae for the 3D and 2D Ising lattices. Therefore, following the Montroll, Potts and Ward's procedure for the 2D Ising lattice,[249] we could define an effective skew-symmetric matrix $A_{eff}$ as.

$$A_{eff} = \begin{bmatrix} 0 & 1 + z_1 e^{i\omega'} & -1 & 1 \\ -1 - z_1 e^{-i\omega'} & 0 & 1 & -1 \\ 1 & -1 & 0 & 1 + z_{2,eff} e^{i\omega} \\ 1 & 1 & -1 - z_{2,eff} e^{-i\omega} & 0 \end{bmatrix}, \quad (5.9)$$

with $z_{2,eff} = \tanh(K' + K'' + K''') = z_2 z_3 z_4$ and $z_4 = \tanh K'''$.

The spin correlation function at finite temperatures of the 3D simple orthorhombic Ising lattices could be expressed effectively as:

$$Z z_1^{-n} z_{2,eff}^{-l} < \sigma_{1,1}\sigma_{1+n.1+l}^{eff} >= (\cosh K \cosh(K' + K'' + K'''))^N \times \sum_{\sigma=\pm1} \prod_{n'=1}^{n} (1 + z_1^{-1}\sigma_{1,n'}\sigma_{1,1+n'}) \prod_{l'=1}^{l} (1 + z_{2,eff}^{-1}\sigma_{l',1+n}\sigma_{1+l',1+n}) \times \prod_{n.n}{}' (1 + z_1\sigma_{\alpha,\beta}\sigma_{\alpha,\beta+1})(1 + z_{2,eff}\sigma_{\alpha,\beta}\sigma_{\alpha+1,\beta}) . \quad (5.10)$$

The formula (5.10) has the form similar to that used by Montroll, Potts and Ward.[249] Following the procedure same as Eqs. (18) – (40) of Montroll, Potts and Ward's paper,[249] one arrives:

$$< \sigma_{1,1}\sigma_{1,2}^{eff} >= \frac{1}{2\pi}\int_{-\pi}^{\pi} e^{i\delta^*(\omega)} d\omega . \quad (5.11)$$

Here δ* (ω) is the function, as expressed by:

$$e^{2i\delta^*} = \frac{(z_1 z_{2,eff}^* e^{i\omega} - 1)(z_1 e^{i\omega} - z_{2,eff}^*)}{(e^{i\omega} - z_1 z_{2,eff}^*)(z_{2,eff}^* e^{i\omega} - z_1)}, \quad (5.12)$$

with

$$z_{2,eff}^* = \frac{1 - z_{2,eff}}{1 + z_{2,eff}} = e^{-2(K' + K'' + K''')} . \quad (5.13)$$

Finally, the correlation could be written as the Toeplitz determinant:[249,253,254]

$$< \sigma_{1,1}\sigma_{1,1+l}^{eff} >= \begin{vmatrix} a_0 & a_1 & a_2 & . & . & . & a_{l-1} \\ a_{-1} & a_0 & a_1 & . & . & . & a_{l-2} \\ a_{-2} & a_{-1} & a_0 & . & . & . & a_{l-3} \\ . & . & . & . & . & . & . \\ . & . & . & . & . & . & . \\ . & . & . & . & . & . & . \\ a_{-l+1} & a_{-l+2} & a_{-l+3} & . & . & . & a_0 \end{vmatrix}, \quad (5.14)$$

where

$$a_r = \frac{1}{2\pi}\int_{-\pi}^{\pi} e^{-ir\omega} e^{i\delta^*(\omega)} d\omega , \quad (5.15)$$

are the coefficients in a series expansion of $e^{i\delta^*}$.

According to the units of the dimensions for the expressions of the correlation functions, the following relations could be realized:

$$< \sigma_{1,1,1}\sigma_{1,1,1+l} >=< \sigma_{1,1}\sigma_{1,1+l}^{eff} >^3 , \quad (5.16)$$

and it yields:

$$I^{\frac{2}{3}} = \lim_{l\to\infty} < \sigma_{1,1,1}\sigma_{1,1,1+l} >^{\frac{1}{3}} = \lim_{l\to\infty} < \sigma_{1,1}\sigma_{1,1+l}^{eff} > . \quad (5.17)$$

Following the procedure of Montroll, Potts and Ward,[249] it is known from the Szego's theorem that if $D_m$ (f) is the determinant of a Toepltiz matrix whose elements are the coefficients in the Laurent expansion of a function f (ω) then:[249,253,254]

$$\lim_{m\to\infty} \frac{D_m(f)}{G(f)^{m+1}} = \exp\left(\sum_{1}^{\infty} nk_n k_n\right), \quad (5.18)$$

where

$$G(f) = \exp\left[\frac{1}{2\pi}\int_{-\pi}^{\pi} \ln f(\omega) d\omega\right], \quad (5.19)$$

and

$$\ln f(\omega) = \sum_{n=-\infty}^{\infty} k_n e^{in\omega} . \quad (5.20)$$

In order to obtain $k_n$, we require the Fourier expansion of:

$$\ln f(\omega) = i\delta^*(\omega) = \frac{1}{2}\ln\frac{(z_1 z_{2,eff}^* e^{i\omega} - 1)(z_1 e^{i\omega} - z_{2,eff}^*)}{(e^{i\omega} - z_1 z_{2,eff}^*)(z_{2,eff}^* e^{i\omega} - z_1)} . \quad (5.21)$$

in two different temperature ranges, for T < $T_c$, $z_{2,eff}^* < z_1 < 1$; for ∞ > T > $T_c$, $z_1 < z_{2,eff}^* < 1$. The spontaneous magnetization below the critical temperature could be derived as:

$$I=\left[1-\frac{(1-z_1^2)^2 z_{2,eff}^{*2}}{z_1^2(1-z_{2,eff}^{*2})^2}\right]^{\frac{3}{8}}=\left[1-\frac{(1-z_1^2)^2(1-z_2^2z_3^2z_4^2)^2}{16z_1^2z_2^2z_3^2z_4^2}\right]^{\frac{3}{8}}$$

$$=\left[1-\frac{16x_1^2x_2^2x_3^2x_4^2}{(1-x_1^2)^2(1-x_2^2x_3^2x_4^2)^2}\right]^{\frac{3}{8}}=\left[1-\sinh^{-2}2K\sinh^{-2}2(K'+K''+K''')\right]^{\frac{3}{8}} \quad (5.22)$$

The spontaneous magnetization I derived in this method is in consistence with that obtained above in the section IV. Similarly, we could prove that for $\infty > T > T_c$, I = 0 since $\sum_1^{\infty} nk_n k_{-n} = -\infty$, as Σ (1/n) is divergent.

### C. Short-range order

From the results above for the correlation and the spontaneous magnetization, we could summarize as follows: The form of the formula for the correlation or the spontaneous magnetization of the 3D simple orthorhombic Ising model would keep to be the same as the cubic power of that for the 2D rectangular Ising model, in the case that the transformations of $z_2 \to z_{2,eff} = z_2 z_3 z_4$ , $z_2^* \to z_{2,eff}^*$ , $x_2 \to x_2 x_3 x_4$ , and $K' \to K'+K''+K'''$ are performed. Furthermore, from the point of view of the dimensions of units, one could expect the transformation of $N \to N^{\frac{3}{2}}$.

Following Wu's work,[252] the correlation function $S_N$ of the 3D simple orthorhombic Ising lattices is re-written as:

$$S_N^{1/3}=\begin{vmatrix} a_0 & a_{-1} & a_{-2} & \cdot & \cdot & \cdot & a_{-N+1} \\ a_1 & a_0 & a_{-1} & \cdot & \cdot & \cdot & a_{-N+2} \\ a_2 & a_1 & a_0 & \cdot & \cdot & \cdot & a_{-N+3} \\ \cdot & \cdot & \cdot & \cdot & \cdot & \cdot & \cdot \\ \cdot & \cdot & \cdot & \cdot & \cdot & \cdot & \cdot \\ \cdot & \cdot & \cdot & \cdot & \cdot & \cdot & \cdot \\ a_{N-1} & a_{N-2} & a_{N-3} & \cdot & \cdot & \cdot & a_0 \end{vmatrix}, \quad (5.23)$$

where

$$a_n = \frac{1}{2\pi}\int_0^{2\pi} \varphi(\theta) e^{-in\theta} d\theta \,, \tag{5.24}$$

with

$$\varphi(\theta) = e^{i\delta *} = \left[\frac{(1-\alpha_1 e^{i\theta})(1-\alpha_2 e^{-i\theta})}{(1-\alpha_1 e^{-i\theta})(1-\alpha_2 e^{i\theta})}\right]^{\frac{1}{2}} . \tag{5.25}$$

Here

$$\alpha_1 = z_1 z^*_{2,eff} = z_1 z^*_2 z^*_3 z^*_4 \,, \tag{5.26a}$$

$$\alpha_2 = z^*_{2,eff} / z_1 = z^*_2 z^*_3 z^*_4 / z_1 \,. \tag{5.26b}$$

For the temperature above the critical point, $\infty > T > T_c$, similar to the Wu's procedure[252] and according to the Szego's theorem,[253,254] the $N \times N$ Toeplitz determinant $R_N$ is given by:

$$\lim_{N\to\infty} (-1)^N R_N = \left[(1-\alpha_1^2)(1-\alpha_2^{-2})(1-\alpha_1/\alpha_2)^2\right]^{\frac{3}{4}} . \tag{5.27}$$

$S_N$ reads as:

$$S_N = (-1)^N R_{N+1} x_N \,. \tag{5.28}$$

The Wiener – Hopf procedure is employed for solving $x_N$.[252] For $\infty > T > T_c$, the desired $x_N$ is found to be as:

$$\begin{aligned} x_N^{1/3} &\doteq \frac{1}{2\pi i}\oint d\xi \cdot \xi^{N-1} P(\xi^{-1}) Q(\xi)^{-1} \\ &= \frac{1}{2\pi i}\oint d\xi \cdot \xi^{N-1} \left[(1-\alpha_1\xi)(1-\alpha_1\xi^{-1})(1-\alpha_2^{-1}\xi)(1-\alpha_2^{-1}\xi^{-1})\right]^{-\frac{1}{2}} \end{aligned} \tag{5.29}$$

Then

$$S_N^{1/3} = \left[(-1)^N R_{N+1}\right]^{1/3} x_N^{1/3} = \left[(1-\alpha_1^2)(1-\alpha_2^{-2})(1-\alpha_1/\alpha_2)^2\right]^{1/4} \cdot x_N^{1/3}. \tag{5.30}$$

Expanding the integrand of the equation and integrating term by term lead to $S_N$ and finally, performing the transformation of $N \to N^{\frac{3}{2}}$ results in:

$$\begin{aligned} S_N &= \pi^{-\frac{3}{2}} N^{-\frac{9}{4}} \alpha_2^{-3N^{3/2}} (1-\alpha_1^2)^{\frac{3}{4}} (1-\alpha_2^{-2})^{-\frac{3}{4}} (1-\alpha_1\alpha_2)^{-\frac{3}{2}} \\ &\times \left[1 + \frac{1}{4} N^{-\frac{3}{2}} A_{1>} + \frac{3}{16} N^{-3} (A_{2>} - \frac{5}{6}) + \frac{15}{64} N^{-\frac{9}{2}} (A_{3>} - \frac{7}{6} A_{1>}) + \cdots \right]^3 \end{aligned} \tag{5.31}$$

The functions and parameters in eqs. (5.28) – (5.31) were defined in section II of Wu's paper.[252]

For temperatures below $T_c$, the spontaneous magnetization of the 3D simple orthorhombic Ising magnet would behave as:

$$S_\infty = (1-\alpha_1^2)^{\frac{3}{4}} (1-\alpha_2^2)^{\frac{3}{4}} (1-\alpha_1\alpha_2)^{-\frac{3}{2}}, \tag{5.32}$$

so that,

$$S_\infty^{\frac{1}{3}} = (1-\alpha_1^2)^{\frac{1}{4}} (1-\alpha_2^2)^{\frac{1}{4}} (1-\alpha_1\alpha_2)^{-\frac{1}{2}} \prod_{n=N}^{\infty} x_{0n}^{1/3}. \tag{5.33}$$

Then one would obtain:

$$\begin{aligned} S_N = (1-\alpha_1^2)^{\frac{3}{4}} (1-\alpha_2^2)^{\frac{3}{4}} (1-\alpha_1\alpha_2)^{-\frac{3}{2}} \Bigg\{ 1 + \frac{1}{2\pi N^3} \alpha_2^{2N^{3/2}} (\alpha_2^{-1} - \alpha_2)^{-2} \Bigg[ 1 + \frac{1}{2N^{3/2}} (-A_{1<} + 4x'_3) \\ + \frac{3}{4N^3} \left( -A_{2<} + A_{1<}^2 - 2x'_3 A_{1<} + 6x'^2_3 - \frac{13}{6} \right) + \cdots \Bigg] \Bigg\}^3 \end{aligned} \tag{5.34}$$

after performing the transformation of $N \to N^{\frac{3}{2}}$. The parameters in eq. (5.34) are the same as those in section III of Wu's paper.[252]

At the critical point, T = T$_c$, one would derive:

$$\left(S_N^{(0)}\right)^{\frac{1}{3}} = e^{\frac{1}{4}} 2^{\frac{1}{12}} A^{-3} N^{-\frac{1}{4}} \left(1 - \frac{1}{64} N^{-2} + \cdots\right). \tag{5.35}$$

Namely,

$$S_N^{(0)} = e^{\frac{3}{4}} 2^{\frac{1}{4}} A^{-9} N^{-\frac{3}{4}} \left(1 - \frac{1}{64} N^{-2} + \cdots\right)^3. \tag{5.36}$$

Performing the transformation of $N \to N^{\frac{3}{2}}$ leads to:

$$S_N^{(0)} = e^{\frac{3}{4}} 2^{\frac{1}{4}} A^{-9} N^{-\frac{9}{8}} \left(1 - \frac{1}{64} N^{-3} + \cdots\right)^3 = e^{\frac{3}{4}} 2^{\frac{1}{4}} A^{-9} N^{-\frac{9}{8}} \left(1 - \frac{3}{64} N^{-3} + \cdots\right). \tag{5.37}$$

Similar to the 2D Ising case,[252] the following relationship is valid

$$S_N \sim (1-\alpha_1^2)^{\frac{3}{4}} (1-\alpha_1\alpha_2)^{-\frac{3}{2}} S_N^{(0)}, \tag{5.38}$$

for the whole temperature range of ∞ > T > 0. More explicitly, at the critical point:

$$S_N \sim (1+\alpha_1)^{\frac{3}{4}} (1-\alpha_1)^{-\frac{3}{4}} S_N^{(0)}, \tag{5.39}$$

as N → ∞. Thus the correlation at the critical point would behave as:

$$S_N \sim e^{\frac{3}{4}} 2^{\frac{1}{4}} A^{-9} (1+\alpha_1)^{\frac{3}{4}} (1-\alpha_1)^{-\frac{3}{4}} N^{-\frac{9}{8}} \left(1 - \frac{3}{64} N^{-3} + \cdots\right), \tag{5.40}$$

approximately for large N.

Therefore, for the 3D Ising lattice, the correlation at the critical point could be written as:

$$\Gamma_c(r) \approx D\left(\frac{a}{r}\right)^{\frac{9}{8}} = D\left(\frac{a}{r}\right)^{d-2+\eta} . \tag{5.41}$$

The critical exponent η is found to be 1/8 for d = 3. The Fourier transformation yields:

$$\hat{\chi}(k, v_c) \approx 4\pi \int_0^{\frac{1}{ka}} D\left(\frac{a}{r}\right)^{\frac{9}{8}} e^{ik\cdot r} r^2 dr \approx \frac{\hat{D}}{|ka|^{\frac{15}{8}}} = \frac{\hat{D}}{|ka|^{2-\eta}} . \tag{5.42}$$

From Eq. (5.42), again, the critical exponent η is 1/8.

## D. True range of the correlation

The true range $\kappa_x$ of the correlation of the 3D simple orthorhombic Ising system could be determined by the procedure similar to that used for the 2D Ising system.[103,247]

$$\begin{aligned} [\kappa_x a]^{\frac{3}{2}} &= -\lim_{n\to\infty} \ln \frac{\lambda_{\max}^-}{\lambda_{\max}^+} = 2(K^* - K' - K'' - K''') = \ln \coth K - 2(K' + K'' + K''') \\ &= \ln \frac{(1-z_2)(1-z_3)(1-z_4)}{z_1(1+z_2)(1+z_3)(1+z_4)} = \ln \frac{x_2 x_3 x_4 (1+x_1)}{(1-x_1)} \end{aligned} \tag{5.43}$$

with $\kappa_x = 1/\xi$, ξ is the correlation length. The power 3/2 for $\kappa_x a$ is added, in accordance with the scale dimension. At the Curie temperature, $\kappa_x \to 0$ or $\xi \to \infty$, namely,

$$\frac{(1-z_2)(1-z_3)(1-z_4)}{z_1(1+z_2)(1+z_3)(1+z_4)} = \frac{x_2 x_3 x_4 (1+x_1)}{(1-x_1)} = 1 . \tag{5.44}$$

For the simple cubic lattice, the relation is reduced to:

$$[\kappa_x a]^{\frac{3}{2}} = -\lim_{n\to\infty} \ln \frac{\lambda^-_{\max}}{\lambda^+_{\max}} = 2(K^* - 3K) = \ln\coth K - 6K$$
$$= \ln \frac{(1-z)^3}{z(1+z)^3} = \ln \frac{x^3(1+x)}{(1-x)} \tag{5.45}$$

At the Curie temperature, $\kappa_x \to 0$ or $\xi \to \infty$, we would have,

$$\frac{(1-z)^3}{z(1+z)^3} = \frac{x^3(1+x)}{(1-x)} = 1. \tag{5.46}$$

It is:

$$z^4 + 4z^3 + 4z - 1 = 0, \tag{5.47}$$

or

$$x^4 + x^3 + x - 1 = 0. \tag{5.48}$$

One of solutions of the equations above leads to the critical point of the simple cubic Ising lattice: $z_c = \sqrt{5} - 2$, or $x_c = \frac{\sqrt{5}-1}{2}$. It is noticed that the values of $z_c$ and $x_c$ for the 3D simple cubic Ising lattice differ, whereas those for the 2D square Ising lattice are the same.[103,247] For the 2D square Ising lattice,[103,247] $z_c = x_c = \sqrt{2} - 1$ is drived from $\frac{(1-z)}{z(1+z)} = \frac{x(1+x)}{(1-x)} = 1$, i.e., $z^2 + 2z - 1 = 0$ and $x^2 + 2x - 1 = 0$. It is worthwhile noting that only the 2D square Ising lattice satisfies the existence of the same values for $z_c$ and $x_c$, since $z = \frac{1-x}{1+x}$ (or $x = \frac{1-z}{1+z}$) is always valid for $z = \tanh K$ and $x = e^{-2K}$ and, if one set $z = x$ one would immediately obtain $x^2 + 2x - 1 = 0$ (or $z^2 + 2z - 1 = 0$).

Near the critical point, the critical behavior of the true range $\kappa_x$ of the correlation could be described by:

$$[\kappa_x a]^{\frac{3}{2}} = 5\ln\frac{\sqrt{5}+1}{2}\left(\frac{T-T_c}{T_c}\right)\left\{1+\left[\frac{\sqrt{5}}{5}\ln\frac{\sqrt{5}+1}{2}-1\right]\left(\frac{T-T_c}{T_c}\right)+o\left[\left(\frac{T-T_c}{T_c}\right)^2\right]\right\} \quad (5.49)$$

Thus the leading term of the true range $\kappa_x$ of the correlation is taken to be:

$$\kappa_x a \propto \left[\frac{T-T_c}{T}\right]^{\frac{2}{3}}. \quad (5.50)$$

The critical exponent $\nu$ is found to equal to 2/3. It is noticed again that that the two golden solutions, $\frac{\sqrt{5}-1}{2}$ and $\frac{\sqrt{5}+1}{2}$ of the equation $x^2 + x - 1 = 0$ appear in the formula for the critical behavior of the true range $\kappa_x$ of the correlation of the 3D simple cubic Ising model, while the two silver solutions of the equation $x^2 + 2x - 1 = 0$ show up in the Fisher's formula for the 2D square Ising model.[103,247] It is interesting to note that if the coordinates of a point in the golden spiral are written as $[r = \varphi^{2\theta/\pi}, \theta]$ as graphed on a polar axis, the golden ratio is related with the golden spiral by $\ln\varphi = \ln\left(\frac{\sqrt{5}+1}{2}\right) = \frac{\pi}{2\theta}\ln r$.

The combination of the critical behaviors of the spin correlation functions results in:

$$\Gamma(r,T) \approx D\left(\frac{a}{r}\right)^{\frac{9}{8}} e^{-\kappa r}\left[1+Q(\kappa,r)\right] = D\left(\frac{a}{r}\right)^{\frac{9}{8}} e^{-\kappa r}\left(1-\frac{3}{64}r^{-3}+\cdots\right). \quad (5.51)$$

In the following sub-section, we shall focus on evaluating the correlation function $\langle s_{ijk}s_{abk}\rangle_{Av}$ between spins of two sites within one of planes.

### E. Procedure for evaluating averages

Following the previous work done by Kaufman and Onsager,[18] we shall first evaluate the correlation function $\langle s_{ijk}s_{abk}\rangle_{Av}$ between spins of two sites within one of planes. We shall employ the approximation that all eigenvalues of **V** are negligible as compared with the largest one, when the power of the eigenvalues is high enough. In order to make use of this approximation, we transform (5.2) by **Ψ** which brings **V** to its diagonal form:

$$< s_{1,1,k} s_{1+a,b,k} >_{Av} = \left\{ \sum_{i=1}^{2^{nl}} \lambda_i^{\ m} \right\}^{-1} \cdot \left\{ \sum_{i=1}^{2^{nl}} \lambda_i^{\ m-a} \cdot (\Psi s_{1,k} V^a s_{b,k} \Psi^{-1})_{ii} \right\}. \qquad (5.52)$$

Similar to Kaufman and Onsager's work,[18] for simplicity of notation, we will no longer differentiate between odd- and even-indexed angles for $\Psi_+$ and $\Psi_-$ and we use at all temperatures the following:

$$< s_{1,1,k} s_{1+a,b,k} >_{Av} \cong \lambda^{-a} \cdot (\Psi_+ s_{1,k} V^a s_{b,k} \Psi_+^{\ -1})_{11}. \qquad (5.53)$$

The quantity **Ψ** has been shown in section III, in terms of the rotation which it reduces in the spinor base $\mathbf{P_1}$, $\mathbf{Q_1}$, $\mathbf{P_2}$, $\mathbf{Q_2}$, …, $\mathbf{P_{nl}}$, $\mathbf{Q_{nl}}$. In section III, we had:

$$\Psi = g \cdot S(TH), \qquad (5.54)$$

where the transformations **g**, **T** and **H** are represented by Eqs. (3.30), (3.26) and (3.36), respectively. The simplest correction function is found to be: [18]

$$< s_1 s_2 >_{AV}^{\ 1/3} = \frac{1}{(nl)^{3/2}} \sum_t \cos \delta_t^* = -\cosh^2 K^* \cdot \Sigma_1 + \sinh^2 K^* \cdot \Sigma_{-1}, \qquad (5.55)$$

with

$$\Sigma_{k-l} = \frac{1}{n \cdot l \cdot o} \times$$
$$\sum_{t_x, t_y, t_z} \left\{ w_x \cos\left[ (k-l) \frac{t_x \pi}{n} + \delta_t' \right] + w_y \cos\left[ (k-l) \frac{t_y \pi}{l} + \delta_t' \right] + w_z \cos\left[ (k-l) \frac{t_z \pi}{o} + \delta_t' \right] \right\}$$

(5.56)

and the general form for the correction function along a row in the 3D lattice is derived as:

$$< s_1 s_{1+k} >_{AV}{}^{1/3} = (-1)^k \left[\cosh^2 K^* \cdot \Delta_k - \sinh^2 K^* \cdot \Delta_{-k}\right], \tag{5.57}$$

with

$$\Delta_k = \begin{vmatrix} \Sigma_1 & \Sigma_2 & \Sigma_3 & \Sigma_4 & \cdot & \cdot & \cdot & \Sigma_k \\ \Sigma_0 & \Sigma_1 & \Sigma_2 & \Sigma_3 & \cdot & \cdot & \cdot & \Sigma_{k-1} \\ \Sigma_{-1} & \Sigma_0 & \Sigma_1 & \Sigma_2 & \cdot & \cdot & \cdot & \Sigma_{k-2} \\ \cdot & \cdot & \cdot & \cdot & \cdot & \cdot & \cdot & \cdot \\ \Sigma_{-k+2} & \Sigma_{-k+3} & \Sigma_{-k+4} & \Sigma_{-k+5} & \cdot & \cdot & \cdot & \Sigma_1 \end{vmatrix}, \tag{5.58}$$

and

$$\Delta_{-k} = \begin{vmatrix} \Sigma_{-1} & \Sigma_{-2} & \Sigma_{-3} & \Sigma_{-4} & \cdot & \cdot & \cdot & \Sigma_{-k} \\ \Sigma_0 & \Sigma_{-1} & \Sigma_{-2} & \Sigma_{-3} & \cdot & \cdot & \cdot & \Sigma_{-k+1} \\ \Sigma_1 & \Sigma_0 & \Sigma_{-1} & \Sigma_{-2} & \cdot & \cdot & \cdot & \Sigma_{-k+2} \\ \cdot & \cdot & \cdot & \cdot & \cdot & \cdot & \cdot & \cdot \\ \Sigma_{k-2} & \Sigma_{k-3} & \Sigma_{k-4} & \Sigma_{k-5} & \cdot & \cdot & \cdot & \Sigma_{-1} \end{vmatrix}, \tag{5.59}$$

Montroll, Potts and Ward[249] proved that the correlations in a row in the form of a single Toeplitz determinant are equivalent to the Onsager-Kaufman results in the form of two Toeplitz determinants. Therefore, for the present 3D system, the results obtained following the Onsager-Kaufman's process are also equivalent to those obtained following Montroll, Potts and Ward's process.

## VI. SUSCEPTIBILITY

The susceptibility of the 3D simple orthorhombic Ising system could be derived by a procedure similar to that developed by Fisher.[98,99] The susceptibility could be evaluated by:

$$\chi_0(T) = \frac{N\mu^2}{k_B T} \sum_{l,m,n} v_{lmn} \omega_{lmn}(T). \tag{6.1}$$

For the 3D simple orthorhombic Ising lattices, one would have from the correlation at the critical point:

$$\omega_{0mn}(T_c) = < s_{1,1,1} s_{1,1+m,1} > \approx \frac{A}{m^{9/8}} \quad (m \to \infty) \tag{6.2}$$

where A is a constant. The relation above could be extended to be:

$$\omega_{lmn}(T_c) \approx \frac{A(\theta,\varphi)}{k^{9/8}} \quad (k \to \infty) \tag{6.3}$$

where $k^2 = l^2 + m^2 + n^2$, A (θ, φ) is similar to A. Similar to the 2D case, one would arrive:

$$\omega_{lmn}(T) \approx A(\theta,\varphi) k^{-\frac{9}{8}} \exp\left[ -kb(1-\frac{T_c}{T})^{\frac{2}{3}} \right]. \tag{6.4}$$

Thus, one could derive the susceptibility:

$$\chi_F(T) \approx \frac{N\mu^2}{k_B T} \int_0^\infty \int_0^{2\pi} \int_0^{\pi} A(\theta,\varphi) k^{-\frac{9}{8}} \exp\left[ -kb(1-\frac{T_c}{T})^{\frac{2}{3}} \right] k^2 \sin\theta d\theta d\varphi dk \, . \tag{6.5}$$

Finally,

$$\chi_F(T) \approx \frac{N\mu^2}{k_B T} \frac{4\pi b^{-\frac{15}{8}} \Gamma(\frac{15}{8})}{(1-\frac{T_c}{T})^{\frac{5}{4}}} \propto \frac{1}{(1-\frac{T_c}{T})^{\frac{5}{4}}} \, . \tag{6.6}$$

Therefore, the critical exponent γ is equal to 5/4 for the 3D Ising model.

## VII. CRITICAL EXPONENTS

The critical exponents of various physical systems have been investigated intensively, since they are the most important factors for understanding the critical behaviors of the continuous phase transitions. The 2D Ising model is among a few examples solved explicitly, which provides the exact values of the critical exponents.[13] Because of the lack of the exact solution for the 3D Ising model, the most reliable information on its critical behavior is provided by renormalization group theory near the critical point.[126-191,208-225] Up to now, the region near the critical point has been explored by various methods of approximation with high precision,[59-225] but the exact mathematical solution for the 3D Ising model is the key for deriving the exact values of the critical exponents.

Fisher and Chen evaluated the validity of hyperscaling in three dimensions for scalar spin systems.[255] By applying a real space version of the Ginzburg criterion, the role of fluctuations and thence the self-consistency of the mean field theory were assessed in a simple fashion for a variety of phase transitions.[256] Based on the concept of the marginal dimensionality d*, the critical behaviors were discussed.[257] When the dimensionality d for a system is larger than the corresponding d*, the mean field theory describes the correct critical behavior,[142] where d* = 4 for the short-range interactions and d* = 3 for uniaxial dipolar ferromagnets or ferroelectrics and for tricritical behavior. When d = d*, the RG equations are exact and the Landau behavior is modified by additional "weak" singular behavior such as logarithmic corrections. One then makes the so-called ε expansion, ε = d* - d, to estimate the critical behavior for d < d*.

Various physical quantities diverge to infinite or converge to zero as the temperature or other variable approaches its critical point. The exponents near the critical temperature are defined as follows: In the region just below the critical point, the spontaneous magnetization is well approximated by a power law,[142-144,152,153]

$$M \propto (T_c - T)^{\beta} . \qquad (7.1)$$

The magnetization at $T_c$ has a critical behavior as:

$$M \sim |h|^{\frac{1}{\delta}}. \quad (7.2)$$

The correlation function has property that if T > $T_c$, it falls off with r, with the asymptotic behavior for large r:

$$\Gamma(r) \sim \exp\left[-\frac{r}{\xi(T)}\right], \quad (7.3)$$

where ξ (T) is the correlation length, which approaches infinity as T → $T_c$.

$$\xi(T) \sim (T - T_c)^{-\upsilon}. \quad (7.4)$$

The correlation function at $T_c$ falls as a power of r in the form:

$$\Gamma_c(r) \sim \frac{1}{r^{d-2+\eta}}. \quad (7.5)$$

The critical behavior of the magnetic susceptibility is described as:

$$\chi \sim (T - T_c)^{-\gamma}. \quad (7.6)$$

Similarly, the critical exponent α controls the critical behavior of the specific heat near the critical temperature:

$$C_h \sim (T - T_c)^{-\alpha}. \quad (7.7)$$

The zero-angle (k → 0) scattering intensity apparently diverges at the critical point as:

$$I_c(k) \sim \frac{1}{k^{2-\eta}}, \text{ (k} \rightarrow 0\text{)}. \quad (7.8)$$

An inverse range parameter $\kappa_1$ (T) is defined to measure the slope of 1/I(k, T) against $k^2$ as k → 0. This again vanishes at the critical point and in zero field it has:

$$\kappa_1(T) \sim (T - T_c)^{\nu}. \quad (7.9)$$

Actually, there are only six critical exponents α, β, γ, δ, η and ν, which are related by the following four scaling laws: [142-144,152,153,258,259]

$$\alpha + 2\beta + \gamma = 2 \quad \text{(Rushbrooke's law);} \tag{7.10a}$$

$$\gamma = \beta(\delta - 1) \quad \text{(Widom's law);} \tag{7.10b}$$

$$\gamma = \nu(2 - \eta) \quad \text{(Fisher's law);} \tag{7.10c}$$

$$\nu d = 2 - \alpha \quad \text{(Josephson's law).} \tag{7.10d}$$

There are several other expressions between the critical exponents, but not independent of these four equations.[255,260,261] Therefore, only two independent parameters exist among these critical exponents.

The value for the critical exponent ν can be evaluated in a simpler manner. The scale dimension [$\kappa_x$] for the true range of the correlation is 1, while the scale dimension [f] for the free energy equals to d. From

$$\xi = 1/\kappa_x \propto |T - T_c|^{-\nu} = \tau^{-\upsilon}. \tag{7.11}$$

Here $\tau = |T - T_c|$. We have

$$f \sim \kappa_x^d \sim |T - T_c|^{\nu d} = \tau^{\upsilon d}. \tag{7.12}$$

On the other hand, the specific heat $C_B$ is determined by:

$$C_B = -T\frac{\partial^2 f}{\partial T^2} \sim \tau^{\nu d - 2} \sim \tau^{-\alpha}. \tag{7.13}$$

One has the relation of $\nu d = 2 - \alpha$, which could be put back to Eq. (7.12) to have:

$$\kappa_x \sim \tau^{\nu} = \tau^{\frac{2-\alpha}{d}}. \tag{7.14}$$

Therefore, we have $\nu = 2/3$ for $d = 3$ and $\alpha = 0$, in comparison of $\nu = 1$ for $d = 2$ and $\alpha = 0$.

According to Ryazanov,[262] three temperature regions: 1) $\tau < - r^{-3/2}$, 2) $- r^{-3/2} < \tau < r^{-3/2}$, 3) $\tau > r^{-3/2}$, exist with different behaviors of the correlation. r represents the distance (p in Ryazanov's paper). Compared with what Ryazanov described for the 2D Ising system,[262] the transformation of $r \to r^{3/2}$ has been performed for the present 3D Ising system. In the second region that is the vicinity of the critical point, the distance to the phase transition point is small than the temperature fluctuations ($\sim 1/\sqrt{N}$, N is the number of particles) in regions with dimensions of order $r^{3/2}$. The temperature fluctuations is defined as $\tau_f = r^{3/2}$ for $d = 3$. It is expected that, the correlation functions depend only on the ratio $\frac{\tau}{\tau_f} = r^{3/2}\tau$. In the region with large values of $r^{3/2}\tau$, the correlation decays like $\exp\left(-r\tau^{2/3}\right)$, also in agreement with $\nu = 2/3$. Near the critical point of the 3D Ising magnet, the spontaneous magnetization behaves as $I \sim \tau^{\frac{3}{8}}$, while the long-range order correlation behaves like $S_\infty \sim \tau^{\frac{3}{4}}$. The relationship between the average temperature and the distance is $\tau \approx \frac{1}{r^{3/2}}$ and the correlation function is of the order of $\tau^{3/4} = \frac{1}{r^{9/8}}$. The correlation function near the critical point is:

$$\Gamma_c(r) \sim \frac{1}{r^{9/8}}, \tag{7.15}$$

and

$$\hat{\chi}(k, v_c) \sim \frac{1}{|ka|^{2-\eta}}, \tag{7.16}$$

in good agreement with the results in the section V. The critical exponent $\eta$ is found to be 1/8. It is expected that just like the 2D case, the correlations for the 3D Ising lattice along the rows and along the diagonals turn out to be different only in the region much above the transition point, where the correlation radius is small and is close to the distance between the nearest neighbors.[262]

The critical exponents $\alpha = 0$, $\beta = 3/8$, $\gamma = 5/4$, $\delta = 13/3$, $\eta = 1/8$ and $\nu = 2/3$ derived based on the two conjectures for the 3D simple orthorhombic Ising lattices satisfy the scaling laws, showing universality behaviors. These putative exact critical exponents are tabulated in Table 1, together with exact values for 1D and 2D Ising lattices, approximate values obtained by the renormalization group and the high temperature series expansion for the 3D Ising lattice, and those of the mean field theory. As what we have done for the critical point, we shall compare first grossly with the data for sixty years and then carefully with those obtained recently. It is clear from Table 1 that the values of the critical exponents for temperatures below $T_c$, taken from Fisher's paper,[103] are not very reliable because of the appearance of the negative critical exponent $\eta$. Thus we should not include this group of the critical exponents in the discussion below. As stated above, the specific heat of the 3D Ising lattice has the same singularity of logarithm as the 2D one, whereas the specific heat of the 4D system as predicted by the mean field theory shows a discontinuous at the critical point. The exact values for the critical exponent $\alpha$ are all equal to zero for the 2D, 3D and 4D (mean field) systems. The small values, but non-zero, of the critical exponent $\alpha$, range from 0.0625 to 0.125, obtained by the renormalization group and the high temperature series expansion, are attributed to the uncertainties of these approximation methods, because of the existence of systematic errors. It is understood that the behavior of the curves with a power law of $\alpha < 0.2$ are very much close to that of logarithms so that the approximation approaches cannot figure it out. The exact values for the critical exponent $\beta$ are 1/8, 3/8 and 1/2 for 2D, 3D and (mean field) 4D systems, respectively. The putative exact critical exponent $\beta$ of $3/8 = 0.375$ is slightly higher than the approximation values of $0.312 \sim 0.340$. The exact values for the critical exponent $\gamma$ are 2, 7/4, 5/4 and 1 for 1D, 2D, 3D and (mean field) 4D systems, respectively. The putative exact critical exponent $\gamma$ of $5/4 = 1.25$ is exactly the same as the approximation values ranging from 1.244 to 1.25, within the errors of 0.48%.

Series for the initial susceptibility at high temperatures provided the smoothest and most regular patterns of behavior of coefficients, which were all found to be positive in sign, and used to estimate the Curie temperatures and critical exponents.[107] One was tempted to the conjecture that the exact value for the critical exponent γ of the 3D Ising lattice is simply γ = 5/4.[95,103] Even if it is not exact it appears to be accurate to within 1/2% and certainly provides an excellent representation of form of the susceptibility coefficients.[83,95,103,107] We are sure with the confidence that the exact critical exponent γ equals to 5/4 for the 3D Ising model. The putative exact critical exponent δ of 13/3 = 4.333… for the 3D Ising lattice is between 15 and 3 for the 2D and 4D (mean field) Ising lattices, which is slightly lower than the approximation values of 4.46 ~ 5.15. The putative exact critical exponent η of 1/8 = 0.125 for the 3D Ising lattice is half that for the 2D Ising lattice, which is slightly larger than the approximation values of 0 ~ 0.055. The putative exact critical exponent ν of 2/3 for the 3D Ising lattice looks very reasonable since it is between 1 and 1/2 for the 2D and 4D (mean field) Ising lattices and is very close to the approximation ones of 0.625 ~ 0.642. It could be concluded that all the putative exact critical exponents for the 3D Ising lattice are located between those for the 2D and 4D (mean field) ones, which are close to those obtained by various approximation methods, if compared grossly.

The critical exponents α = 1/8, β = 5/16, γ = 5/4, δ = 5, η = 0 and ν = 5/8, suggested by Fisher[103] and Domb,[107] which were established by the conjectures based on the results of the series expansions, have been well-accepted by the community for almost forty years. However, as remarked by Domb,[107] there are significant discrepancies in numerical values: There are well illustrated by the formula for η, $2-\eta = d\frac{\delta-1}{\delta+1}$, if one takes δ = 5 as has been estimated for the Ising model of spin 1/2 in 3D, one will find that η = 0. But direct numerical analysis gives η ≈ 1/18, and this is consistent with the result of renormalization group expansions. If one substitutes η ≈ 1/18 into this equation, one will obtain δ ≈ 4.7, which is well outside the confidence limits in the analysis of series expansions. This must be regarded as a serious inconsistency.[107,263] Furthermore, the value of η = 0 in the group of the critical exponents, suggested by Fisher[103] and Domb,[107] gives the same result as what the main field theory predicts, which is certainly not relevant. $\eta$ denotes the deviation from the Ornstein – Zernike behavior, which certainly cannot be zero.[236]

Regarding the actual values of critical exponents, one knows that in 2D, they are all simple integers and fractions. Numerical data suggest a similar result in 3D. It would be difficulty to support any such conclusions from a renormalization group treatment. Nevertheless, as stated by Domb,[107] the possibility is appealing, and hints that there may be simplifying features of the 3D Ising model which remain to be discovered. Our putative exact critical exponents $\alpha = 0$, $\beta = 3/8$, $\gamma = 5/4$, $\delta = 13/3$, $\eta = 1/8$ and $\nu = 2/3$ for the 3D Ising lattices are all simple integers and fractions, which are much simpler than those suggested by Fisher and Domb.[103,107] It is interesting to compare our putative critical exponents further with those for the 1D Ising model. It is impossible to derive the critical exponents $\alpha$ and $\beta$ directly for the 1D Ising model, since there is no spontaneous magnetization at finite temperature. Supposing that the scaling laws are still held for the 1D case, however, one would have $\alpha = 0$ and $\beta = 0$ as derived from $\gamma = 2$, $\delta = \infty$, $\eta = 1$ and $\nu = 2$ (as shown in Table 1).[264] The difference between the critical exponents $\beta$ for the 1D and 2D Ising lattices is 1/8, which is the same as the difference between the critical exponents $\beta$ for the 3D and 4D Ising lattices. The difference between the critical exponents $\beta$ for the 2D and 3D Ising lattices is twice as this value. Similarly, the difference between the critical exponents $\gamma$ for the 2D and 3D Ising lattices is twice as the difference between the critical exponents $\gamma$ for the 1D and 2D Ising lattices. The latter is the same as the difference between the critical exponents $\gamma$ for the 3D and 4D Ising lattices, which equal to 1/4. Indeed, the feature of the nature is very simple, symmetric and beautiful.

Recent advances in the Monte Carlo and the renormalization group techniques have improved the precision of calculations of the critical exponents. If compared precisely, our putative exact values $\alpha = 0$, $\beta = 3/8$, $\gamma = 5/4$, $\delta = 13/3$, $\eta = 1/8$ and $\nu = 2/3$ indeed differ with the values of $\alpha = 0.110(1)$, $\beta = 0.3265(3)$, $\gamma = 1.2372(5)$, $\delta =$ 4.789(2), $\eta$ = 0.0364(5) and $\nu$ = 0.6301(4), well - established in the Pelissetto and Vicari's review,[154] in consideration of the high – precision of simulations. Nowadays, these Pelissetto and Vicari's values are well – accepted by wide community. We could evaluate the difference between the putative exact solution and the approximations by the errors of $\Delta\alpha = |\alpha^{EX} - \alpha^{PV}| / \alpha^{EX}$, $\Delta\beta = |\beta^{EX} - \beta^{PV}| / \beta^{EX}$, …… for all the critical exponents, where the supscripts of EX and PV denote the exact solutions and the Pelissetto and Vicari's values. We find that $\Delta\alpha = \infty$, $\Delta\beta$ = 12.93%, $\Delta\gamma$ = 1.02%, $\Delta\delta$ = 10.51%, $\Delta\eta$ = 70.88% and $\Delta\nu$ = 5.48%, respectively, for these critical exponents. It is evident that our putative exact solution for the critical exponent $\gamma$ is very close to the approximation one, within the error of 1.02%. It is understood that all the differences for the estimates of the critical exponents remained actually arise from the determination of the critical exponent $\alpha$, since there are only two independent parameters among all the six critical exponents. As discussed in details in Section VIII, such differences between the putative exact solutions and the approximations are attributed to the existence of systematical errors of the Monte Carlo and the renormalization group techniques. In the Binder and Luijten's review,[213] the values of $y_t = 1/\nu = 1.588(2)$ and $y_h = 3 - \beta/\nu = 2.482(2)$ are established, in accordance with data in various refernces published in 1980 – 1999. Our putative exact solutions yield $y_t = 1/\nu = 1.5$ and $y_h = 3 - \beta/\nu = 2.4375$, which are very close with the Binder and Luijten's values within the errors of 5.87% and 1.83% respectively. The origins of these errors will be discussed in detals in section VIII.

Note, only exception among all the references is Kaupuzs's paper,[167] which gave the predictions of $\gamma = 5/4$ and $\nu = 2/3$, exactly the same as what we found for the exact solutions. Kaupuzs discussed different perturbation theory treatments of the Ginzburg–Landau phase transition model.[167] The usual perturbation theory was reorganized by appropriate grouping of Feynman diagrams of $\varphi^4$ model with O(n) symmetry.[167] As a result, equations for calculation of the two–point correlation function were obtained which allow to predict possible exact values of critical exponents in two and three dimensions by proving relevant scaling properties of the asymptotic solution at (and near) the criticality.[219,265-267]

It is emphasized that our putative exact critical exponents would represent the behaviors of the system exactly at the critical region, as the critical point could be fixed exactly, which would have physical significances correlated directly with the existence of the fourth curled – up dimension. From the analysis above, it is clear that the estimate of the critical exponent $\alpha$ plays a kay role for deviations between the exact solutions and the approximations. In a deeper understanding, the prediction of a zero critical exponent $\alpha$ reveals the physical significances completely differing with the non-zero critical exponent $\alpha$. As the dimension of the systems alters from 1D to 4D, the critical behaviors should change in a subsequence of continued logic. Namely, all critical indices should vary smoothly with dimensionality.[264] The 1D Ising model shows no ordering at finite temperature. The specific heat of the 2D Ising model behaves logarithmically near the critical point, with a zero critical exponent $\alpha$. The 4D Ising model has a zero critical exponent $\alpha$ also, but with the discontinuous specific heat at the critical point. It is hard for us to understand why the 3D Ising lattice has a power law with a non-zero critical exponent $\alpha$, in case that both the 2D and 4D Ising lattices have the zero critical exponent $\alpha$. Up to date, nobody has succeeded in constructing explicitly a closed form of the eigenvalues as well as the partition function, which leads to a power law with a non-zero critical exponent $\alpha$ for the 3D Ising lattice and can be reduced to a logarithmic singularity with a zero critical exponent for the 2D Ising lattice. From the logic point view of the evolution of physical properties with the dimensionality, it would not be unreasonable to believe that the specific heat of the 3D Ising model has a logarithmic singularity at the critical point. On the other hand, the temperature dependences of the specific heat of the 2D and 3D Ising models, as revealed by various approximation methods,[13,59] have the same trend and the same behaviour near the critical point. The more accurate the method is, the more like logarithmic singularity the specific heat is. As shown in the Newell and Montroll's review,[59] Wakefield's method gave the lowest value for the critical point $1/K_c = 4.497$,[73,74] which had been known to be accurate everywhere except near the critical point.[59] The temperature dependence of the specific heat obtained by the extrapolation of high – and low – temperature expansions of the Wakefield's method looks more like logarithmic singularity than others.[59] It is relevant that the exact solution for the specific heat of the 3D Ising model would behave as the logarithmic singularity at the exact critical point that is further lower than the Wakefield's value. As well-known, the critical exponent $\alpha$ proves considerably harder to calculate than the others, by various theoretical and experimental techniques. Furthermore, it is very hard for the approximation approaches and the experimental data fittings to distinguish the critical behaviors of the logarithmic singularity and the small non-zero critical exponent $\alpha < 0.2$. If one would give up the idea of the existence of the non-zero critical exponent $\alpha$ in the 3D Ising model, and accept the logarithmic singularity, then everything would become much easier than before. According to the Fisher's conjecture of the high-temperature series expansion, the exact value for the critical exponent $\gamma$ of the 3D Ising lattice is simply $\gamma = 5/4$,[103] being accurate to within 1/2%, which is the only parameter that can be accurately determined by high-temperature series expansion theories.[103] In consideration of large differences between the critical points of the high-temperature series expansion and the exact solution, we may give a new conjecture that the determination of the critical exponent $\gamma$ is the most insensitive to the exact location of the critical point. Starting from these two critical exponents $\alpha = 0$ and $\gamma = 5/4$, one would immediately find out the others to be $\beta = 3/8$, $\delta = 13/3$, $\eta = 1/8$ and $\nu = 2/3$. In consideration of the insensitive dependence of the critical exponent $\gamma$ to the exact location of the critical point, we suggest here that the critical exponent $\gamma$ is the most reliable one among the others determined by the renormalization group theory and Monte Carlo simulations. Because other critical exponents depends sensitively on the location of the critical point and because the critical point located by these approximation techniques is far from the exact one, these critical exponents determined also deviate from the exact ones. Nevertheless, starting from the two critical exponents of $\alpha = 0$ and $\gamma = 1.2372$, one easily finds out others as $\beta = 0.3814$, $\delta$

= 4.2438, $\eta$ = 0.1442 and $\nu$ = 2/3. Then, comparing this group of the critical exponents with the putative exact values, one find that the critical exponents $\alpha$ and $\nu$ are the same as the putative exact ones, while the critical exponents $\beta$, $\gamma$, $\delta$ and $\eta$ appear to be accurate to within 1.7%, 1.0%, 2.1% and 15%, respectively. This means that the renormalization group theory and Monte Carlo simulations are still suitable for investigating the critical phenomena, however, it is better to focus only on the high-accurate determination of the critical exponent $\gamma$, since the determination of the critical point and the other critical exponents with high accuracies seems unsuccessful.

It is important to compare the putative exact solution of the critical exponents with the experimental data. In this paragraph, we compare the putative exact critical exponents with the early data collected in Kadanoff et al.'s,[105] Fisher's[103] and Wilson's reviews.[149] The critical exponents $\alpha$ of ferromagnetic iron, $CuK_2Cl_4 \cdot 2H_2O$, is not larger than 0.17, while the specific heat of nickel is fitted by a logarithmic singularity.[105] Actually, it is very difficulty to distinguish the fitting of a power law with $\alpha < 0.2$ with that of logarithms. As indicated by Kadanoff et al.,[105] a set of data for the specific heat of $CoCl_2 \cdot 6H_2O$ can be fitted either by a logarithmic singularity or by $\alpha \lesssim 0.19$. The critical exponents $\beta$ of ferromagnetic iron, nickel, EuS, $YFeO_3$ and $CrBr_2$, determined experimentally, are 0.34 ± 0.02 (or 0.36 ± 0.08), 0.51 ± 0.04 (or 0.33 ± 0.03), 0.33 ± 0.015, 0.55 ± 0.04 (or 0.354 ± 0.005) and 0.365 ± 0.015.[103,105] The values of the critical exponents $\beta$ varies in range of 0.33 ~ 0.55, depending sensitively on the method of the determination, and also on the data range for fitting. The putative exact value for $\beta$ is 3/8 = 0.375, which is very close to the experimental values 0.36 ± 0.08 for iron, 0.365 ± 0.015 for $CrBr_2$, and is not out of the range of 0.33 ~ 0.51 for nickel and 0.354 ~ 0.55 for $YFeO_3$. The experimental data for the critical exponents $\gamma$ of iron, nickel, cobalt, gadolinium are 1.33 ± 0.03, 1.29 ± 0.03 (1.35 ± 0.02), 1.21 ± 0.04, 1.33 (1.16 ± 0.02), respectively.[103,105] The putative exact value for $\gamma$ is 5/4 = 1.25, which is very close to the experimental values 1.29 ± 0.03 for nickel and 1.21 ± 0.04 for cobalt. The critical exponents $\delta$ of nickel, gadolinium, $YFeO_3$ and $CrO_2$ are found experimentally to be 4.2 ± 0.1, 4.0 ± 0.1, 2.8 ± 0.3 and 5.75, respectively.[103,105] The putative exact value for δ is 13/3, very close to the experimental data for nickel and gadolinium. The experimental evidence, notably on iron, indicated $0.2 > \eta \gtrsim 0$,[103,105] which is not inconsistence with the putative exact critical exponent η = 1/8. However, experimental uncertainty for narrow temperature range around the critical temperature precluded drawing any conclusion about the value of η.[105] The critical exponents ν of ferromagnetic iron, antiferromagnetic $Cr_2O_3$, $\alpha$-$Fe_2O_3$ and $KMnF_2$ are 0.64 ± 0.02, 0.67 ± 0.03, 0.63 ± 0.04 and 0.67 ± 0.04,[105] which are very close to the putative exact value 2/3. Furthermore, the putative critical exponents ν is exactly the same as what Wilson accepted in his review article.[149]

There are many factors reducing the accuracies of the experiments, which will be discussed in details in Sec. VIII. For instance, impurities in the magnet sample may affect the value of the critical temperature and in a polycrystal $T_c$ may range in a band which can be of the order of $10^{-4}$. Good measurements then require the use of single crystals of extreme purity and well-defined geometry. From bulk measurements the determinations of critical exponents involve extrapolations to zero internal field values. As stated by Vicentini-Missoni,[268] up to the early 1970's, good data in the critical region are available only on few substances, that is the nickel data of Weiss and Forrer,[269] and those of Kouvel and Comly,[270] the gadolinium data of Graham, Jr.,[271] and the $CrBr_3$ data of Ho and Litster.[272,273] In this paragraph, we compare the exact critical exponents with the data collected in Vicentini-Missoni's review. The data (a) shown in table III of Vicentini-Missoni's chapter[268] were determined by least squares fit of the function $h_{MSG}(x) = E_1\left(\frac{x+x_0}{x_0}\right)\left[1+E_2\left(\frac{x+x_0}{x_0}\right)^{2\beta}\right]^{(\gamma-1)/2\beta}$ to the experimental data;[274-276] in this case β and δ were assumed as the independent exponents and γ is derived using the scaling relation γ = β (δ - 1). The critical exponents obtained by the analysis of Kouvel and Comly's data are as follows: $CrBr_3$: β = 0.364 ± 0.005, δ = 4.32 ± 0.10, γ = 1.21; Gd: β = 0.370 ± 0.010, δ = 4.39 ± 0.10, γ

= 1.25; Ni: $\beta$ = 0.373 ± 0.016, $\delta$ = 4.44 ± 0.18, $\gamma$ = 1.28.[268,270] The critical exponents obtained by the analysis of Weiss and Forrer's data for Ni are: $\beta$ = 0.375 ± 0.013, $\delta$ = 4.48 ± 0.14, $\gamma$ = 1.31.[268,269] The critical exponents derived from several fluids are: $CO_2$: $\beta$ = 0.352 ± 0.008, $\delta$ = 4.47 ± 0.12, $\gamma$ = 1.22; Xe: $\beta$ = 0.35 ± 0.07, $\delta$ = 4.6 ± 0.1, $\gamma$ = 1.26; $He^4$: $\beta$ = 0.355 ± 0.009, $\delta$ = 4.44 ± 0.01, $\gamma$ = 1.24. All the data collected in Vicentini-Missoni's review[268] are in very good agreements with our putative exact solutions. The putative exact critical exponent $\beta$ = 3/8 is exactly the same as the experimental one for Ni and Gd within error bars. The difference between the putative exact critical exponent $\beta$ and the experimental one is 2.9% for $CrBr_3$, 6.1% for $CO_2$, 6.7% for Xe and 5.3% for $He^4$. In consideration of experimental error bars, such difference would reduce to 1.6% for $CrBr_3$, 4.0% for $CO_2$, 0% for Xe and 2.9% for $He^4$. The putative exact value of $\delta$ = 13/3 is exactly the same as the experimental one for $CrBr_3$, Gd and Ni within error bars, also. The difference between the putative exact critical exponent $\delta$ and the experimental one is 3.1% for $CO_2$, 6.2% for Xe and 2.5% for $He^4$. If the experimental error bars were taken into account, such difference would reduce to 0.4% for $CO_2$, 3.8% for Xe and 2.2% for $He^4$.

As shown in Table 1.3 of Kadanoff's chapter,[264] the real fluids show indices close to but not exactly equal to the indices of the 3D Ising model obtained by various approximations. However, the critical indices $\gamma$ = 1.22 ± 0.05, $\delta$ = 4.4 ± 0.2, $\eta$ = 0.123 and 0.65 ± 0.05 of the real fluids are in good agreements with our putative exact values. The differences between them are 2.4% for $\gamma$, 1.5% for $\delta$, 1.6% for $\eta$ and 2.5% for $\nu$, which can be further reduced by taking into account the experimental errors. It can be concluded that the critical indices $\gamma$, $\delta$, $\eta$ and $\nu$ of the real fluids are almost exactly equal to the putative exact ones.

As the temperature T at which two fluid phases are in equilibrium approaches the critical temperature $T_c$, the interfacial tension $\sigma$ is found experimentally to vanish proportionally to a powder of T – Tc: $\sigma \sim (T_c - T)^{\mu}$.[277] The exponent $\mu$ is one of the important and characteristic critical – point indices, with a value that it believed to be universal and is in any event almost certainly in the range $\mu$ = 1.28 ± 0.06. We obtain the putative exact exponent $\mu = 4/3 = 1.3333\ldots$ from the scaling law of $\mu + \nu = 2 - \alpha$ and our exact exponents $\nu = 2/3$ and $\alpha = 0$. Our putative exact exponent $\mu$ and the experimental one coincide, within the experimental errorbars. In fact, the experimental data for the exponent $\mu$ are in range of 1.23 ~ 1.34 for various systems, like, argon, xenon, nitrogen, carbon dioxide, chlorotrifluoromethane, hydrogen, cyclohexane-aniline, cyclohexane-methanol, 3-methylpentane-nitroethane.[277] However, it was emphasized by Buff and Lovett,[278] and Wims et al.[279] that in the measurement of surface tension by capillary rise and equivalent methods, it is not $\mu$ but $\mu - \beta$ that is measured directly, so some of the variability in the value of $\mu$ quoted in Table I of Widom's chapter[277] is a reflection of discrepant value of the assumed $\beta$. The only direction measurement of the interface thickness near the critical point is that on the cyclohexane – methanol system by Huang and Webb,[280] who obtained $\nu = 0.67 \pm 0.02$. This value fits exactly with our putative exact exponent $\nu = 2/3$.

Only were few data reported for the critical exponents of the ferromagnetic transition metals, like Fe, Co and Ni, in last decade.[281,282] Shirane et al. reported the critical exponent $\gamma = 1.333$ for nickel.[281] Seeger et al. obtained the values of $\beta = 0.395(10)$, $\gamma = 1.345(10)$ and $\delta = 4.35(6)$ for the asymptotic critical exponents of nickel,[282] which are close to our putative exact values within the errors of 5.3%, 7.6% and 0.4%. Some experimental results for the critical exponents, most of them published after 1990, were tabulated in Table 7 of the Pelissetto and Vicari's review.[154] It is seen from these most recent data that the critical exponents $\alpha$, $\beta$, $\gamma$, $\eta$ and $\nu$ vary in ranges of 0.077 – 0.12, 0.315 – 0.341, 1.09 – 1.26, 0.03 – 0.058 and 0.60 – 0.70, respectively, where were determined by experiments in the liquid – vapor transition in simple fluids, the mixing transition in multicomponent fluid mixtures and in complex fluids, the transition in a uniaxial magnetic system, the transition in a micellar system and the mixing transition in Coulombic systems.[154,283-293] Again, we emphasize here that the difference between the logarithmic behavior and the character of the non-zero critical exponent $\alpha$ in range of 0.077 – 0.12 cannot be distinguished by experiments, specially, in case of preseting the existence of the non-zero critical exponent α. The putative exact critical exponent γ is in very good agreement with the experimental data,[154,284-286,288,290,291,294-296] specially, which is exactly the same as what was determined in various references.[283,287,292,293] Most of the experimental data for β, η and ν have somewhat larger deviations with the exact values, remaindering us the fact that only is the critical exponent γ the most reliable one for the high – accurate determination of the critical exponents since it depends most insensitively on the accurate location of the critical point. Neverlethess, it is a common fact that the experimental data are less accurate than the theoretical one, which cannot serve as the only standard for judging the correctness of the exact solutions.[154,213]

We tried to check what we found for the putative critical exponents of the 3D Ising models by several criterions: 1) The putative exact critical exponents satisfy the scaling laws and show universality behaviors; 2) The putative exact critical exponents represent the behaviors of the system exactly at the critical region, as the critical point is fixed exactly; 3) The putative exact critical exponents have physical significances, which are correlated directly with the existence of the fourth curled – up dimension; 4) The putative critical exponent γ is exact the same as the conjecture of $\gamma = 5/4$;[83,95,103,107] 5) Our putative exact solutions are in very good agreements with the critical exponents β and δ collected in Vicentini-Missoni's review,[268] which were derived by analysis of the good data in the critical region, which were believed to be available only on few substances; 6) The putative exact critical exponents are almost exactly equal to the critical indices γ, δ, η and ν of the real fluids; 7) The putative exact critical exponents μ and ν coincide with experimental ones for the interfacial tension in the two phase fluids, within experimental error bars; 8) The putative exact critical exponents satisfy the criterion that the critical behaviors should change in a subsequence of continued logic: the critical indices should vary smoothly with dimensionality. All the putative exact critical exponents for the 3D Ising lattice are located between those for the 2D and 4D (mean field) exact critical exponents. The logarithmic behavior of the specific heat verifies that all the systems have the zero critical exponent α, regardless of their dimensions; 9) The putative exact critical exponents are comparable with the approximation values and the experimental data, if compared grossly; 10) The putative exact critical exponents would be very close to the approximations and the experimental data, if one agreed with the existence of the zero critical exponent $\alpha$ and chose the critical exponent $\gamma$ as the most reliable one for the high – accurate determination of the critical exponents; 11) The putative exact critical exponents satisfy the principles of simple, symmetry and beauty with aesthetic appeal, which are all simple integers and fractions. They are much simpler than those suggested by Fisher and Domb.[103,107] Finally, we emphasize that the results of the approximation methods and the experiments cannot serve as the only standard for judging the correctness of the putative exact solutions, but the exact solution can serve for the evaluation of the systematical errors of the approximations and the experiments.

## VIII. DISCUSSION

### A. Scenario of the (3 + 1) – dimensional space framework

It is important to justify the correctness of the present procedure for deriving the exact solution of the 3D simple orthorhombic Ising lattices. To do so, we need to justify the validity of the two Conjectures introduced in Sec. III. The main points of the two Conjectures are: The topologic problem of the 3D Ising system could be solved by introducing an additional rotation in a (3 + 1) - dimensional space with a curled-up dimension attached on the 3D space. The weight factors $w_x$, $w_y$ and $w_z$ on the eigenvectors represent the contributions of $e^{i\frac{t_x \pi}{n}}$, $e^{i\frac{t_y \pi}{l}}$ and $e^{i\frac{t_z \pi}{o}}$ in the 4D space to the energy spectrum of the system. By introducing the two Conjectures, we succeeded in finding out the maximum eigenvalues and the free energy to be same as those of the original 3D Ising model. There should be a natural mechanism for realizing this scenario.

Actually, introducing the new dimensions to our 3D physical system is not a brand-new conception.[297-300] The aim of the early model by Kaluza and Klein to consider a five – dimensional spacetime with one spatial extra dimension was to unify electromagnetism and gravity.[301,302] There exists evidence that convinces us that we live in four noncompact space-dimensions.[303] Because of experimental constraints, the standard model fields cannot propagate into bulk and are forced to lie on a wall, or 3-dimensional brane in the higher dimensions.[304-307] Living in 4 + n noncompact dimensions is also in prefect compatibility with experimental gravity.[303] An effective dimensional reduction occurs without the need of compactifying the fifth dimension, since Kaluza – Klein excitations, which have nonvanishing momentum in the fifth direction, are suppressed near the brane. Thus, even though the Kaluza – Klein modes are light, they almost decouple from matter fields, which are constrained to live on the wall.[303,308] On the other hand, there exist five anomaly free supersymmetric perturbative string theories, known as type-I, type-IIA, type-IIB, SO(32) heterotic and $E_8 \times E_8$ heterotic theories.[297,309] In all of these string theories, besides the four noncompact space-dimensions, more compact dimensions, for instance, a compacted six-dimensional Calabi-Yau space, are needed. The four-dimensional couplings are related to the string mass scale, to the dilation, and to the structure of the extra dimensions mainly in the example of hetereotic theories. These five perturbative string theories are all related to each other by various string dualities (such as T – duality and S – duality) and the (10 + 1) – dimensional M superstring theory could describe these five string theories together with 11-dimensional supergravity.[297,298,309] Nevertheless, even though the compact dimensions maybe too small to detect directly, they still can have profound physical implications. In the present case, introducing our two Conjectures reveals profound physical significances for the 3D Ising system indeed, which comes out automatically from the requirement of solving the topological problem of the 3D Ising lattice. The putative exact results obtained by our procedure are consistent with high – temperature expansions at/near infinite temperature for the 3D Ising model. However, only is the 4D space enough for solving our problem, while the radius of the additional curled-up 4$^{th}$ dimensions presumed to be infinitesimal in the original 3D Ising lattice. Our 4D world remains free for contacting with the (10 + 1) – dimensional world via a 6 – dimensional compact Calabi-Yau manifold plus 1 – dimensional time.

Introducing the additional curled-up dimension supports indirectly that we might live in four noncompact space-dimensions.[303] This conjecture is not inconsistent with Kaluza and Klein's five – dimensional spacetime, and even the super-spring theories.[297,298,309] The evidence for the possibly existence of the dark matter or dark energy in cosmos is still a mystery to scientists.[310,311] The additional term K''' of the energies is included in the expressions Eq. (3.17) and Eq. (3.37) for the eigenvalues and the partition function of the simple orthorhombic Ising lattices. This extra energy term seems to be related directly with the introduction of the additional curled-up dimension, rather than the interactions along three crystallographic axes themselves. Furthermore, the existence of four noncompact space-dimensions provides enough space for the dark matter or dark energy, although we cannot see the fourth curled-up dimension, but which are communicating with or acting on the 3D physical world.

The scenario of the 3D Ising model at different temperatures is illustrated as follows: 1) At infinite temperature, $\kappa = 0$, $w_x = 1$, and $w_y = w_z = \pm\sqrt{\frac{7}{18}}$. There is actually a state without any interactions, because any *finite* interactions lost their actions, in comparison with *infinite* temperature. The configurations are completely random and extremely chaotic. One cannot distinguish any configurations from this completely random phase. This phase could be defined as Phase 1, which is a formless phase of completely random, representing a special state of non-being (non-interaction). 2) At temperature infinitesimal deviates from infinite, the system starts to experience the extremely very much weak interactions. Deviations from complete randomness occur which can be systematically taken into account by means of a series expansion in K or $\kappa = \tanh K$. $\kappa$ becomes non-zero, but infinitesimal and then $w_x = 1$, $w_y \rightarrow 0$ and $w_z \rightarrow 0$, all the configurations of the high temperature series emerge instantaneously and spontaneously from the complete randomness. The system is still quite randomness, with strong quantum fluctuations, but less random than at infinite temperature. This phase could be defined as Phase 2, which is a forming phase of random with detail structure, representing a special state of being. 3)

As temperature is lowered further, at finite temperatures above $T_c$ ($\kappa$ becomes finite, i..e, $\kappa \neq 0$; $w_x = 1$, $w_y = w_z = 0$), a disordering phase is born out of the randomness, which could be defined as Phase 3. 4) Exactly at the critical point, a disordering – ordering transition occurs with infinite correlation length, with singularity of the free energy, the specific heat, etc. This phase can be defined as Phase 4, or the critical phase, which is the origin of singularities occurring at/near the critical point. 5) Below the critical point, Phase 5, or the ordering phase is born out of the critical phase. 6) At zero temperature, the system becomes completely order, which can be defined as Phase 6. From the scenario above for 3D, there actually are five detailed transformations between six phases, evolving from infinite to zero temperature. The putative exact solution reveals the nature of the nature in the disorder and/or random states: The disordering and/or randomness may have different levels and structures! The existence of Phase 2 is an intrinsic character of our 3D world, which does not exist in any models of other dimensions. For instance, in 2D, Phases 2 and 3 coincide each other from temperature near infinite down to the critical point $T_c$ because the radius of the convergence of the high – temperature expansions is the critical point $T_c$. It is a little hard to understand the intricate difference between the Phases 1 and 2. The basic difference between the Phases 1 and 2 is whether the interactions are experienced. The Phase 1 without any clear - seen configurations, but with more random/chaotic, includes everything of the Phase 2 and even other low – temperature phases. The Phase 2 with infinite configurations as described by the high – temperature expansions, with strong quantum fluctuations, but less random/chaotic than the Phase 1, is born out of the Phase 1. These two phases are strange twin: the Phase 1 is empty since nothing can be distinguished from it, while the Phase 2 is a kind of the phase of filling with all of the high – temperature configurations. Phase 3 is out of the strong quantum fluctuations of the Phase 2. However, at the pregnant period of the Phase 3, the Phases 1 and 2 can transform each other to be like a whole, to form a state like quantum vacuum, up on the infinitesimal fluctuation of temperature at the infinite temperature. As long as temperature is lowered to finite, the Phase 3 with completely different configurations is born spontaneously, up on the breaking down of the symmetry and the annihilation of all the high – temperature configurations. Then nobody can return the Phases 1 and 2, because nobody can receive infinite thermal energy to be back infinite temperature. What happen at/near infinite temperature of the 3D Ising model is analogous to the Big Bang at the origin and successive evolutions of our Universe. In the present case, we do not require a singularity point as the origin of the Universe, but only interacting spins in the thermodynamic limit at infinite temperature.

The states of the Phases 1 and 2 are analogous to what Lao Zi described in his famous book "Dao De Jing".[312] Lao Zi was the greatest Chinese philosopher and thinker who lived in c. 585 – 500 B.C. and once was the librarian and archivist of the royal court of Zhou Dynasty. Lao Zi's Thought has been long - living for more than two thousand years, and recently, got to be known by more peoples in the West. The philosophy of Lao Zi is first about the universe, then humanlife and next, politics. Without any models and knowledge in modern physics, Lao Zi tried to understand in deep what the origin of our world was and describe the evolution of the world. The most famous one among his various ruminations is "Non-being is the beginning of the myriad things; being, the mother of them". Simply speaking, "Being was born out of Non-being; Everything was born out of Being". Namely, the Phase 2 at temperature infinitesimal deviates from infinite is born out of the Phase 1 at infinite temperature; all the phases (including disordering and ordering ones) at finite temperature is born out of the Phase 2. However, one might argue, these ruminations seem to have parted far from the domain of the Ising model in 3D, 4D, or anything else. Our point of view is: On the top level, philosophies, sciences, arts and even religions are all correlated, since all of us face to a unique world. We should respect the wisdom of the greatest philosopher and thinker Lao Zi. Interestingly, the success in deriving the putative exact solution of the 3D Ising model for a purpose of understanding the critical properties at the critical point helps to have a clear description on the scenario of the physic world at/near infinite temperature.

One may argue that the 4 - fold for ln Z is not, in principle, mathematically impossible, but it may be physically impossible for ferromagnetic Ising models because the Yang – Lee theorem proved rigorously the absence of zeroes of the partition function except on the imaginary magnetic field axis and together with the existence of a gap in the zero distribution at high enough T (> $T_c$). This implies that the high – temperature expansions of the Ising model converge and fully define $N^{-1}\ln Z$ in the thermodynamic limit, $N \to \infty$, for all $T > T_c$, as is demonstrably so for the 1D and 2D models. However, this judgment based on the Yang – Lee Theorem is not correct, because the Yang – Lee Theorem disregarded the special case of $T = \infty$. It is clearly seen after Theorem 3 in their second paper[9] that "The lattice gas cannot undergo more than one phase transition, which must occur, if at all, at a value of the fugacity equal to $\sigma$, which according to Eq. (24), corresponds to z = 1. The isotherms in the I – H diagram of the corresponding Ising model problem is smooth everywhere except possibly at zero magnetic field (which occurs at z = 1). This is usually believed to be true but was not proved." The most important issue here is that z = 1 corresponds to the possibility of the phase transition. Due to this fact, the Yang – Lee Theorem excluded the occurrence of the phase transition at presence of magnetic fields H, since H = 0 leads to z = 1 in accordance with their Eq. (23) of $z = \exp(-2H/k_BT)$. However, it is clear that Yang and Lee did not discuss the case of infinite temperature. If $T = \infty$, z will equal to one also, providing with the possibility of the occurrence of a phase transition at infinite temperature. Since there is no phase transition at infinite temperature in other dimensions, it is believed here that it should occur in 3D system. Furthermore, the zero distribution as $T \to \infty$ can be quite different with that at finite temperature, just similar to the case that the zero distribution in the thermodynamic limit (volume $V \to \infty$) differs to that for finite volume. It could be true that the behavior of the phase transition at/near infinite temperature differs with that at the critical point. Therefore, the mathematical structure of the free energy Eq. (3.37) is not only mathematically possible, but also physically possible. Nevertheless, all of the facts above suggest that for the 3D Ising lattices, the high temperature series expansion may not be a standard for judging the validity and correctness of the putative exact solution at finite temperatures. This would give an implication that the explicit form of the solutions of any 3D lattice theories may have less direct relation with series expansion of perturbation theories.

It should be emphasized here that for a real 3D system, the 4 - fold for ln Z should be remained for the whole temperature range, while the weight factors can vary in range of [-1, 1]. In consideration of symmetry, for the simple cubic lattice (and also the simple orthorhombic lattices close to it), the roles of the weight factors $w_x$, $w_y$ and $w_z$ can be interchanged without altering the eigenvalues Eq. (3. 17) and the partition function Eq. (3.37). They could interchange their roles (and values) at any time, from the point of view of symmetric. In this way, the system is within the (3 + 1) – dimensional space framework, even in case that anyone of the weights occasionally equals to zero. These lattices as marked by 3D in Fig. 4 show the critical behaviors of a real 3D system. With further decreasing one or two of the three interactions K, K' and K'', the symmetry of the simple orthorhombic lattices decreases, and consequently, the mechanism of interchanging the roles of the weight factors $w_x$, $w_y$ and $w_z$ is gradually weakened to be even forbidden in the 2D or 1D limit. Namely, $w_x = 1$, $w_y \equiv 0$ and $w_z \equiv 0$ send the system back the Onsager's 2D Ising model, and this limit case also corresponds to all the simple orthorhombic lattices (as marked by 2D in Fig. 4) with their critical points lower than the silver solution. The change of such interchange with the symmetry of the system is the origin for the 3D – to – 2D crossover phenomenon, i.e., a gradual crossover between the 3D and 2D behaviors, the 2D behaviors for some simple orthorhombic lattices with less symmetry.

The (3 + 1) dimensional scenario described above for the 3D Ising model might be the intrinsic character of the 3D many – body interacting systems. The physics beyond the (3 + 1) dimensional scenario might be understood in deep as follows: For a nonconservative system, the time – dependent Schrödinger equation is explicity expressed by: $i\eta\frac{\partial}{\partial t}\Psi(r,t) = H(r,t,-i\eta\frac{\partial}{\partial r})\Psi(r,t)$ . For the special case of a conservation system, where H does not depend explicitly on t, a particular solution is $\Psi_n(r,t) = \psi_n(r)\exp(-iE_n t/\eta)$ , where $E_n$ is an eigenvalue and $\psi_n(r)$ is the corresponding eigenfunction of the ordinary Schrödinger equation $H(r,q)\psi(r) = E\psi(r)$. However, according to the relativity theory, any system should be described within the spacetime framework, and the spacetime are closely associated by the Lorentz transformation. We might need to rethink the role of the time on the non-relativistic quantum mechanism. The time should have two roles: one is to evaluate the movements of a particle within the framework of the d – dimensional space; another is to represent the whole system within the (d + 1) – dimensional spacetime. The first role of the time is accounted for by the first - order derivative $\frac{\partial}{\partial t}$, but the second role (the second - order derivative $-\frac{\partial^2}{\partial t^2}$, in accordance with $\nabla^2$ in kinetic energies) is totally neglected in the non-relativistic quantum mechanism. This term of $-\frac{\partial^2}{\partial t^2}$ is eliminated in the famous Schrödinger's equation that plays the role of Newton's law and conservation of energy in classical mechanics. That is, the time-dependant Schrödinger equation is of the first order in time but of the second order with respect to the co-ordinates, hence it is not consistent with relativity. Actually, the Hamiltonian of the whole system (even in the case of non-relativistic) in the (d + 1) – dimensional spacetime is always associated with the time and any spacetime system is always a nonconservative system if taking into account the time evolution. Nomatter where the system is non-relativistic or relativistic, the equation for dynamic of the system should be consistent with relativity. It is our understanding now that when we deal with a Hamiltonian of a non-relativistic system, which does not depend explicitly on t, the second role of the time is actually hiddened. Although this role of the time could be neglected in other non-relativistic systems, it should be taked into account for the 3D many – body interacting system where the Hamiltonian of this non-relativistic system describes the interactions of the spins only in the 3D space (just like accounting only $\nabla^2$ in kinetic energies). The introduction of the extra dimension in the present 3D Ising case might correspond to the second role of the time (just like accounting the effect of $-\frac{\partial^2}{\partial t^2}$), although its appearance is required instantaneously by solving the topologic problem in the 3D Ising system. This implies that the topologic problem in the 3D many - body interacting system might automatically result in the conception of the spacetime. The introduction of the weights might correspond to a mechanism making the nonconservative system in the spacetime to be conservative during the evolution of the time. This work may reveal how the second role of the time could be associated physically with the 3D world (Alternatively, however, one could treat the additional dimension just as a pure mathematic structure, or, a boundary condition).

Furthermore, it is understood that a satisfactory quantum general relativistic theory should take into account simultaneously and properly both the two roles ($\frac{\partial}{\partial t}$ and $-\frac{\partial^2}{\partial t^2}$) of the time. However, it has been a challenge to account properly the second role of the time, since it certainly causes the nonconservation (but at the instant of the evolution of the system, it should be kept to be conservative). This might be the origin of one of the difficulties for establishing a satisfactory quantum general relativistic theory. In recent developments on the quantum gravity theory, in order to study a background independent formulation of M theory, the bulk dynamics was described in terms of causal histories framework in which the time evolution was specified by giving amplitudes to certain local changes of the states.[313,314] In this kind of theory, a new kind of fusion between quantum theory and spacetime was achieved in which states were identified with quantum geometries that represent spacelike surface, and histories were both sequences of states in a Hilbert space and discrete analogues of the causal structures of classical spacetimes. Namely, in order to address the issue of time evolution, one may attach a Hilbert space to each node of the causal set graph in a theory of the causal evolution of the Penrose spin networks.[313,314] In loop quantum gravity, the spin networks are the basis states for the spatial quantum geometry states.[313,314] On the other hand, it is known that the Ising model can be employed to describe the Penrose spin networks used for the quantum theory of gravity, since one could treat exactly equally triangulations and their dual spin networks. The two conjectures in the present work may shed lights on the satisfactory quantum theory of gravity, by illustrating the topologic and causal structures of the spacetime. The four noncompact space-dimensions mentioned above might be thought to be the four spacetime dimensions. Namely, the fourth dimension of the four noncompact space-dimensions might behave as the timelike space-dimension, representing the causal evolution of spin networks. This implies a possibility that a five – dimensional spacetime with four noncompact space-dimensions, mentioned above, may be mathematically treated as a (3 + 2) dimensional spacetime (i.e., three spacelike space-dimensions, one timelike space-dimension and one time dimension). The two time – related dimensions may correspond, respectively, to the two roles of the time: $\frac{\partial}{\partial t}$ and $-\frac{\partial^2}{\partial t^2}$. The second action of the time is hiddened in the Hamiltonian of a 3D conservative system and this additional timelike space-dimension is curled-up in the spacetime.

## B. Approximation technqiues

### 1. General arguments

Next, we need to discuss possible reasons that cause the differences between the putative exact solution and the approximate values obtained by various standard methods, widely accepted by the community, and the differences between the putative exact solution and the experimental data. From the first glimpse, it seems very difficult to imagine that the multitude of separate determinations of these critical exponents throughout the years, by various independent scientists and using completely different techniques (Monte Carlo simulations, high- and low-temperature expansions, renormalization group field theory and experiments) are wrong and all yield (wrong) results that coincide. However, it is understood here that there is no question of which results are wrong or correct, but inexact or exact. Clearly, any approximation results can not be exactly equal to the exact ones. There is no equalization between them. Strictly speaking, it is no significance of comparing the putative exact solution with the approximation values and it does not make sense to against the putative exact solution by well-accepted approximation values. The purpose we present the putative exact solution here is not to address the criticisms on either the approximation or experimental techniques, but to attempt to reveal the truth of the nature. The results obtained by these techniques can be used as valuable references if be adequately considered, but which cannot be applied as the only standard for judging the correctness of the putative exact solution. It is very hard to see any physical insight from those approximation values, whereas the exact solution would have significance of revealing clearly physics beyond them. All of the putative exact critical exponents are derived analytically by simply introducing our first conjecture, namely, the existence of the extra dimension. The putative exact values emerge spontaneously as long as this conjecture is introduced, and they would be corrected if the conjecture were valid. The putative critical exponent of $\alpha = 0$ illustrates the logarithmic singularity of the specific heat at the critical point of the phase transition. The factor of three (or one over three) in the putative critical exponent of $\beta = 3/8$ (or $\nu = 2/3$) comes automatically from this conjecture, which extends the dimensions in the wave-vector space. The putative critical exponents of $\alpha = 0$, $\beta = 3/8$, $\gamma = 5/4$, $\delta = 13/3$, $\eta = 1/8$ and $\nu = 2/3$, show the universality behavior and satisfy the scaling laws. One would find that these values even hidden some intrinsic correlations with the critical exponents of $\alpha = 0$, $\beta = 1/8$, $\gamma = 7/4$, $\delta = 15$, $\eta = 1/4$ and $\nu = 1$ of the 2D Ising model. For instance, both the 2D and 3D Ising models have the critical exponent $\alpha = 0$, with the same logarithmic singularity of the specific heat at the critical point; The critical exponent $\beta$ of the 3D Ising model is exactly three times as that of the 2D model; The critical exponent $\eta$ of the 3D Ising model is exactly half of that of the 2D model; The difference between the critical exponents $\gamma$ for the 2D and 3D Ising lattices is twice as that for the 1D and 2D (or 3D and 4D) Ising lattices (the same is true for $\beta$);…… Most important, the putative critical point of the 3D simple cubic Ising model is located exactly at the golden ratio $x_c = e^{-2K_c} = \frac{\sqrt{5}-1}{2}$, while the critical point of the 2D square Ising model is located exactly at the silver ratio $x_c = e^{-2K_c} = \sqrt{2}-1$. Realizing the facts that the golden ratio and the silver ratio are the two most beautiful numbers in the mathematical world and that intrinsic similarities and correlations exist between them as revealed by the continued fractions and the equations of $x^2 + x - 1 = 0$ and $x^2 + 2x - 1 = 0$, one would believe that no other numbers are more reliable and suitable than the golden ratio for the critical point of the 3D simple cubic Ising model. The continued efforts of the scientists worldwide throughout more than 60 years since Onsager's discovery of the exact solution of the 2D Ising model in 1944, specially the advances in the renormalization group and Monte Carlo simulations since Wilson's discovery of the renormalization group in 1971, contribute greatly to our understanding on the physical behaviors, specially, the critical behaviors, of the 3D Ising model. These previous results of the approximation and experimental techniques provide profound information, which are quite helpful for deriving the exact solution of the 3D Ising model. We would like to believe that the finding of the putative exact solution would improve the development of these techniques. In the following paragraphs, we shall give the explanation why the renormalization group theory and Monte Carlo simulations and other approximation methods cannot yield the exact solution, or a solution close enough to the exact one.

Perturbation expansions have been used widely in astronomy and physics to evaluate the effect of small changes in problems for which exact solutions are available. However, for physical phenomena in which an interaction completely changes the character of the solution, it is necessary to derive substantial numbers of terms of such perturbation expansions, and if possible to estimate the asymptotic behavior of the coefficients. As remarked by Domb,[107] caution must be exercised in using the method of series expansions if wrong conclusions are to be avoided; physical insight into the nature of the expected solution should be invoked wherever possible, and it can be of great help in providing consistency checks; also methods of series analyses should be tested wherever possible on exact closed form solutions.

The quantity F(z) whose critical behavior as a function of z is to be studied must have a power series expansion about the origin z = 0, $F(z)=\sum_{n=0}^{\infty} a_n z^n$, with a finite radius of convergence.[315,316] There are two criterions for the radius of convergence for series expansions.[315-317] 1) If $\lim_{n\to\infty}\left|\frac{a_n}{a_{n+1}}\right| = z_0$, or 2) if $\lim_{n\to\infty}|a_n|^{-1/n} = z_0$, then the series converges for $|z| < z_0$ and diverges for $|z| > z_0$. Correspondingly, there must be at least one singularity (non-analytic point) on the circle of convergence $|z| = z_0$. Unfortunately, the sequence $|a_n|^{-1/n}$ is often slowly converges so that its practical value in estimating $z_0$ from the leading coefficients is rather limited. If all the coefficients $a_n$ are known exactly, in principle, one can analytically continue the function across the z-plane as far as a natural boundary of the function, beyond which it remains undefined. The nature of the coefficients is determined by the singularities of F(z). The singularities nearest the origin will dominate the behavior for large n. If the dominant singularity is on the positive real axis, the coefficients will eventually all have the same sign. Conversely, if the dominant singularity is on the negative real axis, the coefficients will eventually alternate the sign. More irregular behavior of the sign for large n indicates that the dominant singularities are in the complex plane. Since the coefficients are assumed real, the singularities must then occur in complex conjugate pairs.[316]

One difficulty of series expansions is principal: if we are lucky, as the 2D Ising model, the critical point as a physical singularity point will be located exactly on the circle of convergence. However, the most common case is that the existence of a non-physical singularity point with $z < 0$ reduces the circle of convergence so that one can not reach the critical point that is a physical singularity point outside the circle of convergence. It is believed that the 3D Ising model belongs to this category. The Padé approximant method has been applied to overcome such difficulty to get more information outside the circle of convergence. Nevertheless, the radius of the circle of convergence for high temperature expansions has not been proved rigorously yet.

Another difficulty of series expansions is technical one: usually, a finite number of the coefficients $a_n$ can be determined, $a_0$, $a_1, \ldots,$ $a_{n\,max}$. Typically, the calculation is in principle straightforward; however, the increase labor which is necessary for calculating each succeeding coefficient is large. Obviously, a rule of thumb is that computation of $a_{n+1}$ involves at least as much labor as the cumulative calculation of $a_0$, $a_1, \ldots,$ $a_n$. Thus, while there is in principle no limit to the number of calculable coefficients, there is in practices a rather sharp upper bound $a_{n\,max}$ (depending on the details of the model being considered) determined by such practical considerations as time and patience and, at the next level, electronic computer capacity and funding.[315] At present, the upper bound $a_{n\,max}$ of terms in various approximations is around n = 26 (too far from infinite). This is the main reason why almost all the approximation techniques provide the (almost) same inexact results in the 3D Ising case.

It is a fact that all systematic methods for the determination of series coefficients are at some level graphical or diagrammatic. With each coefficient is associated a set of graphs of some given topological type. To each graph corresponds a numerical contribution according to a well-defined rule. To calculate the required coefficient, one simply sums all contributions. As a rule the restrictive embeddings are best in low dimensionality and for rather open lattice structure. For close – packed and in higher dimensionality the renormalization method seems preferable. In any given study, there may be additional considerations favoring one method or another.[315] The early expansions for magnetic systems, especially, the Ising and Heisenberg models which have been important in the study of critical phenomena, were all of the weak (high - temperature) and strong (low – temperature) embedding types. This may partially explain why the high – temperature and low – temperature expansions can give the exact terms for the 2D Ising model, but not for the 3D Ising model. This may also explain why the results of the high – temperature expansions are better and more regular than those of the low – temperature expansions, but worse than those of the renormalization group techniques.

The renormalization group ideas are known to concern with the basic physics of a critical point, namely the long – wavelength fluctuations which are the cause of critical singularities. The starting point in the renormalization group approach is to realize that the most important fluctuations at the critical point do not have a characteristic length. Instead the important fluctuations have *all* wavelengths ranging from the atomic spacing up to the correlation length; close to the critical point the correlation length is much larger than the atomic spacing. Thus, the important wavelengths near the critical point cover many decades.[318] However, the renormalization group techniques are in a certain sense similar to the series expansions, because during the renormalization group procedures, various approximations, such as expansions, perturbations, linearizations, normalizations, etc. are performed. In all the ways, the disadvantages, similar to those of low – and high – temperature expansions, have not been removed completely. The starting point for the $\varepsilon$ expansion is Landau's mean field theory, which is exact apart from logarithms in four dimension ($\varepsilon = 0$). For the simplest (Ising – like) case the critical exponents move in the direction of the 2D values obtained by Onsager, and in three dimensions agree well with high – temperature calculations.

In principle, we can focus our attention only on those methods for which the inclusion of more and more coefficients leads to successive approximation schemes, which appear to converge with reasonable regularity and speed. Extrapolation in principle enables one to draw conclusions about the critical point behavior and to estimate the "errors" involved. However, we have to stress that the error estimated are, unfortunately, in no sense rigorous and only represent a subjective assessment of the rate of convergence of the available numerical data. In principle, one could easily be quite misled by the initial coefficients for there is no mathematical reason why the apparent asymptotic behavior of the first ten to twenty terms, say, should continue to infinity.[316] Indeed, the position is less satisfactory for 3D lattices for which the series converge more slowly; further information is needed to provide direct estimates of critical exponents and amplitudes which can be considered as adequate.[107] We would like to emphasize here: The sufficient and necessary conditions for any approximation result can serve as a standard of judging the correctness of a putative exact solution are that 1) the approximation expansions must be exact and convergent and 2) the variable for such expansions (even exact and convergent ones) must be kept to be very small. The accident victory of low – and high – temperature expansions in 1D and/or 2D cannot be the base of the over – optimism for their validity in 3D. Anyone, who is using the approximation techniques, have to keep in mind that the final state determined by choosing an initiate state plus high order (perturbation or non-perturbation) corrections can be deviate far from the real state, no matter how many terms of these corrections and no matter how precision the techniques can be.

## 2. Low temperature expansions

For low – temperature expansions, the appropriate choice is to define the partition function such that the fully aligned spin – up state is taken as having zero energy, since the low – temperature series is a perturbation expansion about this state. It has been accepted widely by the community that the exact solution of the 3D Ising model must be equal to the exact low temperature expansions. But, unfortunately, the low temperature expansions in 3D are divergent. It is our opinion that nobody can find an exact solution with the close form, which diverges at/near critical point. It is also our opinion that the requirement that the exact expression for spontaneous magnetization must be equal, term by term, to the so – called *exact* low - temperature expansions has reflected the pious hope of the community for a long time.

Compared with the high - temperature expansions, the situation at low temperatures is far less satisfactory.[107,111] As remarked by Domb, the low - temperature series in 3D alternated in sign and it was clear that spurious non-physical singularities were masking the true critical behavior.[107] It is noticed that in the low - temperature series, the leading term is $-2x^6$, while the coefficent for $x^8$ term is zero. The first term with positive coefficent is $14x^{12}$, which follows the term of $-12x^{10}$ (with the same coefficient with the present exact solution). The appearance of plus sign in the low - temperature series expansion seems to be incorrect,[59] and it is likely that it appears to compensate the incorrectness of other terms (specially the leading term). Since there is a masking unphysical singularity, the conjecture of $\beta = 5/16$ based on the low - temperature series expansion is also questionable. Furthermore, the low - temperature expansions, evaluated by systematically overturning spins from the ground state with all spins "up", give the same fundamental leading term for the 2D triangular Ising lattice and the 3D simple cubic lattice, because it relates directly with the Ising model lattice of coordination number q, regardless of what the dimension it is. It is our thought that the fundamental leading term of $-2x^6$ (correct for the 2D triangular Ising lattice) has to disappear for the 3D simple cubic Ising model. The leading term of our putative exact series of the spontaneous magnetization, $-6x^8$, reminds us that we live in four noncompact space-dimensions (as discussion above), which are required intrinsically for dealing with the topologic problem and the non – local behaviors of the 3D Ising model.[303] The introduction of the fourth curled – up dimension realizes its contribution on the spontaneous magnetization, while removing instantaneously the trouble of the dominant singularity on the negative real axis. Such contribution is actually the real effect of the interacting many – body spins in the 3D lattice, which spontaneously leads to the additional contribution of the free energy due to the 3D topologic problem. What the putative exact solution reveals is that the dominant singularity is located on the positive real axis for both 2D and 3D. Furthermore, the variable $x = e^{-2K}$ used for the low- temperature series expansion is small only at extremely low temperatures so that it fits approximately well with the exact solution only at the low temperature range. As shown in Fig. 3(b), the low – temperature series expansion with terms up to the $54^{th}$ order of the simple cubic Ising model numerically fits well with the putative exact one, and oscillates around it, up to $T \approx 0.9\ T_c$, and then deviates from it. It is understood that if one took more terms into account, the low – temperature expansions would fit better with the putative exact solution. The putative exact solution is actually the center of the oscillation of the low – temperature expansions. This could prove indirectly that the present putative exact solution for the spontaneous magnetization (and also the free energy) might be correct. Furthermore, this fact could prove that the low- temperature series expansion could not provide valuable information on the critical region.

Here, one would need to ask why the low - temperature expansions give the same sign in 2D, but the alternated sign in 3D. Why and how does the dominant singularity on the positive real axis for 2D change to be located on the negative real axis for 3D? It is our opinion that all the terms in the correct answer should have the same sign, as in 2D, and in the wrong series they alternate in sign. Guttmann used the method of N-point fits to locate unphysical singularities for various lattices.[107,319] For the single cubic lattice, he found that there is one such singularity on the negative real axis. These singularities lie closer to the origin than the physical singularity thus masking the critical behavior. Domb and Guttmann initiated a configurational analysis of terms of the low temperature series which showed how the spurious singularities arise.[107,320] Starting from the empirical observation that Carley – tree embeddings are far more numerous than those of any other group of connected graphs, they estimated the Carley – tree contribution to the low temperature series and found that in a first approximation, the spurious singularities all lie on the circle $u = u_c$ for $q = 4, 6, 8$ and 12. For $q = 4$ there is only one solution; for $q = 6$ there are two at positive and negative axes; for $q = 8$, there are three in the complex u plane; for $q = 12$, five. Distribution of singularities in the complex u plane for simple cubic lattice was shown in Fig. 21 of ref. 107. Higher order approximations move the spurious singularities in nearer to the origin. It is understood that infinite order approximations would move the spurious singularities to the origin. From the existence of the non-physical singularity points and also its divergence, one could realize that the radius of the circle of convergence for low temperature expansions might be zero. This is also because there is no special point between the zero and the critical point and if the radius were not the critical point, there would a large possibility that it would be reduced to zero. In the following, we shall discuss in detail the radius of the circle of convergence for low temperature expansions.

The irregularity (i.e., the alternated sign) and divergence of the low - temperature expansions in 3D clearly are indeed associated with an unphysical singularity on the negative real axis.[107,319,320] One would expect from the tendency of known terms (up to 54$^{th}$ order (n = 27) [111,238]) that the low - temperature series for the spontaneous magnetization should alter their signs and increases rapidly the coefficients of the terms, all the way up to infinite terms. This obviously leads to the stronger oscillation and divergence of the spontaneous magnetization, specially, at high - temperature region (close to the critical point). It is crucial to detect the radius of the circle of convergence of the low - temperature series. From the coefficients $m_n$ of the low - temperature series,[111,238] one can calculate the ratio of $m_{n+1}/m_n$ in order to evaluate the circle of convergence in accordance with the criterion 1) above.[315-317] From table 2, the calculation results from the last several known terms are as follows: -3.3479826… for n = 21, -3.3621183… for n = 22, -3.3626251… for n = 23, -3.3716472… for n = 24, -3.3741696… for n = 25, - 3.3805110… for n = 26, … Then one can evaluate the difference $\Delta_{(n+1,n)} = (m_{n+2}/m_{n+1} - m_{n+1}/m_n)$ between the neighboring ratios $m_{n+1}/m_n$. It is seen from table 2 that after the initial oscillation between positive and negative values up to $\Delta_{(21,20)}$, it starts to be somehow stable to be small negative values, -0.0141357… for $\Delta_{(22,21)}$, - 0.0005068… for $\Delta_{(23,22)}$, -0.0090221… for $\Delta_{(24,23)}$, -0.0025224… for $\Delta_{(25,24)}$, -0.0063414… for $\Delta_{(26,25)}$, … It would be reasonable to believe all the higher order terms would follow this tendency, i.e., the higher order terms $m_{n+1}/m_n$ would decrease monotonously, by small finite values, with increasing the number n. In consideration of the existence of infinite terms of the exact low - temperature series, it seems reasonable to derive the conclusion that the ratio $m_{n+1}/m_n$ would approach negative infinite as n becomes infinite because small finite values times infinite leads to infinite. For the above criterion 1), one would find that the radius of the circle of convergence of the low - temperature series would be zero. This means that the non-physical singularity on the negative axis would approach the zero when we could take into account all of the infinite terms of the low - temperature series. At least, there would be two possibilities for the circle of convergence of the low temperature series: the zero radius or the non-zero radius. It is our argument that it would be not very reasonable to predict the ratio $m_{n+1}/m_n$ for $n = \infty$ to be about - 3.5 simply by extrapolating plots as a function of 1/n to zero (though we could not exclude fully this possibility), because there are infinite uncalculated points for plots as a function of n.

Furthermore, one could try to evaluate the circle of convergence in accordance with the criterion 2) above.[315-317] It is clear, also from table 2, that the value $(m_n)^{-1/n}$ for magnetization of the 3D Ising model on a simple cubic lattice changes from 1 for n = 0 to 0.352777263… for n = 27. After the initial irregularity, the value $(m_n)^{-1/n}$ decreases monotonously from 0.644137614… for n = 6 to 0.352777263… for n = 27. It would be reasonable to believe that this tendency would be valid for all the terms with n > 27. In consideration of the existence of infinite terms for the low temperature series, it could be thought that the value $(m_n)^{-1/n}$ would decrease steadily down to zero as n approaches infinite. Once again, one reaches the conclusion that the radius of the circle of convergence of the low temperature series could be zero. At least, either the zero radius or the non-zero radius would be true for the circle of convergence of the low - temperature series. Nevertheless, this is opening to be proved rigorously in future.

If one agreed with the statement above for the zero radius of the circle of convergence, one could discuss further the problems of analytic continuation and singularities of the low - temperature series. According to the principle of analytic continuation,[317] a function that is well defined inside its circle of convergence can be continued well beyond its circle of convergence in all directions where singularities are not encountered. The fact that the function is well defined within the circle of convergence is enough to guarantee analytic continuation throughout the remainder of the complex plane unless such continuation is blocked by singularities. In the present case, because the radius of the circle of convergence is zero, singularities of the low temperature series are encountered at origin and thus the function is not well defined within the circle of convergence. Therefore, the function cannot be continued well beyond its circle of convergence in any direction because such continuation is blocked by singularities at origin. This means that if one agreed with the zero radius of the circle of convergence, the so – called *exact* low temperature series would not serve as a standard for judging the correctness of the putative exact solution of the 3D Ising model.

On the other hand, we realize great success of Padé approximants[96,97] as an effective way of deriving information of the critical behaviors up to the critical point, with very high precision. The Padé approximant method has been applied to overcome difficulties to get more information outside the circle of convergence.[96,97] However, doing this way cannot prove/guarantee the continuation or the radius of the circle of convergence of the function, since the Padé approximant can evaluate the series even with the zero radius of the circle of convergence. For instance, the series $F(z) = 1 - 1!\,z + 2!\,z^2 - 3!\,z^3 + 4!\,z^4 - 5!\,z^5 + 6!\,z^6 - \ldots$ The Padé approximant can calculate approximately the value of this series within very high precision. But the radius of the circle of convergence of this series is zero.

The reason for the zero radius of the circle of convergence of the low – temperature series might be: somethings lack in the series as well as in the system. A large possibility is the lack of the extra dimension in the 3D Ising system, which is actually hiddened in accordance with the existence of the topologic problems. This extra dimension should be introducted in order to take into account properly the low – temperature series. Else, the divergent low – temperature series with the zero radius of the circle of convergence are obtained. The inconsistence of the non-relativistic quantum mechanism with the relativity theory, due to the lack of the information of the additional dimension as well as the second action of the time (i.e., $-\frac{\partial^2}{\partial t^2}$) as discussed above, might be the origin of all the troubles (irregularity, divergence, the existence of non-physical singularity point and the zero radius of convergence, etc.) of the low – temperature series expansions when the expansions are employed on the 3D Ising model.

**3. High - temperature expansions**

It is known that the high - temperature series expansion is an exact expansion, which uses the variable κ = tanh K that is small at high temperatures. For the 2D Ising model, it is easy to write down every terms of the high - temperature series expansion, by accounting the loops in the 2D lattice. If one took infinite terms of the high - temperature series expansion into account, one would expect that it would fit exactly with the exact solution from infinite temperature down to the critical point. However, for the 3D Ising model, one first meets the challenge to write down every terms of the high - temperature series expansion, because accounting the polygons becomes very much tedious and extremely difficulty, also because the existence of the crosses and knots in the 3D Ising lattice makes the well – known topologic troubles. For physical phenomena in which an interaction completely changes the character of the solution, it is really necessary to derive substantial numbers of terms of such perturbation expansions, and if possible to estimate the asymptotic behavior of the coefficients. It is believed that one could not achieve the exact information at/near the critical point if one failed in deriving the infinite terms of the high - temperature series expansion. As remarked by Domb,[107] one should take the caution, using the method of series expansions, if wrong conclusions are to be avoided; physical insight into the nature of the expected solution should be invoked wherever possible, and it can be of great help in providing consistency checks; also methods of series analyses should be tested wherever possible on exact closed form solutions. As revealed in Appendixes and discussed below, the situation in the 3D Ising model would be even more pessimistic than the facts above: the putative exact solution could fit well with the high - temperature series expansion only at/near infinite temperature; or from another angle, the high - temperature series expansion could fit well with the putative exact solution only at/near infinite temperature. Once again, there would be two possibilities: 1) the high - temperature series expansion may be inexact at finite temperatue; 2) the putative exact solution may be not correct. In consideration of the possibility of the occurrence of a phase transition at infinite temperature, we would like to believe that the high - temperature series expansion could be invalid at finite temperatues.

In other words, it is our opinion that the (albeit exact) high - temperature series cannot serve as an adequate basis for rejecting a putative exact solution of the 3D Ising model. It is true that the high - temperature expansion of the Ising model converge rigorously and the convergent expansions fully define $N^{-1}\ln Z$ in the thermodynamic limit, $N \to \infty$, for all $T > T_c$, as is demonstrably so for the 2D and 1D models. What we wish to remind is that the convergence and the exactness of the high - temperature expansion series do not equalize to the validity at any temperature range without any conditions. One has to keep in mind that the base of the high temperature expansion series is that K or κ have to be small, i.e., K or $\kappa \to 0$. Although the high - temperature expansion series is valid for all $T > T_c$ in the 2D and 1D models, it does not guarantee that the same thing must happen in 3D. The critical point at 2D is located exactly on the circle of convergence does not guarantee that the same thing must happen in 3D. Actually, in the 2D case, we are extremely lucky, because of the fact that not only the high - temperature expansions but also the low - temperature expansions are exact and convergent, and the critical point is located exactly on the circles of convergence of both the expansions. The free energy can be described by a unique function of expansions for the whole temperature range. However, we are not lucky in 3D, because indeed a non-physical singularity point with $z < 0$ exists. The inexactness, irregularity (i.e., the alternated sign) and divergence of the low temperature expansions in 3D clearly are indeed associated with an unphysical singularity on the negative real axis.[107,319,320] This strongly suggests the existence of such a non-physical singularity point for the high - temperature expansions, since the parameters for the low- and high- temperature expansions are related each other (the definition of $\kappa = \tanh K = (1 - x)/(1 + x)$ with $x = e^{-2K}$). This non-physical singularity point could reduce greatly the radius of the convergence of the (though still exact) high – temperature series. If the radius of the convergence of the high – temperature series were reduced, it would be reduced to be zero (or infinitesimal), namely, a point, because there is no special point between the critical point and infinite temperature. This is also due to the fact that the radius of the convergence of the low – temperature series is zero, according to the discussion above. Although κ looks like convergent as $K \to K_c$, the existence of this non-physical singularity makes such convergence meaningless. Therefore, it could be concluded that the radius of the convergence of either low- or high – temperature series expansions is zero and both the series expansions could be inexact at any finite temperatures. The claim that the high - temperature expansion series is valid for all $T > T_c$ in 3D has not been proved rigorously by anyone. The radius of the convergence of the high – temperature series is the critical temperature of the 2D Ising model was proved to be true by the Onsager's exact solution. Please notice that the Onsager's exact solution served as the standard for judging the validity of the high – temperature series in 2D; not inversely. Only can the exact solution serve as the standard for judging its validity in 3D. A possibility is that for 3D, the high - temperature expansion series is rigorously valid only at its very high temperature limit. The radius $z_0$ of convergence could be not finite, but infinitesimal, i.e., $z_0 \to 0$. The non-physical singularity point could be located on the circle of convergence with its infinitesimal radius $z_0$ (being a point at infinite temperature). The physical singularity point, i.e., the critical point, which is our main interest, could be located outside the circle of convergence for the 3D Ising model. This could be why the high - temperature expansion series cannot locate exactly the critical point of the 3D Ising model, which has shown the results somewhat not better than the renormalization group techniques. The scenario is that all the configurations, used for deriving the high - temperature expansion series (of infinite terms), exist only near infinite temperature (i.e., K or $\kappa \to 0$), in a random fashion, although it is less random than the infinite – temperature state and many (actually infinite) configurations as a kind of microstructures have already emerged. The information of these configurations can be kept in the exact function and the weights in an intriguing way as revealed in this paper.

The main reason for the zero radius of convergence of the series expansions might be: The normal procedure for accounting the terms of these series expansions does not take into account the important hidden intrinsic property of the 3D Ising model. That is, the interacting spins in the thermodynamic limit in the three dimensions intrinsically hide the topologic knots of interacting, which introduce a comparatively higher energy than the simple sum of the normal loops as calculated in other dimensional models. This intrinsic property is a cooperative non-local phenomenon, which cannot be described properly by these approximations taking into account only the local environments. The non-local behaviors exist specially and only in the 3D many body interacting spin system, which can be seen clearly in its complex boundary condition, its topologic problem, and so on. The non-local behaviors might be related with the additional dimension as discussed above. The lack of the information of the additional dimension in the non-relativistic quantum mechanism (which is inconsistent with the relativity theory) might be the origin of all the troubles (the existence of non-physical singularity point and the zero radius of convergence, etc.) of the series expansions when these series expansions are employed on the 3D Ising model described within the framework of the non-relativistic quantum mechanism.

It is understood that the difference between low- and high – temperature series expansions is mainly due to their starting states. The high – temperature series expansion starts from the highly random state where the nonlocal property is negligible, but the low – temperature series expansion initiates from the completely order state where the nonlocal property is extremely strong. One could find a mechanism of spontaneously and simultaneously emerging all the high – temperature configurations at temperature infinitesimal deviates from infinite, whereas it is impossible to find a mechanism of spontaneously emerging all the low – temperature configurations at temperature infinitesimal deviates from zero, because it costs the thermal energy as high as up to the critical point. Therefore, the high – temperature series expansions can be exact at temperature infinitesimal deviates from infinite, whereas the low – temperature series expansions in the previous form cannot be exact at temperature infinitesimal deviates from zero. The situation at low temperatures is worse than at high temperatures, and this may also explain why the low – temperature series expansion is not more reliable than the high – temperature series expansion for predicting the critical point and the critical exponents.

One may ask why the high temperature expansions can be valid for all $T > T_c$ in the 2D and 1D models, but only for $T \to \infty$ in the 3D model. This can be ascribed to differences between the 3D and others. The differences between the 2D and 3D Ising models are indeed evident: Topologic, Symmetric, Dimensionality, Singularity, … The following are several general points on this issue: 1) Types of interactions: It is known that the 3D is specially designed for the validity of the law that the strength of the interaction is inversely proportional to the square of the distance. 2) Types of topologic: The life in high level, like human, fish, …, cannot live in 2D, since a 2D gastrointestinal can divide the body of a 2D fish to separated two parts. Another example is the connection between different points in lattices. The number of the direct connection for communication between two lattice points without any cross to other connections in 1D is two; that in 2D is four; in 3D, infinite. This makes neuro-network of the high – level life possible only in 3D. Next, let us focus on the Ising model, where spins with interactions are located on each lattice point. The intrinsic difference between the 3D and 2D Ising models can be seen clearly by comparing Eqs. (3.3a) and (3.3b) with eq. (14) in Kaufman's paper.[17] The end factors in Eqs. (3.3a) and (3.3b) originate from the boundary condition that the many of the bonds are nonplanar and that these bonds cross over one another with those in other planes. This boundary condition in 3D is much more complicated than that in the 2D case. In order to deal with this complex boundary condition, one needs to introduce the Conjecture 1 to open the topologic knots. As discussed above, this topologic problem hides another related intrinsic property: All the elements in the matrix V are correlated intimately so that the 3D Ising model has the intrinsic nonlocal property, whereas the 2D case lacks such nonlocal behavior. This nonlocal property can be seen also from the form of the additional rotation shown in eq. (3. 6), where the elements K''' of the additional rotation matrix is the mixture of K, K' and K''. This nonlocal property is an intrinsic property of the 3D Ising model, which could cause that any approximation techniques taking into account only the local property cannot be correct for the 3D Ising model, though these approximates work well for other 2D or 1D models. The only exception is the application of the high – temperature expansions at/near infinite temperature, because there is no interaction at infinite temperature and the interaction is comparatively weak at temperature infinitesimal deviates from infinite, in comparison with temperature. Only at this extremely very high - temperature limit, the effect of the nonlocal property can be neglected. This also interprets why the low – temperature expansions diverge in 3D, since such nonlocal property is extremely strong at low temperatures.

Furthermore, a serious problem with the analysis of the 3D Ising model is the presence of confluent singularities, which are extremely weak or non-existent in the 2D model.[111] Both field theory and high – temperature series analysis suggests a value for the confluent exponent not very different from 0.5. Certainly, the 2D and 3D Ising models are intrinsically different. Else, one could find easily the exact solution of the 3D Ising model, immediately after the Onsager's discovery. At least, one can easily locate the critical point as what Kramers and Wannier did. If the critical temperature were indeed, as hoped, the radius of the convergence of the high – temperature series, one should have already located the critical point (almost) exactly by this expansion. The famous group of the critical exponents $\alpha = 1/8$, $\beta = 5/16$, $\gamma = 5/4$, $\delta = 5$, $\eta = 0$ and $\nu = 5/8$, obtained by the high – temperature expansion and suggested by Fisher[103] and Domb,[107] gives the value of $\eta = 0$, which is the same value as what the main field theory predicts, certainly not relevant. The critical exponent $\eta$ denotes the deviation from the Ornstein – Zernike behavior, which certainly cannot be zero. All of these indicate that the success and the optimism of the high – temperature series in other dimensions cannot be planted simply on the 3D case. For 4D, it is well known that the mean field theory, which is believed to be the zero – order approximation, can well describe the behaviors. Any other approximations, better than the mean field theory, of course, can give good results, because they just add higher – order perturbation terms that approach infinitesimal for 4D. To ask further why the 3D differs with other dimensions is analogy to ask why we are living in the 3D world. This is beyond the topic of the present paper. All of these analyses above indicate that both the low- and high-temperature series expansions may not give the exact information at the critical region of the 3D Ising model.

## 4. Monte Carlo simulations

Why can Monte Carlo simulations not give the exact solution? First of all, most of simulations, using computers, are limited by the size effects and the powers of computers. No one can perform the calculations and the simulations on the lattice with the number of the spins or atoms, $N \to \infty$, since the number of the configurations of the Ising model increases in a fashion of $2^N$. In the finite system that is simulated, there is a difficulty that there is not a sharp transition between zero and non-zero magnetization or a sharp peak in the specific heat. Therefore the critical point cannot be located precisely.[212] Furthermore, the absence of a sharp peak in the specific heat hinders to understand its singularity analytically at the critical point. It is known that Monte Carlo simulations are powerful techniques for numerical calculations, which numerically evaluate canonical thermal averages of some observable A by an approximate one, where M states $\{x_\mu\}$ are selected by importance – sampling process.[213] The importance – sampling process consists of the construction of a Markov chain of states ($x_1 \to x_2 \to \ldots x_\mu \to x_{\mu+1} \to \ldots$), where a suitable choice of the transition probability W ($x_\mu \to x_{\mu+1}$) ensures that, for large enough $\mu$, states $x_\mu$ are selected according to the canonical equilibrium probabilities, $P_{eq}$ ($x_\mu$) $\propto$ $\exp\{-H(x_\mu)/k_BT\}$. The limitations of the Monte Carlo simulations were recognized as follows:[213] 1) Only in the limit $M \to \infty$ we can expect to obtain an exact result, while for finite M a "statistical error" is expected. The estimation of this error is a very nontrivial matter, since it depends sensitively on the precise choice of W and subsequently generated states are more or less correlated. It is expected that the "correlation time" diverges in the thermodynamic limit at a second - order phase transition.[321] 2) While the importance sampling method guarantees that, for $\mu \to \infty$, the states are selected according to $P_{eq}$ ($x_\mu$), but for choices of $\mu$ that are not large enough there is still some "memory" of the (arbitrary) initial state with which the Markov chain was started. The non-equilibrium relaxation time for the system has relaxed from the initial state toward the correct thermal equilibrium is divergent at a second – order phase transition in the thermodynamic limit.[322] 3) For the realization of the Markov chain, (pseudo-)random numbers are used both for constructing a trail state $x_\mu$' from a given state $x_\mu$ and for the decision whether or not to accept the trial configuration as a new configuration. For instance, in the Metropolis algorithm, this is done if the transition probability W exceeds a random number that is uniformly distributed in the interval from zero to one.[193-202] Evidently, it is necessary to carefully test the quality of the random numbers since bad random numbers indeed cause systematic errors. However, this is again a nontrivial matter, since there is no unique way of testing random – number generators, and there is no absolute guarantee that a random – number generator that has passed all the standard tests does not yield random numbers leading to systematic errors in a particular application.[234,323-326] 4) Monte Carlo methods apply to system of finite size only, and the results of calculations near a phase transition are affected by finite size and boundaries. The finite size of the simulated lattice, typically a (hyper-)cubic lattice of linear dimension L with periodic boundary conditions in all lattice directions, causes a systematic rounding and shifting of the critical singularities. This is because singularities of the free energy can only develop in the thermodynamic limit $L \to \infty$. This remark is particularly obvious for the correlation length $\xi$, which cannot diverge toward infinite in a finite simulation box, so that serious finite – size effects must be expected when the correlation length $\xi$ has grown to a size comparable to L.[235,327-329] Therefore, the results obtained by the Monte Carlo simulations depend sensitively on the numbers of Monte Carlo steps, the linear dimension L of the simulation box with periodic boundary conditions, the non-equilibrium relaxation time of the system and the quality of the random – number generators, etc.. Actually, the correlation length diverges to infinite at the critical point of a second – order phase transition, the physical fluctuations in the magnetization become very large near the critical point. These fluctuations cannot be entirely suppressed by the importance sampling. In order to obtain good average, the Monte Carlo simulations should be run for an inordinately long time, and to reach the exact solution, unfortunately, for infinite time. Most important, any approximation method cannot prove exact information at/near the critical point, since whenever the thermodynamic functions have an essential singularity it is difficult to perform any computation by successive approximation because the convergence of approximation by analytic functions in such cases is notoriously slow.[13] Even if we continued to work with the smallest lattices possible, the inclusion of longer-range interactions/correlations would force us to bigger systems, and there is a limit to what could be done even with computers many times faster than those available today. Nowadays, calculations of the Ising models have been performed on lattice sizes of L = 256 ~ 5888, far to say infinite. Up to data, most of 3D simulations with the lattice size lager than 4800 are short runs, which could not produce well – equilibrated configurations at the critical point.[184,185,213] In a normal case, increasing the lattice sizes would lower the estimates of the critical point,[170,212,213] which would push the values collected in the Binder and Luijten's review[213] toward to our putative exact solution. For the 2D Ising model, the Monte Carlo simulations give much better results,[213] not only because much larger lattice sizes can be dealt with and but also because there is no topologic problem of the crosses and the knots. Although the combination of the Monte Carlo simulations and the renormalization group techniques reduces the calculation time and give better results by allowing us to study much larger than those possibly by the direct summation methods,[178,179,330] the difficulties above of the Monte Carlo simulations are not removed essentially. In any cases of utilizing the Monte Carlo simulations, the accuracy of the calculations can be improved by increasing either the lattice size or the running time of importance sampling, it is always not practical explicitly to count the states of lattices, or to determine the critical parameters, with more than a handful of lattice sites on any computer, no matter how powerful.[212] Fortunately and misfortunately, this combination intermingle both the advantages and disadvantages of the renormalization group techniques.

## 5. Renormalization group techniques

Why can the renormalization group techniques not give the exact solution? The renormalization group techniques developed by Wilson and others divide roughly into two categories:[141-144,149,152-160,192-202,208-225] 1) The real-space renormalization group techniques, which are close in spirit to the originate idea of Kadanoff,[210,243,331] which allow one to simplify calculations of the critical exponents in the critical regime, without ever working out the partition function. 2) The field theoretical or k – space renormalization group techniques, as developed by pursuing the analogy between statistical mechanics and quantum field theory. In the former case, we are actually concerned with the construction of new models from old by averaging dynamical variables of the old model to form the block variable of the new one. In the latter case, we are actually concerned with changing the parameters of the Landau – Ginzburg model to experimentally more accessible quantities. Superficially, it seems that there is no connection between the categories of the renormalization group techniques. However, at the deeper level there is a close connection, because the basic idea of them is the same and because there are connections between $\phi$ and a set of block variables.[212] It is believed that in both the two categories, performing the renormalization reduces the degrees of freedom, while losing the information of the system.

For the real-space renormalization techniques, the final results depend sensitively on how to divide Kadanoff blocks, define the effective Hamiltonian, determine the details of the block variables, and calculate approximately the partial trace.[141-144,149,152-160,208-225] The larger the Kadanoff blocks, the more accurate the calculations, but the more complicated the procedure. The cumulant expansion during the calculation of the partial trace also causes the uncertainty of the results. The more terms remain, the more accurate the final results. The final results would approach the exact solution if and only if the size of the Kadanoff blocks were chose to be infinite and the infinite terms of the expansion were remained. This is impossible to be done, because much more variables would emerge with increasing the size of the Kadanoff blocks and taking into account more terms of the expansion, which make the calculations become extremely difficult. Calculating the critical exponent ν relates the divergence of the correlation length ($\xi \to \infty$) to the temperature, while the calculation of the critical exponents η and δ must work in the limit of $n \to \infty$ where n stands for the number of iterations of the renormalization transformation. For the calculation of the remaining three ($T \neq T_c$) critical exponents β, γ and α, the difficulties arise from the fact that in the regime very close to the critical fixed point, many measurable quantities change very drastically in response to not only small temperature variations near $T_c$ but also small changes in the parameters appearing in the effective Hamiltonian, remembering that the effective Hamiltonian is temperature – dependent. The critical exponent α proves considerably harder to calculate than the others, because the specific heat is not in any simple way related to the block variables.[212] Moreover, when we renormalize, we knows very little about how T and $T_c$ vary, but just skip the problem by eliminating $T - T_c$ in the favor of the correlation length. All of these difficulties above imply that it is hard to locate exactly the critical point by the real-space renormalization techniques. One could remember that the values of the critical point, well – established in the two recent review articles,[154,213] vary in a very large deviation. Furthermore, a scientist, who does not know what all the values of the exact solutions of the critical point and the critical exponents that are correlated strongly, may choose one of these parameters (though it is an inexact one) as the standard to determine others. This indeed makes serious problems on the accuracy of the calculations, since the high – accurate critical exponents must be determined only at the very narrow temperature interval near to the critical point and the high – accurate critical point can be determined only when the critical exponents used during the calculations are exact.

For the field theoretical or k – space renormalization techniques,[141-144,149,152-160,208-225] as one follows the lines of the original works,[126,127,142] one easily finds out that a series of approximations made could make serious problems for the renormalization group calculations. In the first step, high order contributions to the initial Hamiltonian, which are proportional to $|\vec{s}|^6$, $|\vec{s}|^8$, etc., are neglected for adoption of a continuous local variable or spin $\vec{s}$ with a magnitude constrained by a weight factor for each individual variable $\vec{s}_x$. This point is questionable, especially, for the spin 1/2 Ising models where the strong constraint s = ± 1 might still play a special role. These high order contributions could affect seriously the construction of the reduced Hamiltonian in momentum space for the spin 1/2 Ising models.[143] The high momentum cut - off during the definition of a renormalization group by partial trace over high momentum variables may cause troubles also, since at the critical point all of the spin components with long – and short – wavelengthes (or low- and high- momentums) become comparably dominant. In the final and crucial computational step of realizing the renormalizing group operator by a perturbation expansion treating the coupling constant u (in the field theoretic language) as small parameters, which leads to a graphical formulation abounding in Feynman type integrals, the big problem is that the small parameter is not, in fact, such coupling constant, but rather, the dimensional difference $\in = 4 - d$.[143] For the 3D Ising model, $\in = 1$, cannot be treated as a small parameter. It seems that the series expansion for n = 1 and d = 3 at order $\in^2$ gives the best match of the critical exponent $\gamma$ with the exact solution. Furthermore, the introduction of Feynman graphical techniques, which was originally developed for quantum electrodynamics, makes other serious problems of approximations, because iterating many times a set of non-linear recursion relations obtained by the linearization to leading order (or even several leading orders) in the coupling constant certainly cause uncertainty of the final results and, because no body knows how to account the contributions of all the infinite Feynman graphs that certainly become important in the same order at the critical point of the second – order phase transition.

Although the renormalization group theory is described in mathematical terms, it is not rigorous. Besides we have to make several assumptions: We deal with formal series expansions without knowing anything about their convergence or divergence and the term limit is used without having defined a metric.[236] A serious problem with the renormalization group transformation in real – space or otherwise is that there is no guarantee that they will exhibit fixed points.[149] For some renormalization group transformation, iteration of a critical point does not lead to a fixed point, presumably yielding instead interactions with increasingly long-range forces.[236,332] There is no known principle for avoiding this possibility and a simple approximation to a transformation can misleadingly give a fixed point even when the full transformation cannot.[144-146] Nothing has been known yet about how the absence of a fixed point would be manifested in the Monte Carlo renormalization group computations. It can be concluded from all the facts above that the high-precision calculations of both the real – space and the field theory renormalization group techniques cannot give the exact solution, because of the existence of the systematical errors.

We still need to give the further explanation why the high precision calculations of the renormalization group theory and Monte Carlo simulations cannot give the exact solutions for the critical point and the critical exponents (except for $\gamma$). The key factor is that the systematical errors exist seriously in these techniques, which are caused by their disadvantages discussed above. These systematical errors of these approximation techniques are related directly to the physical conceptions/pictures at the first beginning and the neglects of some important factors during the procedures, etc. For instance, the linearization during the calculations is not very reliable since the non-linear terms could become dominant near/at the critical point. Therefore, the systematical errors of the renormalization group theory and Monte Carlo simulations are intrinsic, which cannot be removed by the efforts of improving technically the precision. This means that estimates of the critical point and the critical exponents by various theories and experimental techniques can become more precise, but not sufficiently being more accurate. The situation here is similar to that everyone uses a gun of the same kind with a very high "so-called accuracy" (actually only a high precision) and a high systematic error, which can shoot 9 points with a high precision, but never shoot exactly 10 points on the target. However, anyone who is willing to take a step back would find the sky is still blue: The renormalization group theory and Monte Carlo simulations are still powerful techniques for the study of the critical phenomena as long as the simulations focus only on the critical exponent γ and, one should keep in mind that there are systematic errors for the critical point and other critical exponents.

**6. Experimental techniques**

Why can the experimental techniques not give the exact solution? First of all, the critical regions are very narrow, $\frac{\Delta T}{T_c} = \frac{T - T_c}{T_c} \lesssim 10^{-2} \sim 10^{-1}$.[105] The temperature difference ΔT can be measured much more accurately than the temperature T itself. The major experimental uncertainty is the relative location of $T_c$ itself.[105] For instance, in specific heat measurement, a rounded peak is often observed, making the precise location of the critical point uncertain. In resonance experiments, the lines of both the order and the disorder phases often overlap in a small temperature region close to the critical point. Measurements in applied magnetic field require the extrapolation to zero field to fix the critical point, making further uncertain. The determinations of critical exponents for bulk even involve extrapolations to zero internal field values. The choice of the critical region and the critical point are certainly interdependent and affect the evaluation of the critical exponents. Furthermore, the domains often exist in ferromagnets and the domain walls serve to break up the long - range correlations so essential to the critical behaviors. For the experiments, it is essential to ensure that the domains should be larger than the theoretical coherence length, but it is infinite at the critical point. The existence of long - range interactions and magnetic anisotropy in the real materials may affect the critical behaviors. Impurities in the magnet sample may seriously affect the value of the critical temperature and thus good measurements require the use of single crystals of extreme purity and well-defined geometry. Last but not least, the spontaneous magnetostriction alters the lattice parameters of the materials at the critical regime. This effect should be included in the analysis of precise data, in order to compare with a theory that has fixed lattice constants. In every case, it is necessary to be convinced either that the transition is indeed second order or that the latent heat in a first order transition is too small to change the critical behaviors under study, since the connection between the critical effects and lattice size and shape can in some case make the transition the first order.[105] It is well – known that the accuracy of experiments is less than that of theories. As stated by Vicentini-Missoni,[268] good data in the critical region are available only on few substances. All of the factors above block the accurate determination of the critical exponents in the real materials, though the accuracy of the experimental techniques has been improved greatly in past several decades.

## 7. More general discussions

We still need to give the further explanation why the multitude of separate determinations of these critical exponents throughout the years, by various independent scientists and using completely different techniques (Monte Carlo simulations, high- and low-temperature expansions, renormalization group field theory and experiments) coincide. Superficially, all of these different techniques are independent each other. In the deeper level, they are related and connected closely. Nowadays, it is believed that the field theory renormalization group technique gives the highest accurate estimates among all these approximation techniques.[154,212,213] The coincidence of the results of the Gell-Mann and Low approach with the Kadanoff – Wilson method, seems at first sight accidental. As mentioned above, the field - theory renormalization group technique is connected with the real – space renormalization group technique, since they share the general idea of the renormalization group and there are the connections between the variables of them.[212] Di Castro and Jona-Lasinio[333] made an effort to underline the deep conceptual unity which is implicit in the various aspects of the renormalization group idea by making explicit the connections between the Kadanoff – Wilson and the Gell-Mann-Low approach. Indeed, with exception of the spaces where the renormalization group techniques are performed, the concepts as well as the processes of the two kinds of the techniques are very similar and closely related each other. For instance, the model Hamiltonian, describing either a relativistic field – theory model in d – 1 space and one time dimension or a classic ferromagnet in d – space dimension of Ising type for n = 1, is actually the same;[333,334] dividing the Kadanoff blocks in the real – space corresponds to cutting off the momentums in the k – space; both methods lead to an asymptotically scaling invariant theory, with equal critical exponents at least in the first few orders in the ε-expansion; …. The Monte Carlo technique and its related renormalization group techniques share the disadvantages of the Monte Carlo process. Even the low-temperature and high-temperature series expansions are related closely with the field theories.[142] Studying the possible phases of interacting constituents in a high or low temperature equalizes in the field theoretic language to study strongly cut - off field theories as what the field theory renormalization technique does for the critical point.[142] It has been well-known that the low-temperature expansions have the lowest precise among all the theoretical techniques, while the accuracy of the experiments is lower than that of the theories. It could be very normal that although separate determinations of these critical exponents were carried out independently, during the processes of determinations and publications, scientists would like to refer more or less to the data published or well – established. For instance, a scientist, who is working on the Monte Carlo technique, may believe that it is advisable to proceed in steps, using the values of the critical exponents from the field – theory renormalization as initial guesses to obtain a good first estimate of his own calculation, in order to avoid ambiguities with the fitting procedure;[213] a scientist, who is doing the experiments, may be asked to compare (or match) with the well – established theoretical values; almost all scientists in the field have pre-set the existence of a non-zero value of the critical exponent α for fitting the experimental data or calculating the critical parameters; almost all scientists in the field use the scaling laws to determine other remained critical exponents from part of the calculated critical exponents, in most cases, including the critical exponent α (pre-set already to be non-zero); ... In the deepest level, all of these theoretical techniques have the same troubles with neglecting the high order terms, the size effects and the knots' effects, etc.. Evidently, all of these terms/effects, which might be negligible in other cases, become comparable with the leading terms at/near the critical point of the second – order phase transition in the thermodynamic limit for the 3D case. Such neglects indeed results in the systematical errors. Furthermore, it could be a fact that typically these approximation calculations are in principle straightforward; however, the increased labor necessary for calculating each succeeding coefficient is large.[315] As normal, computation of the coefficient $a_{n+1}$ in series methods should involve at least as much labor as the cumulative calculation of $a_0$, $a_1$, ..., $a_n$. Thus, while there is in principle no limit to the number of calculable coefficients, in practices, there is a rather sharp upper bound $a_{n\ max}$ (depending on the details of the model being considered) determined by such practical considerations as time and patience and even, at the next level, electronic computer capacity and funding. [315] Typically, the first few coefficients are trivial and no special methodology is necessary. However, in higher orders, the bookkeeping is extremely involved; and, scrupulous accuracy is necessary in determining the coefficients, since the extrapolative analysis of the computed coefficients makes apparent critical – point behavior exceedingly sensitive to tiny fractional changes in the last few available coefficients. For these reasons, it could become of paramount importance to have a well – defined, systematic procedure for computing coefficients, which incorporates as many short cuts as possible and minimizes the opportunity for careless error. However, the position is less satisfactory for 3D lattices for which the series converge more slowly than those for 2D ones.[107] This fact is true not only for the series methods, like low – and high – temperature series expansions, but also for the renormalization group techniques that actually involve the spirit of the series methods. The existence of the sharp upper bound in determining the coefficients could answer why the multitude of separate determinations of the critical exponents throughout the years, by various independent scientists and using completely different techniques coincide.

For six decades since Onsager's solution of the 2D Ising model was reported in 1944, the exact calculation of properties of the 3D version has proved hopelessly difficult. Onsager himself realized immediately that the 3D Ising model cannot be solved exactly by using only the procedure he developed for the 2D version. The application of the algebraic method to the 3D problem is seriously hindered at an early stage, because the operators of interest generate a much large Lie algebra being very difficult to be dealt with.[59] Even it seems impossible to locate exactly the critical point of the 3D Ising model simply by the dual transformation, which was used by Kramers and Wannier[14,15] to locate exactly the critical point of the 2D Ising model. This is because the symmetry for the dual transformation of the 2D Ising model is broken down by the introduction of the third dimension. The combinatorial method of counting the closed graph, developed by Kac and Ward,[60] cannot be generated in any obvious way to the 3D problem, since it introduces some problems in topology that have not been rigorously solved. Actually, their success to simplify the procedure to re-derive the Onsager's results was in large measure due to knowing the answer and they were, in fact, guided by this knowledge.[104,335] Feynman provided the key technical formulation of the needed missing lemma,[104,335] the so-called Feynman's conjecture, which eventually was proved by Sherman,[240,241] making the Kac – Ward method completely rigorous. It has been rather puzzling that the two methods for finding the exact solutions for the Ising problem, namely the algebraic method of Onsager and the combinatorial method employing Pfaffians, have exactly the same range of application, although they appear so different in approach. As marked by Hurst,[336] problems which yield to one method yield to the other, whilst problems which are not tractable by one approach also fail to be exactly solved by the other, although the reasons for this failure appears to have completely different mathematic origins. On the one hand, Ising problems which cannot be solved by the Pfaffian method are characterized by the appearance or crossed bonds which produce unwanted negative signs in the combinatorial generating functions, and such crossed bonds are usually manifestations of the topological structure of the lattice being investigated, i.e., the 3D simple cubic lattice. On the other hand, the Onsager approach breaks down because the Lie algebra encountered in the process of the solution cannot be decomposed into sufficiently simple algebra. It is usually stated that such more complicated algebra occur only when the corresponding lattice has crossed bonds. Barahona and Istrial proved that the general, spin glass 3D Ising model belongs to a class of problems that theorists believe will remain unsolved forever, by translating the Ising model into terms of graph theory.[337-340] What they proved was that computing the energy states for the general, spin glass 3D Ising model is what computer scientists call an NP-complete problem — one of a class of recalcitrant calculations that theorists believe can be solved only by arduous brute – force computations. This is because the 3D lattices are inherently nonplanar and any nonplanar graph throws up a barrier of computational intractability. Following the fundamental results of Onsager,[13] Kac and Ward,[60,335] Feynman,[341] Kasteleyn,[245,246,342] Temperley,[343] Hurst and Green,[244] and Barahona,[337] Istrail showed that the essential ingredient in the NP-completeness of the Ising model is nonplanarity. This criterion includes two-dimensional models with next nearest-neighbor interactions in addition to the nearest-neighbor kind, which researchers had found as vexing to solve as their three-dimensional cousins. Every nonplanar lattice in two or three dimensions, as Istrail showed, contains a subdivision of an infinite graph he called the "basic Kuratowskian".[338] For the basic Kuratowskian with weights –1, 0, and 1 assigned to the edges, the problem of computing a minimum weight "cut" (i.e., set of edges joining vertices in opposite states) is NP-complete. The calculation of the partition function for four spin glass 3D Ising models with {-J, 0, +J} interactions, with {-J, 0} interactions, with {0, +J} interactions, and with {-J, +J} interactions is NP-complete, since their crystal lattice is non-planar. Istrail claimed that the problem falls into the 'computationally intractable' class of conundrums that are too complex to be solved on any realistic timescale. NP-completeness, however, does not mean things are completely hopeless. The complexity results bars algorithms only from solving all instances of the problem in polynomial time.[340] Moreover, such NP-completeness from the point view of computer sciences cannot be fully used to judge the advances in mathematics, which are benefit to uncover the exact solution. Finally, as Istrail noted, it might still be possible to find exact answers for some special cases of the Ising model and in particular, Ising's original, ferromagnetic 3D model, in which all coupling constants are equal (and positive), may turn out to be simple enough to solve within polynomial time.[339,340] With fortunate, what we attempt to exactly solve here is exactly the only possibility opening for exact answers, as Istrail indicated, the ferromagnetic 3D Ising model. The key to solve all the algebraic, combinatorial and topologic problems listed above for solving exactly the 3D Ising model is the introduction of our first Conjecture. The large Lie algebra can be decomposed into sufficiently simple algebra, by introducing the additional rotation in the physical space with higher dimensions, while it also serves to open the crosses/knots to solve the combinatorial and topologic problems. In fact, it is our hope that the putative exact solution of the 3D Ising model reported in this work would provide the keys to efficient algorithms for solving thousands of other computational problems, ranging from factoring large numbers to the notorious traveling salesman problem.[340]

The key point is how to properly judge the correctness of a putative exact solution. In the case that no body knows the standard of such judgment, the proper steps are to judge the correctness of the assumption/conception/conjecture and the deriving procedure. If there were nothing wrong for the assumption/conception/conjecture and the deriving procedure, one would accept the correctness of the final results. Any theories (even those as great as Einstein's general relativity) should have their own assumptions, the conceptions, the conjectures or whatever the starting points. One should allow the existence of such the starting points. In the present case, the only starting points are the two conjectures. The conjecture 1 is based on the well – known fact in topologic, namely, the knots in a 3D space can be opened by a rotation in a 4D space, which is introduced to serve to deal with the well – known topologic problem as well as the non-local property of the 3D model. As mentioned above, the introduction of the additional dimension is not contradictory with the existence of four noncompact space-dimensions, as introduced in Kaluza and Klein's theory for unifying electromagnetism and gravity, and also string theories. Alternatively, one could also treat the additional dimension just as a pure mathematic structure, or, a boundary condition. The introduction of the weights on the eigenvectors, as stated in the conjecture 2, is a very common technique in either physics or mathematics. The detailed calculation of the weights could provide a mechanism for the high – temperature series expansion being exact at temperature infinitesimal deviates from infinite, even in case that its radius of the convergence is reduced to be zero/infinitesimal. After introduction of the two conjectures, the deriving procedure simply follows those used by Onsager, Kaufman, Yang, Fisher, et al. If one would not point out the incorrectness of the two conjectures and the deriving procedure, one would have to accept the correctness of the final solution as the immediate consequence of the conjectures and the procedure.

The principles for judging the correctness of a theory usually are: 1) Self-consistency; 2) Compatibility; 3) Simplicity; 4) Consistency with experiments. The present work is self-consistent, compatible with the exact solution of the 2D Ising model. The present procedure is very simple and elegant, only by the introduction of the two conjectures, while most steps of the procedure directly follow what others employed for the 2D Ising model. That is, employing new initial conditions as the least as possible, deriving new final results as the richest as possible. A good theory indeed! The results obtained are in consistent with the results of experiments carefully performed, although usually the precision of experiments on the critical phenomena is comparatively low.

### C. Symmetry, Uniqueness and Beauty

Finally, we need to check the symmetry and uniqueness of the putative exact solution. For the rectangular lattice, the critical temperatures determined by $K^* = K'$ or $K'^* = K$ coincide. When one interchanges the roles of K and K' (in case of K ≠ K'), however, the eigenvalues and the specific heat of the 2D system could be different by a factor.[13,17] The dual transformation is valid for the rectangular lattice, from which one can also derive another condition of $\sinh 2K \cdot \sinh 2K' = 1$ for the critical temperature.[13,17] For the simple orthorhombic lattices, $K^* = K' + K'' + \frac{K'K''}{K}$ is one of the relations for their Curie temperatures. The dual transformation is held for interchanging the roles of K and $K'+K''+\frac{K'K''}{K}$, i.e., $\left[K'+K''+\frac{K'K''}{K}\right]^*=K$, from which the condition of $\sinh 2K\cdot\sinh 2(K'+K''+\frac{K'K''}{K})=1$ can be derived for the critical temperature. Similarly, when one interchanges the roles of K and $K'+K''+\frac{K'K''}{K}$, the eigenvalues and the specific heat of the 3D system could be different by a factor. It is noticed that the dual transformation is held only for interchanging the roles of K and $K'+K''+\frac{K'K''}{K}$, not for interchanging the roles of K and K' (or K''). This makes the situation of the 3D system very much complicated. For the simple cubic lattice, K = K' = K'', the critical temperatures determined by the procedures setting one of the three crystallographic axes as the starting point of the diagonalization (or the standard axis of the procedure) coincide, certainly, to be at the golden ratio. However, for the simple orthorhombic Ising model with less symmetry, although the form of $KK^*=KK'+KK''+K'K''$ is very symmetric, if we interchanged the role of K with K' (or K'') at the beginning of the procedure, the final results would be different. For instance, if we selected the axis of K' to define K'* as $e^{-2K'}\equiv\tanh K'^*$, the critical temperature of the simple orthorhombic Ising model would be determined by the relation of $K'K'^*=KK'+KK''+K'K''$, which differs indeed with that determined by $KK^*=KK'+KK''+K'K''$. Because K' is smaller than K, the former critical temperature is higher than the latter. Namely, the critical temperature depends on which crystallographic axis of the lattice is set as the standard axis of the procedure. The critical temperature derived from $KK^*=KK'+KK''+K'K''$ is the lowest, as K is the largest one among K, K' and K''. It is a bewilderment that we have not succeeded in equalizing the critical points or other physical quantities, for the situation with any difference between K, K' and K'', which are obtained by setting different axes as the standard one. It would be assured as the following discussion that lack of uniqueness of the solution is ascribed to the intrinsic character of the 3D Ising lattice. From the condition for the critical temperature of the rectangular lattice, $K^*=K'$ (or $K'^*=K$), we have $\frac{K^*}{K}=\frac{K'}{K}$ (or $\frac{K'^*}{K'} = \frac{K}{K'}$) and $K^* K'^* = KK'$. One could map the points in the subspace of $\frac{K'}{K} < 1$ in the parametric axis of $\frac{K'}{K}$ one to one into the subspace of $\frac{K}{K'} > 1$ in the parametric axis of $\frac{K}{K'}$. As considered along the parametric axis of $\frac{K'}{K}$, the duality transformation is held for the parameters in two subspace separated by the point $\frac{K'}{K} = 1$, at which the silver solution is located for the square Ising lattice (the most symmetric one in the 2D system). Note that the silver ratio is actually the largest solution for the 2D Ising system, if we always set the larger one from K and K' as the standard axis. This means that any solution higher than the silver ratio would be forbidden for the 2D Ising lattice, if we started our procedure in this way. From the conditions for the critical temperature of the simple orthorhombic Ising lattices, we have $\frac{K^*}{K} = \frac{K'}{K} + \frac{K''}{K} + \frac{K'K''}{K^2}$ and $KK^* = K'K'^* = K''K''^*$. The condition of $KK^* = K'K'^* = K''K''^*$ for the 3D Ising system differs with $K^* K'^* = KK'$ for the 2D Ising one. But the duality transformation is held for the parameters in two subspaces separated by the curve of $\frac{K'}{K} + \frac{K''}{K} + \frac{K'K''}{K^2} = 1$ in the parametric plane $\frac{K'}{K} \sim \frac{K''}{K}$. Actually, all the points at the curve of $\frac{K'}{K} + \frac{K''}{K} + \frac{K'K''}{K^2} = 1$ (the dashed curve in Fig. 4) correspond to the simple orthorhombic Ising lattices with the critical temperature of the silver solution. However, the golden solution for the simple cubic Ising lattice (the most symmetric one in the 3D system) is not located at the line of $\frac{K'}{K} + \frac{K''}{K} + \frac{K'K''}{K^2} = 1$, but at the point (1, 1) (the star in Fig. 4) in the parametric plane $\frac{K'}{K} \sim \frac{K''}{K}$. The subspace determined by $\frac{K'}{K} + \frac{K''}{K} + \frac{K'K''}{K^2} < 1 \cap \frac{K'}{K} > 0 \cap \frac{K''}{K} > 0$ does not coincide exactly with the one determined by $0 < \frac{K'}{K} < 1 \cap 0 < \frac{K''}{K} < 1$ that can be mapped fully one to one into the subspace determined by $\frac{K''}{K'} < \frac{K}{K'} \cap \frac{K}{K'} > 1 \cap \frac{K''}{K'} > 0$ in the parametric plane $\frac{K}{K'} \sim \frac{K''}{K'}$. This illustrates clearly that lack of uniqueness of the solution is the intrinsic character of the 3D Ising lattice. Even if the term of the additional rotation $\frac{K'K''}{K}$ were not included in our calculation, lack of unification of the solution would be still true. At the very beginning of our procedure, the conditions of K ≥ K' and K ≥ K'' (i.e., $0 \leq \frac{K'}{K} \leq 1 \cap 0 \leq \frac{K''}{K} \leq 1$ in the parametric plane $\frac{K'}{K} \sim \frac{K''}{K}$), are fixed to be held, which are found to be very important, because if one loosed them, the solution would be not unique, physically meaningless. Just taken as a reasonable explanation, we must keep that K''' is not larger than one of K' and K'', in consideration with that the additional rotation is performed in a curled-up dimension. At the beginning of the diagonalization procedure, we have to set up only the largest one among K, K' and K'' as the standard axis for definition of K*, also because the solution obtained in this way is the lowest one. In this way, the points in the area of $0 \leq \frac{K'}{K} \leq 1 \cap 0 \leq \frac{K''}{K} \leq 1$ in the parametric plane $\frac{K'}{K} \sim \frac{K''}{K}$ (see Fig. 4) can represent fully all the possible parameters for the 3D simple orthorhombic Ising lattices. It is obvious that the simple cubic lattice with the highest symmetry has the highest critical temperature, at the golden ratio. In other word, if the 3D lattice is asymmetry with any difference between K, K' and K'', the critical temperature will be lower than the golden ratio, for the largest one among K, K' and K'' is always set as the standard axis. This means that any solution higher than the golden ratio is forbidden for the 3D simple orthorhombic Ising lattices, because of lack of physical significance. It is reasonable since the simple cubic lattice with the highest symmetry is the system with the most evident characters of the three dimensions and the 3D lattice with less symmetry is more or less closer to the 1D or 2D one. As shown in Fig. 1, if one fixes K and sets K' = K'', the critical point decreases with decreasing K' and K'' to disappear as no ordering occurs in the 1D system. However, on the other hand, if one fixed K' = K, the critical point of the simple tetragonal lattices would decrease from the golden ratio of the simple cubic lattice to the silver ratio of the square lattice with decreasing K''.

One of the most remarkable aspects of the golden ratio is the proportion inherent in it. The golden ratio always surprisingly appears at the crosspoint of the simple and complex, the classic geometry and irregular geometry.[344] Most people are familiar with the golden ratio, since it is one of the most ubiquitous irrational numbers known to man, actually, the most irrational number ever, and it is related to the beauty of the nature, such as the golden section, the golden angle, the golden ellipse, the golden triangle, the golden rectangle, the pentagram, the golden spiral, ... The famous Fibonacci sequence and the golden ratio are intertwined with each other. The golden ratio occurs in the structure of both plants and animals and the most well known example is the nautilus shell. The system trends to equilibrium at the state with minimizing the cost of the free energy. It is thought that the system with the golden ratio may stand for such state.[344] Various ways the golden ratio appears in real life may originate from one essential source: the competition between the interaction energy and the thermal activity balances at the critical temperature of the golden ratio for the most symmetric 3D Ising lattice. The natural symmetry is the most important for physical properties of the system. It has been noticed that the golden ratio solution appears for the Curie temperature of the 2D rectangular Ising lattice with K' = 3K, which has less symmetry than the square Ising lattice with the silver ratio solution and therefore, such the golden ratio solution can be excluded from the 2D system as discussed above. It is clear that this most beautiful solution of the golden ratio corresponds only to the critical temperature of the most symmetric 3D simple cubic Ising system. The natural symmetry is the most important for physical properties of the system. The most symmetric, the most beautiful. This is the nature of our symmetric three dimensions, the nature of the world we are living.

## IX. CONCLUSIONS

A putative exact solution of the 3D Ising model on the simple orthorhombic lattices has been derived explicitly. The partition function of the 3D simple orthorhombic Ising model has been evaluated by the spinor analysis, by introducing the two conjectures employing an additional rotation in the fourth curled-up dimension and the weight factors on the eigenvectors. The partition function of the 3D simple orthorhombic Ising model have been dealt within a (3 + 1) - dimensional framework with different weight factors on the eigenvectors. The relation of $KK^* = KK' + KK'' + K'K''$ or $\sinh 2K \cdot \sinh 2(K' + K'' + \frac{K'K''}{K}) = 1$ would be valid for the critical temperature of the simple orthorhombic Ising model, if the two conjectures would be true. For the simple cubic Ising lattice, the putative critical point is located exactly at the golden ratio $x_c = e^{-2K_c} = \frac{\sqrt{5}-1}{2}$, as derived from $K^* = 3K$ or $\sinh 2K \cdot \sinh 6K = 1$. The specific heat of the simple orthorhombic Ising lattices shows a logarithmic singularity at the critical point of the phase transition, however, also based on the validity of the conjectures. The putative exact value for the critical temperature is lower than all the approximation values, as it should be. Because we always set the largest one among K, K' and K'' as the standard axis for the definition of K*, any solution higher than the golden (or silver) ratio would be forbidden for the 3D (or 2D) Ising lattices. The golden (or silver) ratio is the largest solution for the critical temperature of the 3D (or 2D) Ising systems, corresponding the most symmetric simple cubic (or square) lattice. The natural symmetry is the most important for physical properties of the system. The most symmetric, the most beautiful. The spontaneous magnetization of the 3D simple orthorhombic Ising ferromagnet has been derived by the perturbation procedure, following the introduction of the conjectures. The spin correlation functions have been discussed on the terms of the Pfaffians, by defining the effective skew-symmetric matrix $A_{eff}$. The true range $\kappa_x$ of the correlation, and the susceptibility of the 3D simple orthorhombic Ising system have been determined by the procedures similar to those used for the two-dimensional Ising system. The putative critical exponents for the 3D simple orthorhombic Ising lattices have been derived explicitly to be $\alpha = 0$, $\beta = 3/8$, $\gamma = 5/4$, $\delta = 13/3$, $\eta = 1/8$ and $\nu = 2/3$, showing the universality behavior and satisfying the scaling laws. These exact values for the critical exponents of the 3D Ising lattice are located between those for the 2D and (mean field) 4D Ising ones. These critical exponents are close to the approximation values and the experimental data. The exact solutions have been judged by several criterions. The reasons for the deviations of the approximation results and the experimental data from the putative exact solutions are interpreted. The simple cubic lattice with the highest symmetry is with the most evident characters of the three dimensions and, the 3D lattice with less symmetry is more or less closer to the 1D or 2D one. The 3D – to – 2D crossover phenomenon differs with the 2D – to – 1D crossover phenomenon and there is a gradual crossover of the exponents from the 3D values to the 2D ones. Special attentions have been also paid on the extra energy caused by the introduction of the fourth curled-up dimension, the chaotic states at/near infinite temperature as revealed by the introduction of the weight factors of the eigenvectors. The physics beyond the conjectures and the existence of the extra dimension are discussed, with a rethought on the quantum mechanism. The present work would not only have a big significance in statistic physics and condensed matter physics, but also fill the gap between the fields of the quantum field theory, the cosmology theory, high-energy particle physics, graph theory and computer sciences. Our results with aesthetic appeal reveal the nature of the nature: simple, symmetry and beautiful.

## ACKNOWLEDGEMENTS

The author appreciates very much the continued supports of the National Natural Science Foundation of China since 1990 (under grant numbers 59001452, 59371015, 19474052, 59421001, 59725103, 59871054, 59831010, 50171070, 10274087, 10674139, 50331030, and 50332020) and the supports of the Sciences and Technology Commission of Shenyang since 1994. He is grateful to Fei Yang for understanding, encouragement, support and convesation.

## Appendix A:

The effects of the rotations $\omega_{2t_y}$ and $\omega_{2t_z}$ with their weight factors $w_y$ and $w_z$ are discussed as follows. It is not surprising that things can come out of nothing in nature. The addition of the fourth curled-up dimension expands the 3D physical world to the higher dimensional world so that it is necessary to introduce the weights $w_y$ and $w_z$ make it be effectively equalized with the original system. The weights defined in the Conjecture 2 vary in range of [-1, 1]. They can be equal to 0, ±1, and any values between 0 and ±1. They could interchange their roles (and values) at any time, from the point of view of symmetric. In this way, the system studied is always (3 + 1) – dimensional, even in case that anyone of the weights occasionally equals to zero. A "zero" weight factor $w_y$ or $w_z$ is still acting with other dimensions, is not more mystical than the Euler's equation $e^{i\pi} + 1 = 0$: an imaginary number interacting with real numbers to produce nothing.

The partition function of the 3D simple cubic Ising lattice could be expressed as (3.63):

$$N^{-1} \ln Z = \ln 2 + \frac{1}{2(2\pi)^4} \int_{-\pi}^{\pi}\int_{-\pi}^{\pi}\int_{-\pi}^{\pi}\int_{-\pi}^{\pi} \ln\left[\cosh 2K \cosh 6K - \sinh 2K \cos \omega' - \sinh 6K (w_x \cos \omega_x + w_y \cos \omega_y + w_z \cos \omega_z)\right] d\omega' d\omega_x d\omega_y d\omega_z \tag{A.1}$$

where $w_x = 1$ and

$$w_y = w_z = \pm\sqrt{\sum_{i=0}^{\infty} b_i \kappa_i^{2i}} , \tag{A.2}$$

with $b_0$ = 7/18, $b_1$ = -4025/216, $b_2$ = -62125/432, $b_3$ = -315237349/31104, $b_4$ = -196961527937/186624, $b_5$ = -81949884191959/746496, $b_6$ = -159843718723121207/13436928, $b_7$ = -887867077613442477677/644972544, $b_8$ = -216261883802726526301599/1289945088, $b_9$ = -2730834093284969018142818253/139314069504, $b_{10}$ = -931292714680130608913105509313/278628139008, …

*Ansatz 1*: All of the coefficients $b_i$ for the terms embodied in $w_y$ and $w_z$ are negative for i ≥ 1.

There is no mathematical trouble for the introduction of the weights $w_x$, $w_y$ and $w_z$, which is used to embody some information of the high temperature terms. Clearly, the close form of the free energy has been found for the 3D Ising model from T = 0 to any finite temperatures and also at the infinite temperature ($\kappa = 0$), with exception of near the infinite temperature ($\kappa \to 0$). Actually, in principle, we could account all of the infinite terms for near the infinite temperature, based on the law of the high – temperature expansion (though extremely difficulty). The coefficients $b_i$ can be determined exactly to higher orders as long as the corresponding terms of the high temperature expansion are determined. One would estimate how the higher – order terms look like from these existed terms. All of the coefficients for the terms up to $b_{10}$, except for $b_0$, embodied in $w_y$ and $w_z$ are negative. *Ansatz 1* is introduced to extend this tendency to all the terms inside the square root. The ratio of $b_{i+1}/b_i$ varies in range of 70 – 171 for $i \geq 2$. One could expect that all the coefficients $b_i$ for $i > 10$ are negative, with the ratios of $b_{i+1}/b_i$ in order of hundreds. Actually, these facts can be proved by evaluating the higher – order coefficients $b_i$ approximately with high precisions, even without having information of higher – order terms of the high temperature expansion. This is because in the procedure of determining each coefficient $b_i$, there is always a term, which does not relate with any lower order coefficients, but appears to compensate most of contribution of the high – order terms of the high temperature expansion. Furthermore, the terms that relate directly with lower order coefficients are dominant for the higher order coefficient $b_i$. Thus, near the infinite temperature ($\kappa \to 0$), the effects of the high order terms (higher than 22nd) are extremely small, which can be neglected in a certain sense. Therefore, these infinite terms of the high temperature expansion can be embodied into the close square form of the weight factors.

One would estimate how the weights look like from these existed terms, supposing the *Ansatz 1 is* true. It is noticed that the weights $w_y = w_z = \pm\sqrt{\frac{7}{18}}$ at infinite temperature as $\kappa = 0$. The weights $w_y$ and $w_z$ approach zero identically when the temperature of the system becomes infinitesimal below infinite (i.e., $\kappa \to \neq 0$ and infinitesimal), because if *Ansatz 1* were true, the truncated sums in (A.2) would become negative making the square root to be imagine, physically no meaningful. This indicates that the weights $w_y$ and $w_z$ are always equal to zero at any finite temperatures.

By the following procedure, we will show that at/near infinite temperature, the exact solution for the partition function of the 3D simple cubic Ising lattice above fits exactly to the high temperature series expansion.[80,93,107,111] The formula (A.1) can be expressed as:

$$\ln\lambda - \ln 2(\cosh 2K \cosh 6K)^{\frac{1}{2}} = \frac{1}{2(2\pi)^4}\int_0^{\pi}\int_0^{\pi}\int_0^{\pi}\int_0^{\pi} \ln[1 - 2\kappa_{3D}\cos\omega' - 2\kappa'_{3D}(w_x \cos\omega_x + w_y\cos\omega_y + w_z\cos\omega_z)]d\omega' d\omega_x d\omega_y d\omega_z \quad (A.3)$$

Here $\lambda = Z^{1/N}$,

$$\kappa_{3D} = \frac{\tanh 2K}{2\cosh 6K} = \frac{\kappa(1-3\kappa^2+3\kappa^4-\kappa^6)}{1+16\kappa^2+30\kappa^4+16\kappa^6+\kappa^8}$$
$$= \kappa - 19\kappa^3 + 277\kappa^5 - 3879\kappa^7 + 54057\kappa^9 - 752955\kappa^{11} + 10487357\kappa^{13} - 146070095\kappa^{15} + 2034494033\kappa^{17} - 28336846435\kappa^{19} + 394681356133\kappa^{21}\cdots \quad (A.4)$$

and

$$\kappa'_{3D} = \frac{\tanh 6K}{2\cosh 2K} = \frac{\kappa(3+7\kappa^2-7\kappa^4-3\kappa^6)}{1+16\kappa^2+30\kappa^4+16\kappa^6+\kappa^8}$$
$$= 3\kappa - 41\kappa^3 + 559\kappa^5 - 7765\kappa^7 + 108123\kappa^9 - 1505921\kappa^{11} + 20974727\kappa^{13} - 292140205\kappa^{15} + 4068988083\kappa^{17} - 56673692889\kappa^{19} + 789362712287\kappa^{21}\cdots \quad (A.5)$$

with $\kappa = \tanh K$. Then one can calculate immediately the high order terms $\kappa_{3D}^2$, $\kappa_{3D}^4$, $\kappa_{3D}^6$, $\kappa_{3D}^8$,… ; $\kappa'^2_{3D}$, $\kappa'^4_{3D}$, $\kappa'^6_{3D}$, $\kappa'^8_{3D}$,… ; $\kappa_{3D}^2\kappa'^2_{3D}$, $\kappa_{3D}^4\kappa'^2_{3D}$, $\kappa_{3D}^6\kappa'^2_{3D}$, …; $\kappa_{3D}^2\kappa'^4_{3D}$, $\kappa_{3D}^4\kappa'^4_{3D}$, $\kappa_{3D}^6\kappa'^4_{3D}$, …; ……

The left side of the equation above for the partition function is re-written as:

$$\ln\lambda - \ln\left[2\cosh^3 K\left(\frac{1+16\kappa^2+30\kappa^4+16\kappa^6+\kappa^8}{1-\kappa^2}\right)^{\frac{1}{2}}\right] =$$

$$\ln\lambda - \ln\left[2\cosh^3 K\left(1+\frac{17}{2}\kappa^2 - \frac{101}{8}\kappa^4 + \frac{2221}{16}\kappa^6 - \frac{157133}{128}\kappa^8 + \frac{3128095}{256}\kappa^{10}\right.\right.$$
$$-\frac{132058577}{1024}\kappa^{12} + \frac{2909991333}{2048}\kappa^{14} - \frac{529380209469}{32768}\kappa^{16} + \frac{12331364457995}{65536}\kappa^{18}$$
$$\left.\left.-\frac{585509987761867}{262144}\kappa^{20} + \frac{155271445442468377}{5767168}\kappa^{22} + \cdots\right)\right]$$

(A.6)

Expanding the logarithmic function on the right hand of the expression for the partition function by $\ln(1-x) = -\sum_{n=1}^{\infty}\frac{x^n}{n}$ yields:

$$-\sum_{n=1}^{\infty}\frac{1}{n}\cdot\sum_{p+q+r+s=n}\frac{n!}{p!q!r!s!}\cdot 2^n \kappa_{3D}^{p}\kappa'^{(q+r+s)}_{3D}\cdot f^{(r+s)}\cdot\cos^p\omega'\cos^q\omega_x\cos^r\omega_y\cos^s\omega_z$$

(A.7)

and then integrating each term results in:

$$-\left\{\frac{1}{2}(\kappa_{3D}^2+\kappa'^2_{3D}) + f^2\kappa'^2_{3D}\right\}$$
$$-\left\{\left[\frac{3}{4}(\kappa_{3D}^4+\kappa'^4_{3D}) + 3\kappa_{3D}^2\kappa'^2_{3D}\right] + 6f^2(\kappa_{3D}^2\kappa'^2_{3D}+\kappa'^4_{3D}) + 4.5f^4\kappa'^4_{3D}\right\}$$
$$-\left\{\left[\frac{5}{3}(\kappa_{3D}^6+\kappa'^6_{3D}) + 15(\kappa_{3D}^4\kappa'^2_{3D}+\kappa_{3D}^2\kappa'^4_{3D})\right] + f^2\left[30(\kappa_{3D}^4\kappa'^2_{3D}+\kappa'^6_{3D}) + 120\kappa_{3D}^2\kappa'^4_{3D}\right]\right.$$
$$\left.+90f^4(\kappa_{3D}^2\kappa'^4_{3D}+\kappa'^6_{3D}) + \frac{100}{3}f^6\kappa'^6_{3D}\right\}$$
$$-\left\{\left[\frac{35}{8}(\kappa_{3D}^8+\kappa'^8_{3D}) + 70(\kappa_{3D}^6\kappa'^2_{3D}+\kappa_{3D}^2\kappa'^6_{3D}) + \frac{315}{2}\kappa_{3D}^4\kappa'^4_{3D}\right]\right.$$
$$+f^2\left[140(\kappa_{3D}^6\kappa'^2_{3D}+\kappa'^8_{3D}) + 1260(\kappa_{3D}^4\kappa'^4_{3D}+\kappa_{3D}^2\kappa'^6_{3D})\right]$$
$$\left.+f^4\left[945(\kappa_{3D}^4\kappa'^4_{3D}+\kappa'^8_{3D}) + 3780\kappa_{3D}^2\kappa'^6_{3D}\right] + 1400f^6(\kappa_{3D}^2\kappa'^6_{3D}+\kappa'^8_{3D}) + 306.25f^8\kappa'^8_{3D}\right\}$$

$$
\begin{aligned}
&-\Bigg\{\left[\frac{63}{5}(\kappa_{3D}^{10}+\kappa'^{10}_{3D})+315(\kappa_{3D}^{8}\kappa'^{2}_{3D}+\kappa_{3D}^{2}\kappa'^{8}_{3D})+1260(\kappa_{3D}^{6}\kappa'^{4}_{3D}+\kappa_{3D}^{4}\kappa'^{6}_{3D})\right]\\
&+f^{2}\left[630(\kappa_{3D}^{8}\kappa'^{2}_{3D}+\kappa'^{10}_{3D})+10080(\kappa_{3D}^{6}\kappa'^{4}_{3D}+\kappa_{3D}^{2}\kappa'^{8}_{3D})+22680\kappa_{3D}^{4}\kappa'^{6}_{3D}\right]\\
&+f^{4}\left[7560(\kappa_{3D}^{6}\kappa'^{4}_{3D}+\kappa'^{10}_{3D})+68040(\kappa_{3D}^{4}\kappa'^{6}_{3D}+\kappa_{3D}^{2}\kappa'^{8}_{3D})\right]\\
&+f^{6}\left[25200(\kappa_{3D}^{4}\kappa'^{6}_{3D}+\kappa'^{10}_{3D})+100800\kappa_{3D}^{2}\kappa'^{8}_{3D}\right]\\
&+22050f^{8}(\kappa_{3D}^{2}\kappa'^{8}_{3D}+\kappa'^{10}_{3D})+3175.2f^{10}\kappa'^{10}_{3D}\Bigg\}\\
&-\Bigg\{\Bigg[\frac{77}{2}(\kappa_{3D}^{12}+\kappa'^{12}_{3D})+1386(\kappa_{3D}^{10}\kappa'^{2}_{3D}+\kappa_{3D}^{2}\kappa'^{10}_{3D})\\
&+8662.5(\kappa_{3D}^{8}\kappa'^{4}_{3D}+\kappa_{3D}^{4}\kappa'^{8}_{3D})+15400\kappa_{3D}^{6}\kappa'^{6}_{3D}\Bigg]\\
&+f^{2}\left[2772(\kappa_{3D}^{10}\kappa'^{2}_{3D}+\kappa'^{12}_{3D})+69300(\kappa_{3D}^{8}\kappa'^{4}_{3D}+\kappa_{3D}^{2}\kappa'^{10}_{3D})+277200(\kappa_{3D}^{6}\kappa'^{6}_{3D}+\kappa_{3D}^{4}\kappa'^{8}_{3D})\right]\\
&+f^{4}\left[51975(\kappa_{3D}^{8}\kappa'^{4}_{3D}+\kappa'^{12}_{3D})+831600(\kappa_{3D}^{6}\kappa'^{6}_{3D}+\kappa_{3D}^{2}\kappa'^{10}_{3D})+1871100\kappa_{3D}^{4}\kappa'^{8}_{3D}\right]\\
&+f^{6}\left[308000(\kappa_{3D}^{6}\kappa'^{6}_{3D}+\kappa'^{12}_{3D})+2772000(\kappa_{3D}^{4}\kappa'^{8}_{3D}+\kappa_{3D}^{2}\kappa'^{10}_{3D})\right]\\
&+f^{8}\left[606375(\kappa_{3D}^{4}\kappa'^{8}_{3D}+\kappa'^{12}_{3D})+2425500\kappa_{3D}^{2}\kappa'^{10}_{3D}\right]\\
&+349272f^{10}(\kappa_{3D}^{2}\kappa'^{10}_{3D}+\kappa'^{12}_{3D})+35574f^{12}\kappa'^{12}_{3D}\Bigg\}\\
&-\Bigg\{\Bigg[\frac{858}{7}(\kappa_{3D}^{14}+\kappa'^{14}_{3D})+6006(\kappa_{3D}^{12}\kappa'^{2}_{3D}+\kappa_{3D}^{2}\kappa'^{12}_{3D})\\
&+54054(\kappa_{3D}^{10}\kappa'^{4}_{3D}+\kappa_{3D}^{4}\kappa'^{10}_{3D})+150150(\kappa_{3D}^{8}\kappa'^{6}_{3D}+\kappa_{3D}^{6}\kappa'^{8}_{3D})\Bigg]\\
&+f^{2}\Big[12012(\kappa_{3D}^{12}\kappa'^{2}_{3D}+\kappa'^{14}_{3D})+432432(\kappa_{3D}^{10}\kappa'^{4}_{3D}+\kappa_{3D}^{2}\kappa'^{12}_{3D})\\
&+2702700(\kappa_{3D}^{8}\kappa'^{6}_{3D}+\kappa_{3D}^{4}\kappa'^{10}_{3D})+4804800\kappa_{3D}^{6}\kappa'^{8}_{3D}\Big]\\
&+f^{4}\Big[324324(\kappa_{3D}^{10}\kappa'^{4}_{3D}+\kappa'^{14}_{3D})+8108100(\kappa_{3D}^{8}\kappa'^{6}_{3D}+\kappa_{3D}^{2}\kappa'^{12}_{3D})\\
&+32432400(\kappa_{3D}^{6}\kappa'^{8}_{3D}+\kappa_{3D}^{4}\kappa'^{10}_{3D})\Big]\\
&+f^{6}\left[3003000(\kappa_{3D}^{8}\kappa'^{6}_{3D}+\kappa'^{14}_{3D})+48048000(\kappa_{3D}^{6}\kappa'^{8}_{3D}+\kappa_{3D}^{2}\kappa'^{12}_{3D})+108108000\kappa_{3D}^{4}\kappa'^{10}_{3D}\right]\\
&+f^{8}\left[10510500(\kappa_{3D}^{6}\kappa'^{8}_{3D}+\kappa'^{14}_{3D})+94594500(\kappa_{3D}^{4}\kappa'^{10}_{3D}+\kappa_{3D}^{2}\kappa'^{12}_{3D})\right]\\
&+f^{10}\left[13621608(\kappa_{3D}^{4}\kappa'^{10}_{3D}+\kappa'^{14}_{3D})+54486432\kappa_{3D}^{2}\kappa'^{12}_{3D}\right]\\
&+5549544f^{12}(\kappa_{3D}^{2}\kappa'^{12}_{3D}+\kappa'^{14}_{3D})+420665\frac{1}{7}f^{14}\kappa'^{14}_{3D}\Bigg\}
\end{aligned}
$$

$$
-\left\{\left[\frac{6435}{16}(\kappa_{3D}^{16}+\kappa'^{16}_{3D})+25740(\kappa_{3D}^{14}\kappa'^{2}_{3D}+\kappa_{3D}^{2}\kappa'^{14}_{3D})\right.\right.
$$

$$
\left.+315315(\kappa_{3D}^{12}\kappa'^{4}_{3D}+\kappa_{3D}^{4}\kappa'^{12}_{3D})+1261260(\kappa_{3D}^{10}\kappa'^{6}_{3D}+\kappa_{3D}^{6}\kappa'^{10}_{3D})+\frac{7882875}{4}\kappa_{3D}^{8}\kappa'^{8}_{3D}\right]
$$

$$
+f^{2}\left[51480(\kappa_{3D}^{14}\kappa'^{2}_{3D}+\kappa'^{16}_{3D})+2522520(\kappa_{3D}^{12}\kappa'^{4}_{3D}+\kappa_{3D}^{2}\kappa'^{14}_{3D})\right.
$$

$$
\left.+22702680(\kappa_{3D}^{10}\kappa'^{6}_{3D}+\kappa_{3D}^{4}\kappa'^{12}_{3D})+63063000(\kappa_{3D}^{8}\kappa'^{8}_{3D}+\kappa_{3D}^{6}\kappa'^{10}_{3D})\right]
$$

$$
+f^{4}\left[1891890(\kappa_{3D}^{12}\kappa'^{4}_{3D}+\kappa'^{16}_{3D})+68108040(\kappa_{3D}^{10}\kappa'^{6}_{3D}+\kappa_{3D}^{2}\kappa'^{14}_{3D})\right.
$$

$$
\left.+425675250(\kappa_{3D}^{8}\kappa'^{8}_{3D}+\kappa_{3D}^{4}\kappa'^{12}_{3D})+756756000\kappa_{3D}^{6}\kappa'^{10}_{3D}\right]
$$

$$
+f^{6}\left[25225200(\kappa_{3D}^{10}\kappa'^{6}_{3D}+\kappa'^{16}_{3D})+630630000(\kappa_{3D}^{8}\kappa'^{8}_{3D}+\kappa_{3D}^{2}\kappa'^{14}_{3D})\right.
$$

$$
\left.+2522520000(\kappa_{3D}^{6}\kappa'^{10}_{3D}+\kappa_{3D}^{4}\kappa'^{12}_{3D})\right]
$$

$$
+f^{8}\left[137950312.5(\kappa_{3D}^{8}\kappa'^{8}_{3D}+\kappa'^{16}_{3D})+2207205000(\kappa_{3D}^{6}\kappa'^{10}_{3D}+\kappa_{3D}^{2}\kappa'^{14}_{3D})\right.
$$

$$
\left.+4966211250\kappa_{3D}^{4}\kappa'^{12}_{3D}\right]
$$

$$
+f^{10}\left[317837520(\kappa_{3D}^{6}\kappa'^{10}_{3D}+\kappa'^{16}_{3D})+2860537680(\kappa_{3D}^{4}\kappa'^{12}_{3D}+\kappa_{3D}^{2}\kappa'^{14}_{3D})\right]
$$

$$
+f^{12}\left[291351060(\kappa_{3D}^{4}\kappa'^{12}_{3D}+\kappa'^{16}_{3D})+1165404240\kappa_{3D}^{2}\kappa'^{14}_{3D}\right]
$$

$$
\left.+88339680f^{14}(\kappa_{3D}^{2}\kappa'^{14}_{3D}+\kappa'^{16}_{3D})+5176153.125f^{16}\kappa'^{16}_{3D}\right\}
$$

$$
-\left\{\left[\frac{12155}{9}(\kappa_{3D}^{18}+\kappa'^{18}_{3D})+109395(\kappa_{3D}^{16}\kappa'^{2}_{3D}+\kappa_{3D}^{2}\kappa'^{16}_{3D})+1750320(\kappa_{3D}^{14}\kappa'^{4}_{3D}+\kappa_{3D}^{4}\kappa'^{14}_{3D})\right.\right.
$$

$$
\left.+9529520(\kappa_{3D}^{12}\kappa'^{6}_{3D}+\kappa_{3D}^{6}\kappa'^{12}_{3D})+21441420(\kappa_{3D}^{10}\kappa'^{8}_{3D}+\kappa_{3D}^{8}\kappa'^{10}_{3D})\right]
$$

$$
+f^{2}\left[218790(\kappa_{3D}^{16}\kappa'^{2}_{3D}+\kappa'^{18}_{3D})+14002560(\kappa_{3D}^{14}\kappa'^{4}_{3D}+\kappa_{3D}^{2}\kappa'^{16}_{3D})\right.
$$

$$
+171531360(\kappa_{3D}^{12}\kappa'^{6}_{3D}+\kappa_{3D}^{4}\kappa'^{14}_{3D})+686125440(\kappa_{3D}^{10}\kappa'^{8}_{3D}+\kappa_{3D}^{6}\kappa'^{12}_{3D})
$$

$$
\left.+1072071000\kappa_{3D}^{8}\kappa'^{10}_{3D}\right]
$$

$$
+f^{4}\left[10501920(\kappa_{3D}^{14}\kappa'^{4}_{3D}+\kappa'^{18}_{3D})+514594080(\kappa_{3D}^{12}\kappa'^{6}_{3D}+\kappa_{3D}^{2}\kappa'^{16}_{3D})\right.
$$

$$
\left.+4631346720(\kappa_{3D}^{10}\kappa'^{8}_{3D}+\kappa_{3D}^{4}\kappa'^{14}_{3D})+12864852000(\kappa_{3D}^{8}\kappa'^{10}_{3D}+\kappa_{3D}^{6}\kappa'^{12}_{3D})\right]
$$

$$
+f^{6}\left[190590400(\kappa_{3D}^{12}\kappa'^{6}_{3D}+\kappa'^{18}_{3D})+6861254400(\kappa_{3D}^{10}\kappa'^{8}_{3D}+\kappa_{3D}^{2}\kappa'^{16}_{3D})\right.
$$

$$
\left.+42882840000(\kappa_{3D}^{8}\kappa'^{10}_{3D}+\kappa_{3D}^{4}\kappa'^{14}_{3D})+76236160000\kappa_{3D}^{6}\kappa'^{12}_{3D}\right]
$$

$$
+f^{8}\left[1500899400(\kappa_{3D}^{10}\kappa'^{8}_{3D}+\kappa'^{18}_{3D})+37522485000(\kappa_{3D}^{8}\kappa'^{10}_{3D}+\kappa_{3D}^{2}\kappa'^{16}_{3D})\right.
$$

$$
\left.+150089940000(\kappa_{3D}^{6}\kappa'^{12}_{3D}+\kappa_{3D}^{4}\kappa'^{14}_{3D})\right]
$$

$$
+f^{10}\left[5403237840(\kappa_{3D}^{8}\kappa'^{10}_{3D}+\kappa'^{18}_{3D})+86451805440(\kappa_{3D}^{6}\kappa'^{12}_{3D}+\kappa_{3D}^{2}\kappa'^{16}_{3D})\right.
$$

$$
\left.+194516562240\kappa_{3D}^{4}\kappa'^{14}_{3D}\right]
$$

$$
+f^{12}\left[8805276480(\kappa_{3D}^{6}\kappa'^{12}_{3D}+\kappa'^{18}_{3D})+79247488320(\kappa_{3D}^{4}\kappa'^{14}_{3D}+\kappa_{3D}^{2}\kappa'^{16}_{3D})\right]
$$

$$
+f^{14}\left[6007098240(\kappa_{3D}^{4}\kappa'^{14}_{3D}+\kappa'^{18}_{3D})+24028392960\kappa_{3D}^{2}\kappa'^{16}_{3D}\right]
$$

$$
\left.+1407913650f^{16}(\kappa_{3D}^{2}\kappa'^{16}_{3D}+\kappa'^{18}_{3D})+65664011\frac{1}{9}f^{18}\kappa'^{18}_{3D}\right\}
$$

$$
\begin{aligned}
&-\Big\{\Big[4618.9(\kappa_{3D}^{20}+\kappa'^{20}_{3D})+461890(\kappa_{3D}^{18}\kappa'^{2}_{3D}+\kappa_{3D}^{2}\kappa'^{18}_{3D})\\
&+\frac{18706545}{2}(\kappa_{3D}^{16}\kappa'^{4}_{3D}+\kappa_{3D}^{4}\kappa'^{16}_{3D})+66512160(\kappa_{3D}^{14}\kappa'^{6}_{3D}+\kappa_{3D}^{6}\kappa'^{14}_{3D})\\
&+203693490(\kappa_{3D}^{12}\kappa'^{8}_{3D}+\kappa_{3D}^{8}\kappa'^{12}_{3D})+\frac{1466593128}{5}\kappa_{3D}^{10}\kappa'^{10}_{3D}\Big]\\
&+f^{2}\Big[923780(\kappa_{3D}^{18}\kappa'^{2}_{3D}+\kappa'^{20}_{3D})+74826180(\kappa_{3D}^{16}\kappa'^{4}_{3D}+\kappa_{3D}^{2}\kappa'^{18}_{3D})\\
&+1197218880(\kappa_{3D}^{14}\kappa'^{6}_{3D}+\kappa_{3D}^{4}\kappa'^{16}_{3D})+6518191680(\kappa_{3D}^{12}\kappa'^{8}_{3D}+\kappa_{3D}^{6}\kappa'^{14}_{3D})\\
&+14665931280(\kappa_{3D}^{10}\kappa'^{10}_{3D}+\kappa_{3D}^{8}\kappa'^{12}_{3D})\Big]\\
&+f^{4}\Big[56119635(\kappa_{3D}^{16}\kappa'^{4}_{3D}+\kappa'^{20}_{3D})+3591656640(\kappa_{3D}^{14}\kappa'^{6}_{3D}+\kappa_{3D}^{2}\kappa'^{18}_{3D})\\
&+43997793840(\kappa_{3D}^{12}\kappa'^{8}_{3D}+\kappa_{3D}^{4}\kappa'^{16}_{3D})+175991175360(\kappa_{3D}^{10}\kappa'^{10}_{3D}+\kappa_{3D}^{6}\kappa'^{14}_{3D})\\
&+274986211500\kappa_{3D}^{8}\kappa'^{12}_{3D}\Big]\\
&+f^{6}\Big[1330243200(\kappa_{3D}^{14}\kappa'^{6}_{3D}+\kappa'^{20}_{3D})+65181916800(\kappa_{3D}^{12}\kappa'^{8}_{3D}+\kappa_{3D}^{2}\kappa'^{18}_{3D})\\
&+586637251200(\kappa_{3D}^{10}\kappa'^{10}_{3D}+\kappa_{3D}^{4}\kappa'^{16}_{3D})+1629547920000(\kappa_{3D}^{8}\kappa'^{12}_{3D}+\kappa_{3D}^{6}\kappa'^{14}_{3D})\Big]\\
&+f^{8}\Big[14258544300(\kappa_{3D}^{12}\kappa'^{8}_{3D}+\kappa'^{20}_{3D})+513307594800(\kappa_{3D}^{10}\kappa'^{10}_{3D}+\kappa_{3D}^{2}\kappa'^{18}_{3D})\\
&+3208172467500(\kappa_{3D}^{8}\kappa'^{12}_{3D}+\kappa_{3D}^{4}\kappa'^{16}_{3D})+5703417720000\kappa_{3D}^{6}\kappa'^{14}_{3D}\Big]\\
&+f^{10}\Big[73916293651.2(\kappa_{3D}^{10}\kappa'^{10}_{3D}+\kappa'^{20}_{3D})+1847907341280(\kappa_{3D}^{8}\kappa'^{12}_{3D}+\kappa_{3D}^{2}\kappa'^{18}_{3D})\\
&+7391629365120(\kappa_{3D}^{6}\kappa'^{14}_{3D}+\kappa_{3D}^{4}\kappa'^{16}_{3D})\Big]\\
&+\mathrm{f}^{12}\Big[188212784760(\kappa_{3D}^{8}\kappa'^{12}_{3D}+\kappa'^{20}_{3D})+3011404556160(\kappa_{3D}^{6}\kappa'^{14}_{3D}+\kappa_{3D}^{2}\kappa'^{18}_{3D})\\
&+6775660251360\kappa_{3D}^{4}\kappa'^{16}_{3D}\Big]\\
&+f^{14}\Big[228269733120(\kappa_{3D}^{6}\kappa'^{14}_{3D}+\kappa'^{20}_{3D})+2054427598080(\kappa_{3D}^{4}\kappa'^{16}_{3D}+\kappa_{3D}^{2}\kappa'^{18}_{3D})\Big]\\
&+f^{16}\Big[120376617075(\kappa_{3D}^{4}\kappa'^{16}_{3D}+\kappa'^{20}_{3D})+481506468300\kappa_{3D}^{2}\kappa'^{18}_{3D}\Big]\\
&+22457091800f^{18}(\kappa_{3D}^{2}\kappa'^{18}_{3D}+\kappa'^{20}_{3D})+853369488.4f^{20}\kappa'^{20}_{3D}\Big\}
\end{aligned}
$$

$$
\begin{aligned}
&-\Bigg\{\Bigg[\frac{176358}{11}(\kappa_{3D}^{22}+\kappa'^{22}_{3D})+1939938(\kappa_{3D}^{20}\kappa'^{2}_{3D}+\kappa_{3D}^{2}\kappa'^{20}_{3D})\\
&+48498450(\kappa_{3D}^{18}\kappa'^{4}_{3D}+\kappa_{3D}^{4}\kappa'^{18}_{3D})+436486050(\kappa_{3D}^{16}\kappa'^{6}_{3D}+\kappa_{3D}^{6}\kappa'^{16}_{3D})\\
&+1745944200(\kappa_{3D}^{14}\kappa'^{8}_{3D}+\kappa_{3D}^{8}\kappa'^{14}_{3D})+3422050632(\kappa_{3D}^{12}\kappa'^{10}_{3D}+\kappa_{3D}^{10}\kappa'^{12}_{3D})\Bigg]\\
&+f^{2}\Big[3879876(\kappa_{3D}^{20}\kappa'^{2}_{3D}+\kappa'^{22}_{3D})+387987600(\kappa_{3D}^{18}\kappa'^{4}_{3D}+\kappa_{3D}^{2}\kappa'^{20}_{3D})\\
&+7856748900(\kappa_{3D}^{16}\kappa'^{6}_{3D}+\kappa_{3D}^{4}\kappa'^{18}_{3D})+55870214400(\kappa_{3D}^{14}\kappa'^{8}_{3D}+\kappa_{3D}^{6}\kappa'^{16}_{3D})\\
&+171102531600(\kappa_{3D}^{12}\kappa'^{10}_{3D}+\kappa_{3D}^{8}\kappa'^{14}_{3D})+246387645504\kappa_{3D}^{10}\kappa'^{12}_{3D}\Big]\\
&+f^{4}\Big[290990700(\kappa_{3D}^{18}\kappa'^{4}_{3D}+\kappa'^{22}_{3D})+23570246700(\kappa_{3D}^{16}\kappa'^{6}_{3D}+\kappa_{3D}^{2}\kappa'^{20}_{3D})\\
&+377123947200(\kappa_{3D}^{14}\kappa'^{8}_{3D}+\kappa_{3D}^{4}\kappa'^{18}_{3D})+2053230379200(\kappa_{3D}^{12}\kappa'^{10}_{3D}+\kappa_{3D}^{6}\kappa'^{16}_{3D})\\
&+4619768353200(\kappa_{3D}^{10}\kappa'^{12}_{3D}+\kappa_{3D}^{8}\kappa'^{14}_{3D})\Big]\\
&+f^{6}\Big[8729721000(\kappa_{3D}^{16}\kappa'^{6}_{3D}+\kappa'^{22}_{3D})+558702144000(\kappa_{3D}^{14}\kappa'^{8}_{3D}+\kappa_{3D}^{2}\kappa'^{20}_{3D})\\
&+6844101264000(\kappa_{3D}^{12}\kappa'^{10}_{3D}+\kappa_{3D}^{4}\kappa'^{18}_{3D})+27376405056000(\kappa_{3D}^{10}\kappa'^{12}_{3D}+\kappa_{3D}^{6}\kappa'^{16}_{3D})\\
&+42775632900000\kappa_{3D}^{8}\kappa'^{14}_{3D}\Big]\\
&+f^{8}\Big[122216094000(\kappa_{3D}^{14}\kappa'^{8}_{3D}+\kappa'^{22}_{3D})+5988588606000(\kappa_{3D}^{12}\kappa'^{10}_{3D}+\kappa_{3D}^{2}\kappa'^{20}_{3D})\\
&+53897297454000(\kappa_{3D}^{10}\kappa'^{12}_{3D}+\kappa_{3D}^{4}\kappa'^{18}_{3D})+149714715150000(\kappa_{3D}^{8}\kappa'^{14}_{3D}+\kappa_{3D}^{6}\kappa'^{16}_{3D})\Big]\\
&+f^{10}\Big[862356759264(\kappa_{3D}^{12}\kappa'^{10}_{3D}+\kappa'^{22}_{3D})+31044843333504(\kappa_{3D}^{10}\kappa'^{12}_{3D}+\kappa_{3D}^{2}\kappa'^{20}_{3D})\\
&+194030270834400(\kappa_{3D}^{8}\kappa'^{14}_{3D}+\kappa_{3D}^{4}\kappa'^{18}_{3D})+344942703705600\kappa_{3D}^{6}\kappa'^{16}_{3D}\Big]\\
&+\mathrm{f}^{12}\Big[3161974783968(\kappa_{3D}^{10}\kappa'^{12}_{3D}+\kappa'^{22}_{3D})+79049369599200(\kappa_{3D}^{8}\kappa'^{14}_{3D}+\kappa_{3D}^{2}\kappa'^{20}_{3D})\\
&+316197478396800(\kappa_{3D}^{6}\kappa'^{16}_{3D}+\kappa_{3D}^{4}\kappa'^{18}_{3D})\Big]\\
&+f^{14}\Big[5992080494400(\kappa_{3D}^{8}\kappa'^{14}_{3D}+\kappa'^{22}_{3D})+95873287910400(\kappa_{3D}^{6}\kappa'^{16}_{3D}+\kappa_{3D}^{2}\kappa'^{20}_{3D})\\
&+215714897798400\kappa_{3D}^{4}\kappa'^{18}_{3D}\Big]\\
&+f^{16}\Big[5617575463500(\kappa_{3D}^{6}\kappa'^{16}_{3D}+\kappa'^{22}_{3D})+50558179171500(\kappa_{3D}^{4}\kappa'^{18}_{3D}+\kappa_{3D}^{2}\kappa'^{20}_{3D})\Big]\\
&+f^{18}\Big[2357994639000(\kappa_{3D}^{4}\kappa'^{18}_{3D}+\kappa'^{22}_{3D})+9431978556000\kappa_{3D}^{2}\kappa'^{20}_{3D}\Big]\\
&+358415185128f^{20}(\kappa_{3D}^{2}\kappa'^{20}_{3D}+\kappa'^{22}_{3D})+11309870605\frac{1}{11}f^{22}\kappa'^{22}_{3D}\Bigg\}
\end{aligned}
\tag{A.8}
$$

with $f \equiv w_y = w_z$. One needs to put the expansions for $\kappa_{3D}$ and $\kappa'_{3D}$ (and also high order terms) into the expression (A.8). The algebra is straightforward and one arrives:

$$-\frac{17}{2}\kappa^2+\frac{207}{4}\kappa^4-\frac{2573}{6}\kappa^6+\frac{39135}{8}\kappa^8-\frac{504377}{10}\kappa^{10}+\frac{7589331}{12}\kappa^{12}$$
$$-\frac{97218677}{14}\kappa^{14}+\frac{1488554079}{16}\kappa^{16}-\frac{18520701857}{18}\kappa^{18}+\frac{295407516807}{20}\kappa^{20}$$
$$-\frac{3466155691037}{22}\kappa^{22}+......$$

(A.9)

Re-writing this formula in the logarithmic form yields:

$$\ln\left[1-\frac{17}{2}\kappa^2+\frac{703}{8}\kappa^4-\frac{15537}{16}\kappa^6+\frac{1531259}{128}\kappa^8-\frac{37568183}{256}\kappa^{10}+\frac{1944333867}{1024}\kappa^{12}\right.$$
$$-\frac{49369925777}{2048}\kappa^{14}+\frac{10486210475347}{32768}\kappa^{16}-\frac{270077234947067}{65536}\kappa^{18}$$
$$\left.+\frac{14620934319382209}{262144}\kappa^{20}-\frac{377787247047993783}{524288}\kappa^{22}+\cdots\right]$$

(A.10)

Then, from (A.6) and (A.10), one derives:

$$\ln\lambda=\ln\left[2\cosh^3K\left(1+\frac{17}{2}\kappa^2-\frac{101}{8}\kappa^4+\frac{2221}{16}\kappa^6-\frac{157133}{128}\kappa^8+\frac{3128095}{256}\kappa^{10}-\frac{132058577}{1024}\kappa^{12}+\cdots\right)\right.$$
$$\left.\times\left(1-\frac{17}{2}\kappa^2+\frac{703}{8}\kappa^4-\frac{15537}{16}\kappa^6+\frac{1531259}{128}\kappa^8-\frac{37568183}{256}\kappa^{10}+\frac{1944333867}{1024}\kappa^{12}-\cdots\right)\right]$$

$$=\ln\left[2\cosh^3K\left(1+3\kappa^4+22\kappa^6+192\kappa^8+2046\kappa^{10}+24853\kappa^{12}+329334\kappa^{14}\right.\right.$$
$$\left.\left.+4649601\kappa^{16}+68884356\kappa^{18}+1059830112\kappa^{20}+16809862992\kappa^{22}\cdots\right)\right]$$

(A.11)

The result equals, term by term, to the high temperature series expansion at its high – temperature limit:[80,93,107,111]

$$\lambda=Z^{1/N}=2\cosh^3K\left[1+3\kappa^4+22\kappa^6+192\kappa^8+2046\kappa^{10}+24853\kappa^{12}+329334\kappa^{14}\right.$$
$$\left.+4649601\kappa^{16}+68884356\kappa^{18}+1059830112\kappa^{20}+16809862992\kappa^{22}\cdots\right].$$

(A.12)

For finite temperatures, the partition function can be expansed as:

$$\lambda = Z^{1/N} = 2\cosh^3 K\Big[1+\frac{7}{2}\kappa^2+\frac{87}{8}\kappa^4+\frac{3613}{48}\kappa^6+\frac{170209}{384}\kappa^8+\frac{929761}{256}\kappa^{10}$$
$$+\frac{318741323}{9216}\kappa^{12}+\frac{6705494087}{18432}\kappa^{14}+\frac{411207879769}{98304}\kappa^{16}+\frac{270525364951805}{5308416}\kappa^{18}$$
$$+\frac{13788821925530329}{21233664}\kappa^{20}+\frac{121443302317649443}{14155776}\kappa^{22}\cdots\Big]$$

(A.13)

From another angle of view, one would be able to realize that the high - temperature series expansion could fit well with the putative exact solution only at/near infinite temperature, if one believed the putative exact solution were correct. Possibly, the high temperature series expansion would be valid only at/near infinite temperature. However, the appearance of fraction numbers in (A.13) suggests that $\kappa_{3D}$ and $\kappa'_{3D}$, instead of κ, might be a good basis of series expansion for the exact solution.

**Appendix B:**

Similarly, one could prove the result of the high temperature series expansion fits well with the putative exact solution only at/near infinite temperature, also for the simple orthorhombic Ising lattices. According to Eq. (3.37), the partition function for the simple orthorhombic Ising lattices could be expressed as:

$$N^{-1}\ln Z = \ln 2 + \frac{1}{2(2\pi)^4}\int_{-\pi}^{\pi}\int_{-\pi}^{\pi}\int_{-\pi}^{\pi}\int_{-\pi}^{\pi}\ln\big[\cosh 2K\cosh 2(K'+K''+K''') - \sinh 2K\cos\omega'$$
$$-\sinh 2(K'+K''+K''')(w_x\cos\omega_x + w_y\cos\omega_y + w_z\cos\omega_z)\big]d\omega' d\omega_x d\omega_y d\omega_z$$

(B.1)

with $w_x = 1$ and:

$$w_y = w_z = \pm\sqrt{A_0 - A_1 - A_2 - \cdots}, \qquad \text{(B.2)}$$

where the leading terms are as follows:

$$A_0 = \frac{\kappa_4^2 + 2\kappa_2\kappa_3 + 2\kappa_3\kappa_4 + 2\kappa_4\kappa_2}{2(\kappa_2 + \kappa_3 + \kappa_4)^2}, \tag{B.3a}$$

$$\begin{aligned} A_1 = -\frac{1}{8(\kappa_2+\kappa_3+\kappa_4)^3} \big[ & 23\kappa_4^5 + 115\kappa_2\kappa_4^4 + 115\kappa_3\kappa_4^4 + 32\kappa_2^4\kappa_4 + 32\kappa_3^4\kappa_4 + 32\kappa_2^4\kappa_3 \\ & + 32\kappa_2\kappa_3^4 + 128\kappa_2^3\kappa_3^2 + 128\kappa_2^2\kappa_3^3 + 136\kappa_2^3\kappa_4^2 + 136\kappa_3^3\kappa_4^2 + 196\kappa_2^2\kappa_4^3 + 196\kappa_3^2\kappa_4^3 \\ & + 596\kappa_2\kappa_3^2\kappa_4^2 + 596\kappa_2^2\kappa_3\kappa_4^2 + 528\kappa_2^2\kappa_3^2\kappa_4 + 460\kappa_2\kappa_3\kappa_4^3 + 272\kappa_2^3\kappa_3\kappa_4 + 272\kappa_2\kappa_3^3\kappa_4 \big] \end{aligned} \tag{B.3b}$$

$$\begin{aligned} A_2 = -\frac{1}{16(\kappa_2+\kappa_3+\kappa_4)^4} \big[ & 472\kappa_2^6\kappa_4^2 + 472\kappa_3^6\kappa_4^2 + 464\kappa_2^6\kappa_3^2 + 464\kappa_2^2\kappa_3^6 + 1040\kappa_2^3\kappa_3^5 \\ & + 1040\kappa_2^5\kappa_3^3 + 752\kappa_2^5\kappa_4^3 + 752\kappa_3^5\kappa_4^3 + 1376\kappa_2^4\kappa_3^4 + 472\kappa_2^4\kappa_4^4 + 472\kappa_3^4\kappa_4^4 + 944\kappa_2^6\kappa_3\kappa_4 \\ & + 944\kappa_2\kappa_3^6\kappa_4 + 2320\kappa_2\kappa_3^5\kappa_4^2 + 2320\kappa_2^5\kappa_3\kappa_4^2 + 2608\kappa_2^5\kappa_3^2\kappa_4 + 2608\kappa_2^2\kappa_3^5\kappa_4 + 52\kappa_2^2\kappa_3^2\kappa_4^4 \\ & + 4712\kappa_2^2\kappa_3^4\kappa_4^2 + 4712\kappa_2^4\kappa_3^2\kappa_4^2 + 454\kappa_2\kappa_3^3\kappa_4^4 + 454\kappa_2^3\kappa_3\kappa_4^4 + 2304\kappa_2^4\kappa_3\kappa_4^3 \\ & + 2304\kappa_2\kappa_3^4\kappa_4^3 + 4400\kappa_2^3\kappa_3^4\kappa_4 + 4400\kappa_2^4\kappa_3^3\kappa_4 + 6096\kappa_2^3\kappa_3^3\kappa_4^2 + 3280\kappa_2^2\kappa_3^3\kappa_4^3 \\ & + 3280\kappa_2^3\kappa_3^2\kappa_4^3 + 112\kappa_2^7\kappa_4 + 112\kappa_3^7\kappa_4 + 112\kappa_2^7\kappa_3 + 112\kappa_2\kappa_3^7 - 33\kappa_4^8 - 172\kappa_2\kappa_4^7 \\ & - 172\kappa_3\kappa_4^7 - 309\kappa_2^2\kappa_4^6 - 309\kappa_3^2\kappa_4^6 - 90\kappa_2^3\kappa_4^5 - 90\kappa_3^3\kappa_4^5 - 744\kappa_2\kappa_3\kappa_4^6 - 934\kappa_2\kappa_3^2\kappa_4^5 \\ & - 934\kappa_2^2\kappa_3\kappa_4^5 - 460\kappa_1^2\kappa_4^6 - 2760\kappa_1^2\kappa_2\kappa_4^5 - 2760\kappa_1^2\kappa_3\kappa_4^5 - 544\kappa_1^2\kappa_2^5\kappa_3 - 544\kappa_1^2\kappa_2\kappa_3^5 \\ & - 544\kappa_1^2\kappa_2^5\kappa_4 - 544\kappa_1^2\kappa_3^5\kappa_4 - 6252\kappa_1^2\kappa_2^2\kappa_4^4 - 6252\kappa_1^2\kappa_3^2\kappa_4^4 - 3136\kappa_1^2\kappa_2^4\kappa_3^2 - 3136\kappa_1^2\kappa_2^2\kappa_3^4 \\ & - 3200\kappa_1^2\kappa_2^4\kappa_4^2 - 3200\kappa_1^2\kappa_3^4\kappa_4^2 - 5184\kappa_1^2\kappa_2^3\kappa_3^3 - 6608\kappa_1^2\kappa_2^3\kappa_4^3 - 6608\kappa_1^2\kappa_3^3\kappa_4^3 \\ & - 34208\kappa_1^2\kappa_2^2\kappa_3^2\kappa_4^2 - 19824\kappa_1^2\kappa_2^3\kappa_3\kappa_4^2 - 19824\kappa_1^2\kappa_2\kappa_3^3\kappa_4^2 - 18400\kappa_1^2\kappa_2^3\kappa_3^2\kappa_4 \\ & - 18400\kappa_1^2\kappa_2^2\kappa_3^3\kappa_4 - 6400\kappa_1^2\kappa_2^4\kappa_3\kappa_4 - 6400\kappa_1^2\kappa_2\kappa_3^4\kappa_4 - 25008\kappa_1^2\kappa_2^2\kappa_3\kappa_4^3 \\ & - 25008\kappa_1^2\kappa_2\kappa_3^2\kappa_4^3 - 13800\kappa_1^2\kappa_2\kappa_3\kappa_4^4 \big] \end{aligned} \tag{B.3c}$$

Here $\kappa_1 = \tanh K$, $\kappa_2 = \tanh K'$, $\kappa_3 = \tanh K''$ and $\kappa_4 = \tanh K'''$.

The formula (B.1) is reduced to be:

$$\begin{aligned} \ln\lambda - \ln 2(\cosh 2K \cosh 2(K' + K'' + K'''))^{\frac{1}{2}} = \frac{1}{2(2\pi)^4} \int_0^\pi\int_0^\pi\int_0^\pi\int_0^\pi & \ln[1 - 2\kappa_{3D}\cos\omega' \\ - 2\kappa'_{3D}(w_x\cos\omega_x + w_y\cos\omega_y + w_z\cos\omega_z)] & \, d\omega' d\omega_x d\omega_y d\omega_z \end{aligned} \tag{B.4}$$

Here

$$\kappa_{3D} = \frac{\tanh 2K}{2\cosh 2(K'+K''+K''')} \tag{B.5}$$
$$= \kappa_1(1-\kappa_2^2-\kappa_3^2-\kappa_4^2+\kappa_2^2\kappa_3^2+\kappa_3^2\kappa_4^2+\kappa_2^2\kappa_4^2-\kappa_2^2\kappa_3^2\kappa_4^2)\Big/\Gamma \; .$$

and

$$\kappa'_{3D} = \frac{\tanh 2(K'+K''+K''')}{2\cosh 2K}$$
$$= \frac{(1-\kappa_1^2)}{\Gamma}\cdot\left(\kappa_2+\kappa_3+\kappa_4+\kappa_2\kappa_3^2+\kappa_2\kappa_4^2+\kappa_2^2\kappa_3+\kappa_3\kappa_4^2+\kappa_2^2\kappa_4+\kappa_3^2\kappa_4\right. \tag{B.6}$$
$$\left.+\kappa_2\kappa_3^2\kappa_4^2+\kappa_2^2\kappa_3\kappa_4^2+\kappa_2^2\kappa_3^2\kappa_4+4\kappa_2\kappa_3\kappa_4\right)$$

with

$$\Gamma = 1+F_2+F_4+F_6+F_8 \tag{B.7}$$

$$F_2 = \kappa_1^2+\kappa_2^2+\kappa_3^2+\kappa_4^2+4\kappa_2\kappa_3+4\kappa_3\kappa_4+4\kappa_2\kappa_4 \tag{B.8a}$$

$$F_4 = \kappa_1^2\kappa_2^2+\kappa_1^2\kappa_3^2+\kappa_1^2\kappa_4^2+\kappa_2^2\kappa_3^2+\kappa_3^2\kappa_4^2+\kappa_2^2\kappa_4^2+4\kappa_1^2\kappa_2\kappa_3$$
$$+4\kappa_1^2\kappa_3\kappa_4+4\kappa_1^2\kappa_2\kappa_4+4\kappa_2\kappa_3\kappa_4^2+4\kappa_2\kappa_3^2\kappa_4+4\kappa_2^2\kappa_3\kappa_4 \tag{B.8b}$$

$$F_6 = \kappa_1^2\kappa_2^2\kappa_3^2+\kappa_1^2\kappa_3^2\kappa_4^2+\kappa_1^2\kappa_2^2\kappa_4^2+\kappa_2^2\kappa_3^2\kappa_4^2+4\kappa_1^2\kappa_2\kappa_3\kappa_4^2+4\kappa_1^2\kappa_2\kappa_3^2\kappa_4+4\kappa_1^2\kappa_2^2\kappa_3\kappa_4$$

(B.8c)

$$F_8 = \kappa_1^2\kappa_2^2\kappa_3^2\kappa_4^2 \tag{B.8d}$$

Employing the formula of $1/(1+x) = 1 - x + x^2 - x^3 + x^4 - \ldots$ to expand $1/\Gamma$, one obtains:

$$\frac{1}{\Gamma} = 1-F_2+(F_2^2-F_4)+(2F_2F_4-F_6-F_2^3)+(F_4^2+2F_2F_6+F_2^4-F_8-3F_2^2F_4)+\ldots$$

(B.9)

Then one has:

$$\kappa_{3D} = G_1 - (G_1F_2 + G_3) + (G_1F_2^2 - G_1F_4 + G_3F_2 + G_5)$$
$$+ (2G_1F_2F_4 - G_1F_6 - G_1F_2^3 - G_3F_2^2 + G_3F_4 - G_5F_2 - G_7)$$
$$+ (G_1F_4^2 + 2G_1F_2F_6 + G_1F_2^4 - G_1F_8 - 3G_1F_2^2F_4 - 2G_3F_2F_4 + G_3F_6 + G_3F_2^3 + G_5F_2^2 - G_5F_4 + G_7F_2)$$

(B.10)

with

$$G_1 = \kappa_1 \quad \text{(B.11a)}$$

$$G_3 = \kappa_1\kappa_2^2 + \kappa_1\kappa_3^2 + \kappa_1\kappa_4^2 \quad \text{(B.11b)}$$

$$G_5 = \kappa_1\kappa_2^2\kappa_3^2 + \kappa_1\kappa_3^2\kappa_4^2 + \kappa_1\kappa_2^2\kappa_4^2 \quad \text{(B.11c)}$$

$$G_7 = \kappa_1\kappa_2^2\kappa_3^2\kappa_4^2 \quad \text{(B.11d)}$$

and

$$\kappa'_{3D} = H_1 - (H_1F_2 - H_3) + (H_1F_2^2 - H_1F_4 - H_3F_2 + H_5)$$
$$+ (2H_1F_2F_4 - H_1F_6 - H_1F_2^3 + H_3F_2^2 - H_3F_4 - H_5F_2 + H_7)$$
$$+ (H_1F_4^2 + 2H_1F_2F_6 + H_1F_2^4 - H_1F_8 - 3H_1F_2^2F_4 + 2H_3F_2F_4 - H_3F_6 - H_3F_2^3 + H_5F_2^2 - H_5F_4 - H_7F_2)$$

(B.12)

with

$$H_1 = \kappa_2 + \kappa_3 + \kappa_4 \quad \text{(B.13a)}$$

$$H_3 = \kappa_2\kappa_3^2 + \kappa_2\kappa_4^2 + \kappa_2^2\kappa_3 + \kappa_3\kappa_4^2 + \kappa_2^2\kappa_4 + \kappa_3^2\kappa_4$$
$$+ 4\kappa_2\kappa_3\kappa_4 - \kappa_1^2\kappa_2 - \kappa_1^2\kappa_3 - \kappa_1^2\kappa_4 \quad \text{(B.13b)}$$

$$H_5 = \kappa_2\kappa_3^2\kappa_4^2 + \kappa_2^2\kappa_3\kappa_4^2 + \kappa_2^2\kappa_3^2\kappa_4 - \kappa_1^2\kappa_2\kappa_3^2 - \kappa_1^2\kappa_2^2\kappa_3 - \kappa_1^2\kappa_2\kappa_4^2 - \kappa_1^2\kappa_2^2\kappa_4 - \kappa_1^2\kappa_3\kappa_4^2 - \kappa_1^2\kappa_3^2\kappa_4 - 4\kappa_1^2\kappa_2\kappa_3\kappa_4 \tag{B.13c}$$

$$H_7 = -\kappa_1^2\kappa_2\kappa_3^2\kappa_4^2 - \kappa_1^2\kappa_2^2\kappa_3\kappa_4^2 - \kappa_1^2\kappa_2^2\kappa_3^2\kappa_4 \tag{B.13d}$$

Then one can calculate the high order terms $\kappa_{3D}^2$, $\kappa_{3D}^4$, $\kappa_{3D}^6$, … ; $\kappa'^2_{3D}$, $\kappa'^4_{3D}$, $\kappa'^6_{3D}$,… ; $\kappa_{3D}^2\kappa'^2_{3D}$, $\kappa_{3D}^4\kappa'^2_{3D}$, …; $\kappa_{3D}^2\kappa'^4_{3D}$, $\kappa_{3D}^4\kappa'^4_{3D}$, …; ……

The left side of the equation above for the partition function is re-written as:

$$\ln\lambda - \ln\left[2\cosh K\cosh K'\cosh K''\left(\frac{\Gamma}{1-\kappa_4^2}\right)^{\frac{1}{2}}\right] = \ln\lambda - \ln\left[2\cosh K\cosh K'\cosh K''\left(1+\Omega_1+\Omega_2+\Omega_3\cdots\right)\right] \tag{B.14}$$

with

$$\Omega_1 = \frac{1}{2}(F_2 + \kappa_4^2) \tag{B.15a}$$

$$\Omega_2 = \frac{1}{2}F_4 + \frac{1}{4}F_2\kappa_4^2 + \frac{3}{8}\kappa_4^4 - \frac{1}{8}F_2^2 \tag{B.15b}$$

$$\Omega_3 = \frac{1}{2}F_6 + \frac{1}{4}F_4\kappa_4^2 + \frac{3}{16}F_2\kappa_4^4 + \frac{5}{16}\kappa_4^6 - \frac{1}{4}F_2F_4 - \frac{1}{16}F_2^2\kappa_4^2 + \frac{1}{16}F_2^3 \tag{B.15c}$$

Expanding the expression the right hand of the expression for the partition function and integrating each term yields:

$$-\left[\frac{1}{2}(\kappa_{3D}^2 + \kappa'^2_{3D}) + f^2\kappa'^2_{3D}\right]$$
$$-\left[\frac{3}{4}(\kappa_{3D}^4 + \kappa'^4_{3D}) + 3\kappa_{3D}^2\kappa'^2_{3D} + 6f^2(\kappa_{3D}^2\kappa'^2_{3D} + \kappa'^4_{3D}) + 4.5f^4\kappa'^4_{3D}\right]$$
$$-\left[\frac{5}{3}(\kappa_{3D}^6 + \kappa'^6_{3D}) + 15(\kappa_{3D}^4\kappa'^2_{3D} + \kappa_{3D}^2\kappa'^4_{3D}) + 30f^2(\kappa_{3D}^4\kappa'^2_{3D} + \kappa'^6_{3D}) + 120f^2\kappa_{3D}^2\kappa'^4_{3D} + 90f^4(\kappa_{3D}^2\kappa'^4_{3D} + \kappa'^6_{3D}) + \frac{100}{3}f^6\kappa'^6_{3D}\right]$$
$$-\ldots\ldots$$

(B.16)

Putting the expansions for $\kappa_{3D}$ and $\kappa'_{3D}$ (and also high order terms) into the expression (B.18), one arrives after a little algebra at:

$$-\Omega_1+\left[(\kappa_1^2\kappa_2^2+\kappa_2^2\kappa_3^2+\kappa_1^2\kappa_3^2)+\frac{1}{2}\Omega_1^2-\Omega_2\right]$$
$$+\left[(16\kappa_1^2\kappa_2^2\kappa_3^2+\kappa_1^4\kappa_2^2+\kappa_1^2\kappa_2^4+\kappa_2^4\kappa_3^2+\kappa_2^2\kappa_3^4+\kappa_3^4\kappa_1^2+\kappa_3^2\kappa_1^4)+\Omega_1\Omega_2-\Omega_3-\frac{1}{3}\Omega_1^3\right]$$

(B.17)

Re-writing this formula in the logarithmic form results in:

$$\ln(1+\Lambda_1+\Lambda_2+\Lambda_3\cdots) \tag{B.18}$$

with

$$\Lambda_1=-\Omega_1 \tag{B.19a}$$

$$\Lambda_2=(\kappa_1^2\kappa_2^2+\kappa_2^2\kappa_3^2+\kappa_1^2\kappa_3^2)+\Omega_1^2-\Omega_2 \tag{B.19b}$$

$$\begin{aligned}\Lambda_3&=(16\kappa_1^2\kappa_2^2\kappa_3^2+\kappa_1^4\kappa_2^2+\kappa_1^2\kappa_2^4+\kappa_2^4\kappa_3^2+\kappa_2^2\kappa_3^4+\kappa_3^4\kappa_1^2+\kappa_3^2\kappa_1^4)\\&+2\Omega_1\Omega_2-\Omega_3-\Omega_1^3-\Omega_1(\kappa_1^2\kappa_2^2+\kappa_2^2\kappa_3^2+\kappa_1^2\kappa_3^2)\end{aligned} \tag{B.19c}$$

Then one obtains from (B.14) and (B.18):

$$\begin{aligned}\ln\lambda&=\ln[2\cosh K\cosh K'\cosh K''(1+\Omega_1+\Omega_2+\Omega_3+\cdots)(1+\Lambda_1+\Lambda_2+\Lambda_3\cdots)]\\&=\ln\{2\cosh K\cosh K'\cosh K''[1+(\kappa_1^2\kappa_2^2+\kappa_2^2\kappa_3^2+\kappa_3^2\kappa_1^2)\\&+(16\kappa_1^2\kappa_2^2\kappa_3^2+\kappa_1^4\kappa_2^2+\kappa_1^2\kappa_2^4+\kappa_2^4\kappa_3^2+\kappa_2^2\kappa_3^4+\kappa_3^4\kappa_1^2+\kappa_3^2\kappa_1^4)+\cdots\cdots]\}\end{aligned}$$

(B.20)

This result fits well with the high temperature series expansion at high temperature limit:[93]

$$\lambda = Z^{1/N} = 2\cosh K \cosh K' \cosh K'' \big[1 + (\kappa_1^2\kappa_2^2 + \kappa_2^2\kappa_3^2 + \kappa_3^2\kappa_1^2) \\ + (16\kappa_1^2\kappa_2^2\kappa_3^2 + \kappa_1^4\kappa_2^2 + \kappa_1^2\kappa_2^4 + \kappa_2^4\kappa_3^2 + \kappa_2^2\kappa_3^4 + \kappa_3^4\kappa_1^2 + \kappa_3^2\kappa_1^4) + \cdots\cdots\big]. \tag{B.21}$$

Evidently, (B.3a) – (B.3c) return back the first three terms $b_0$, $b_1$ and $b_2$ in (A.2) as K = K' = K'' (and thus $\kappa_1 = \kappa_2 = \kappa_3 = \kappa_4 = \kappa$). One could try to add more terms $A_3$, $A_4$, … Certainly, such calculations would be very much laborious, tedious, and extremely difficulty. Nonetheless, the first three terms, look impressive, seem enough to illustrate clearly that the same procedure for (A. 2) can be generalized to the general cases of K', K'' and K''', and to reveal some physical significance as well as some symmetries among the parameters $\kappa_1$, $\kappa_2$, $\kappa_3$ and $\kappa_4$. The addition of the higher terms $A_3$, $A_4$, … would not add more evident physical significance. All the efforts are devoted to reveal a possibility of embodying elegantly the opening form of infinite terms of the high - temperature expansion into the square roots of the weights. These weights can vanish, when the temperature is lowered to deviate from the high - temperature limit.

The weights $w_y$ and $w_z$ in the form of the square root just embody the opening form of the high temperature expansion in infinite terms into a closed form. This is a novel and elegant method of dealing with the problem of infinite. The combination of this closed form with the close form of the function in the 4 - fold integral produces the closed – form expressions for the free energy of the 3D Ising model from T = 0 to any finite temperatures and also at infinite temperature.

Table 1 The putative exact critical exponents for the 3D Ising lattice, together with the exact values for the 2D Ising lattice, the approximate values obtained by the Monte Carlo renormalization group (PV-MC),[154] the renormalization group (RG) with the ε expansion to order $\varepsilon^2$, the high temperature series expansion (SE) for the 3D one,[103,142] and those of the mean field (MF) theory. PV taken from Pelissetto and Vicari's review,[154] WK from Wilson and Kogut's review,[142] F from Fisher's series expansion.[103] Notice that Domb's values in ref. [107] are same as Fisher's ($T > T_c$).[103]

| Ising | α | β | γ | δ | η | ν |
|---|---|---|---|---|---|---|
| 1D Exact | — | — | 2 | ∞ | 1 | 2 |
| 2D Exact | 0 | $\frac{1}{8}$ | $\frac{7}{4}$ | 15 | $\frac{1}{4}$ | 1 |
| 3D Exact | 0 | $\frac{3}{8}$ | $\frac{5}{4}$ | $\frac{13}{3}$ | $\frac{1}{8}$ | $\frac{2}{3}$ |
| 4D MF | 0 | $\frac{1}{2}$ | 1 | 3 | 0 | $\frac{1}{2}$ |
| 3D WK-RG | 0.077 | 0.340 | 1.244 | 4.46 | 0.037 | 0.626 |
| 3D PV-MC | 0.110 | 0.3265 | 1.2372 | 4.789 | 0.0364 | 0.6301 |
| 3D WK-SE | 0.125 ±0.015 | 0.312 ±0.003 | 1.250 ±0.003 | 5.150 ±0.02 | 0.055 ±0.010 | 0.642 ±0.003 |
| 3D F($T<T_c$) | $\frac{1}{16}$ | $\frac{5}{16}$ | $\frac{21}{16}$ | $\frac{26}{5}$ | $-\frac{1}{31}$ | $\frac{31}{48}$ |
| 3D F($T>T_c$) | $\frac{1}{8}$ | $\frac{5}{16}$ | $\frac{5}{4}$ | 5 | 0 | $\frac{5}{8}$ |

Table 2 The low temperature series coefficients $m_n$, the ratio $m_{n+1}/m_n$, the difference $\Delta_{(n+1,n)} = (m_{n+2}/m_{n+1} - m_{n+1}/m_n)$ between the neighboring ratios $m_{n+1}/m_n$ and the $(m_n)^{-1/n}$ for magnetization of the three – dimensional Ising model on a simple cubic lattice. The values for low temperature series coefficients $m_n$ are taken from table VI of ref. [238].

| n | $m_n$ | $m_{n+1}/m_n$ | $\Delta_{(n+1,n)} =$ $(m_{n+2}/m_{n+1} - m_{n+1}/m_n)$ | $(m_n)^{-1/n}$ |
|---|---|---|---|---|
| 0 | 1 | 0 | — | 1 |
| 1 | 0 | — | — | 0 |
| 2 | 0 | -∞ | ∞ | 0 |
| 3 | -2 | 0 | -∞ | 0.793700525984099… |
| 4 | 0 | -∞ | ∞ | 0 |
| 5 | -12 | -1.166666666666666… | -5.261904761904761… | 0.608364341893205… |
| 6 | 14 | -6.428571428571428… | 4.295238095238095… | 0.644137614709094… |
| 7 | -90 | -2.133333333333333… | -1.991666666666666… | 0.525802320771714… |
| 8 | 192 | - 4.125 | 1.412878787878787… | 0.518307324814038… |
| 9 | -792 | -2.712121212121212… | -0.880057558828508… | 0.476342601582425… |
| 10 | 2148 | -3.592178770949720… | 0.577404276392955… | 0.464297742761535… |
| 11 | -7716 | -3.014774494556765… | -0.403332289038798… | 0.443200975243094… |
| 12 | 23262 | -3.418106783595563… | 0.248094709946303… | 0.432626129059729… |
| 13 | -79512 | -3.170012073649260… | -0.188903079452852… | 0.419801165692409… |
| 14 | 252054 | -3.358915153102113… | 0.106577644774961… | 0.411319244095722… |
| 15 | -846628 | -3.252337508327151… | -0.090946724218825… | 0.402550612240097… |
| 16 | 2753520 | -3.343284232545977… | 0.044154987939316… | 0.395828431381549… |
| 17 | -9205800 | -3.299129244606661… | -0.045677618649371… | 0.389358196365129… |
| 18 | 30371124 | -3.344806863256032… | 0.016620758351184… | 0.383979226771075… |
| 19 | -101585544 | -3.328186104904847… | -0.024389916227353… | 0.378955133136635… |
| 20 | 338095596 | -3.352576021132200… | 0.004593441138821… | 0.374582966156199… |
| 21 | -1133491188 | -3.347982579993378… | -0.014135718971191… | 0.370541820614560… |
| 22 | 3794908752 | -3.362118298964570… | -0.00050678373868… | 0.366928763492823… |
| 23 | -12758932158 | -3.362625082703257… | -0.009022115314501… | 0.363594050587216… |
| 24 | 42903505303 | -3.371647198017758… | -0.00252237410530… | 0.360561596174962… |
| 25 | -144655483440 | -3.374169572123065… | -0.006341432681172… | 0.357755734601860… |
| 26 | 488092130664 | -3.380511004804237… | | 0.355174876597187… |
| 27 | -1650000819068 | | | 0.352777263173715… |

Figure captions:

Fig. 1 Temperature dependence of the specific heat C for the 3D simple orthorhombic Ising lattices with K' = K'' = K, 0.5 K, 0.1 K and 0.0001 K (from right to left).

Fig. 2 Plots of γ ~ K of the simple cubic Ising lattice for different values of $\omega_{2t_x} = \pi$, $3\pi/4$, $\pi/2$, $\pi/4$ and 0 (from top to bottom), neglecting the effects of $\omega_{2t_y}$ and $\omega_{2t_z}$.

Fig. 3 Temperature dependence of the spontaneous magnetization I for (a) several simple orthorhombic lattices with K' = K'' = K, 0.5 K, 0.1 K and 0.0001 K (from right to left) and (b) simple cubic Ising lattice (solid curve) in comparison with that (dashed curve) obtained by Yang for the square Ising model,[20] and the result of the low – temperature series expansion (the dotted curve with terms up to the 52nd order one;[111,238] the dash - dot curve with terms up to the 54th order[238] of the simple cubic Ising lattice.

Fig. 4 Phase diagram in the parametric plane $\frac{K'}{K} \sim \frac{K''}{K}$ of the simple orthorhombic Ising lattices. The golden solution for the simple cubic Ising lattice is located at the star point (1, 1). The dashed curve of $\frac{K'}{K} + \frac{K''}{K} + \frac{K'K''}{K^2} = 1$ corresponds to the points with the critical temperature of the silver solution. The district between the dashed curve and the dash dot one of $\frac{K'}{K} + \frac{K''}{K} + \frac{K'K''}{K^2} \approx 1.39$ is for the 3D to 2D crossover phenomenon. All the points below the dashed curve have the 2D critical exponent, while all the points above the dash dot curve behave as a real 3D system.

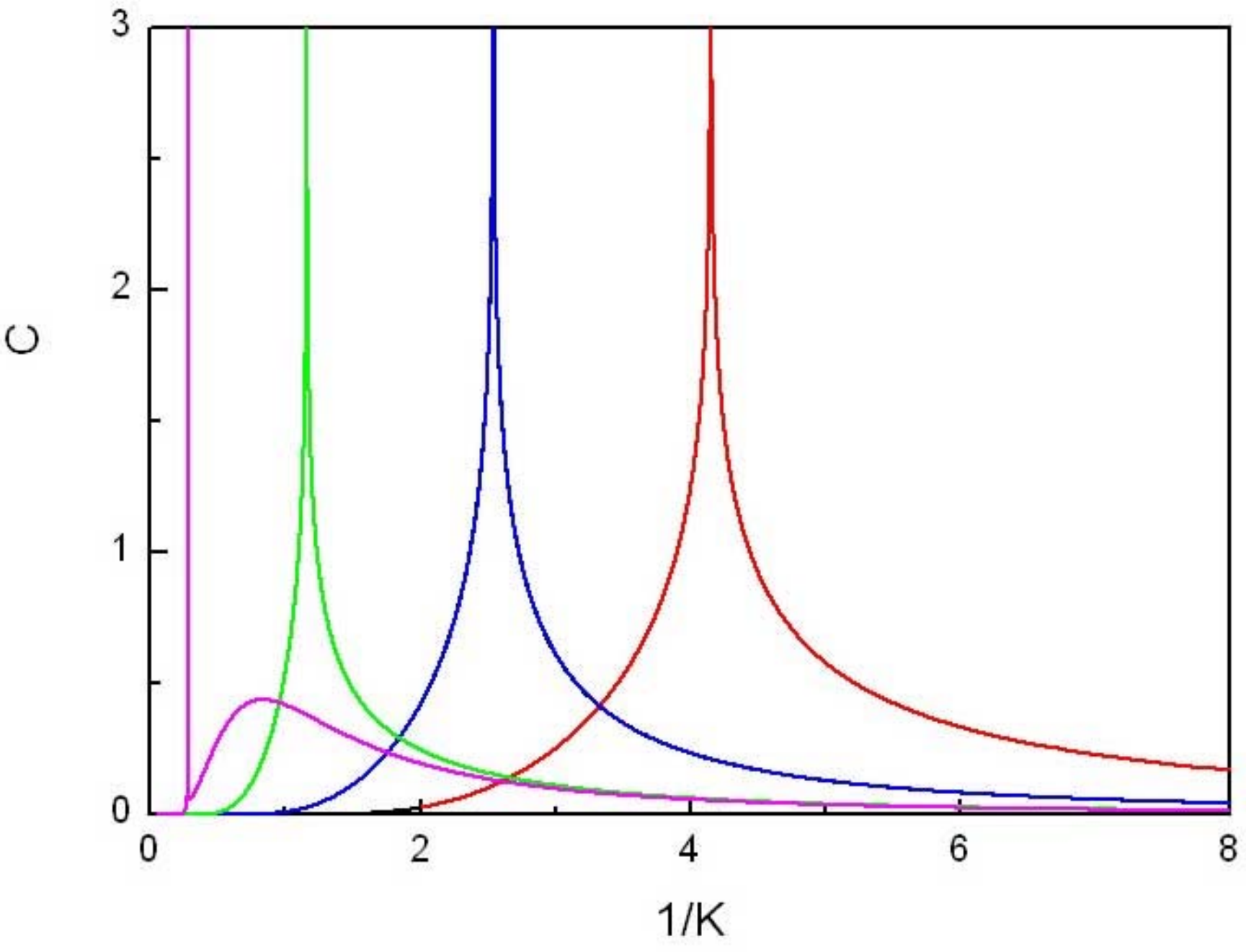

Fig. 1

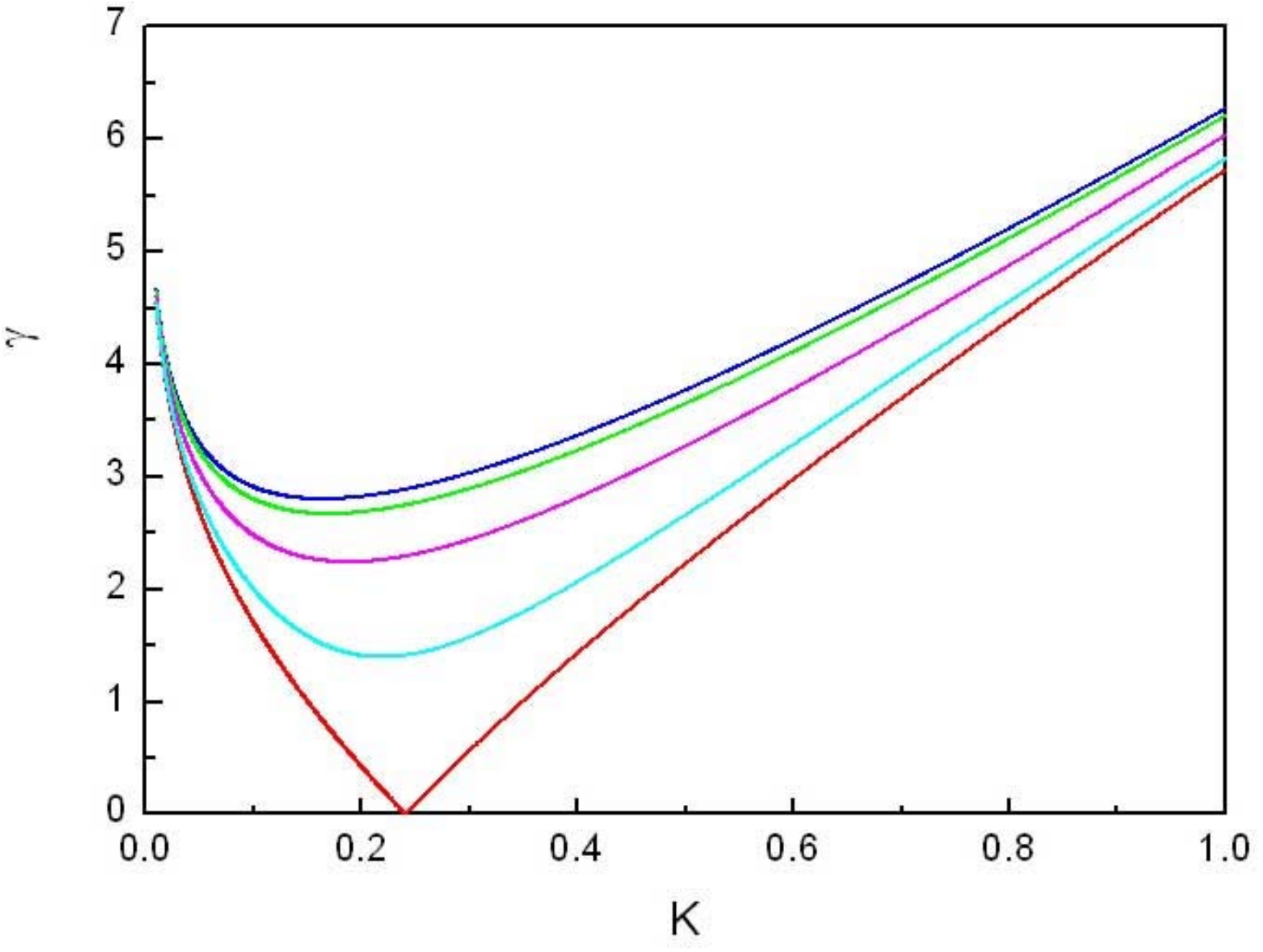

Fig. 2

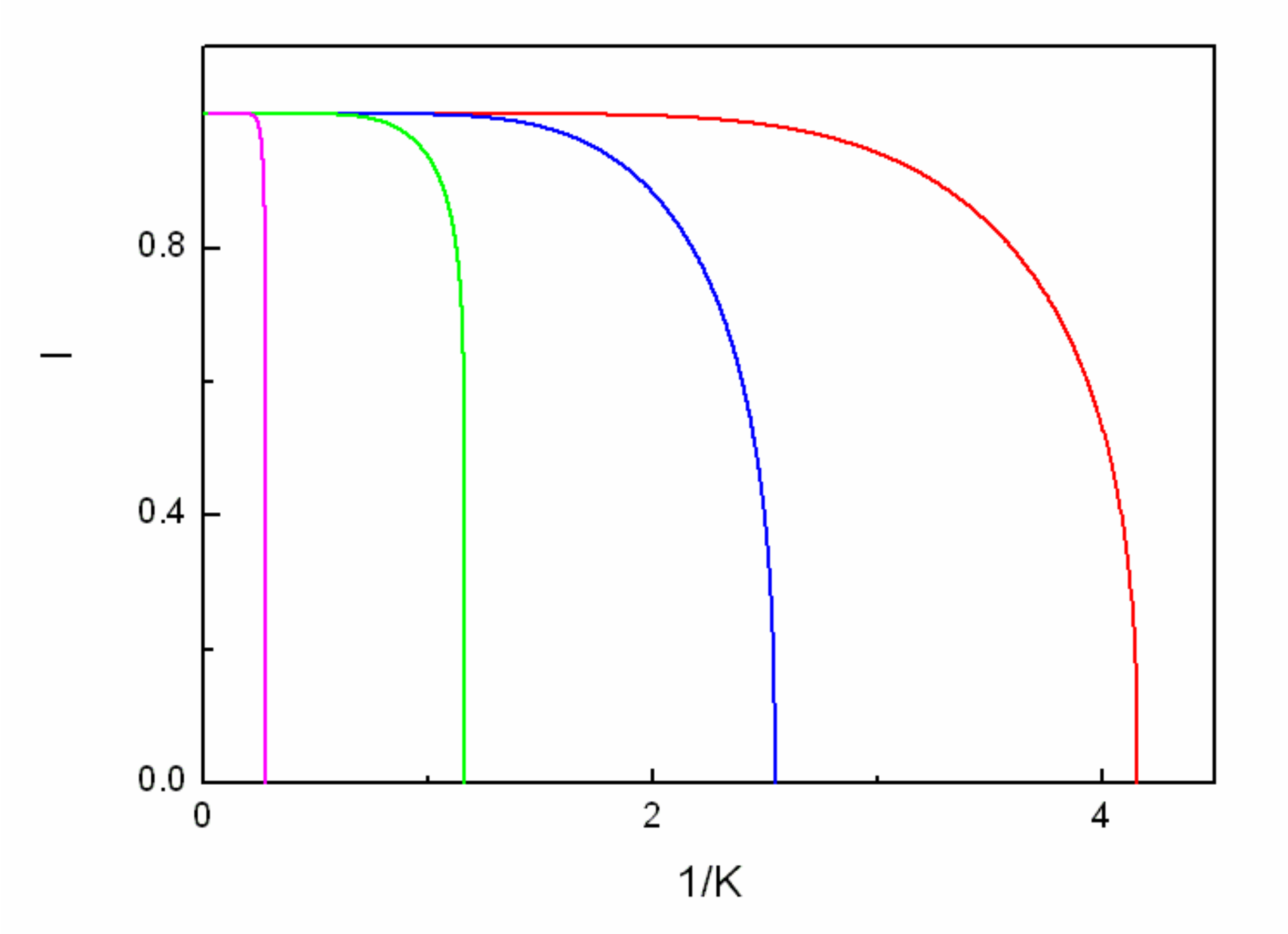

Fig. 3a

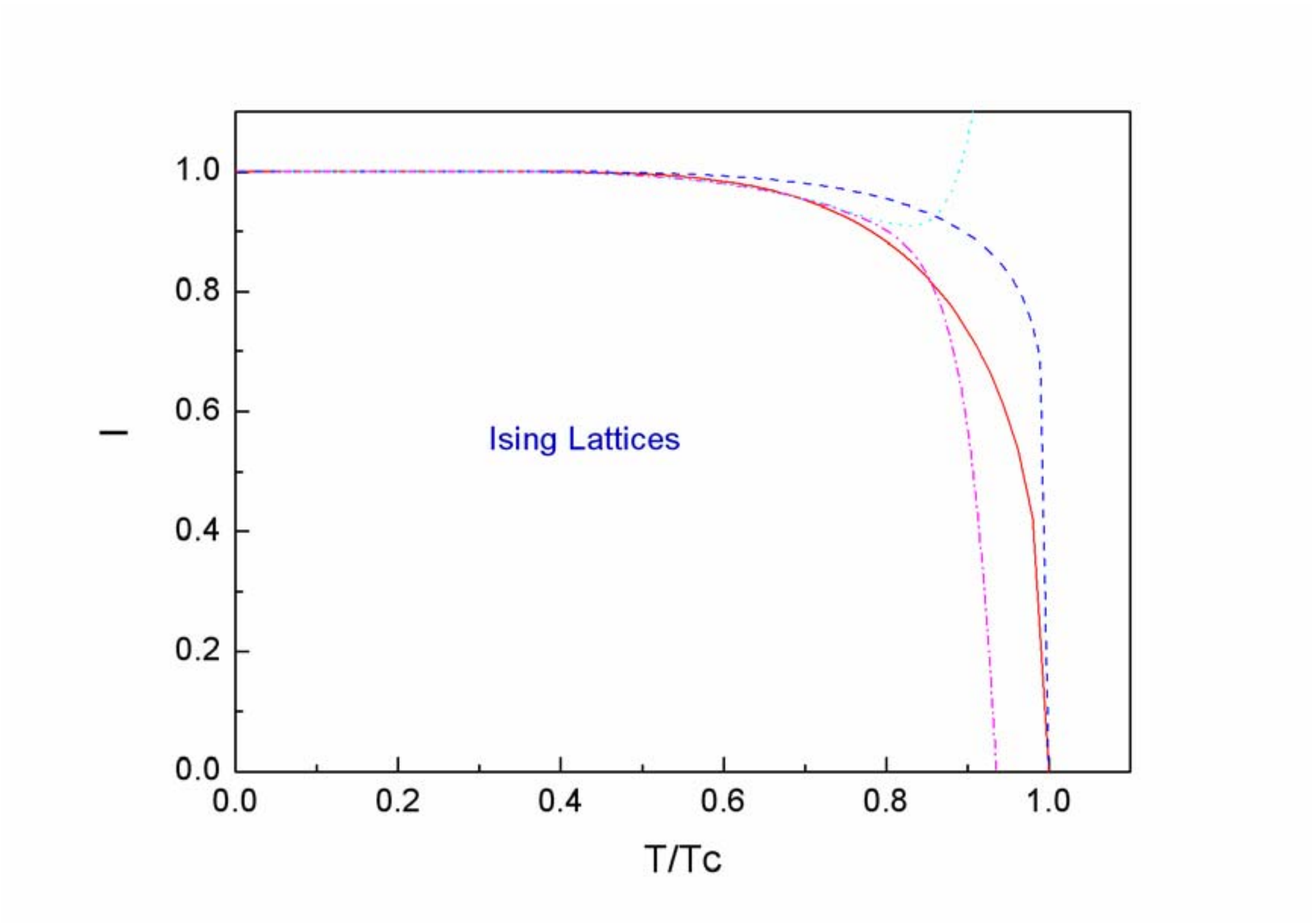

Fig. 3b

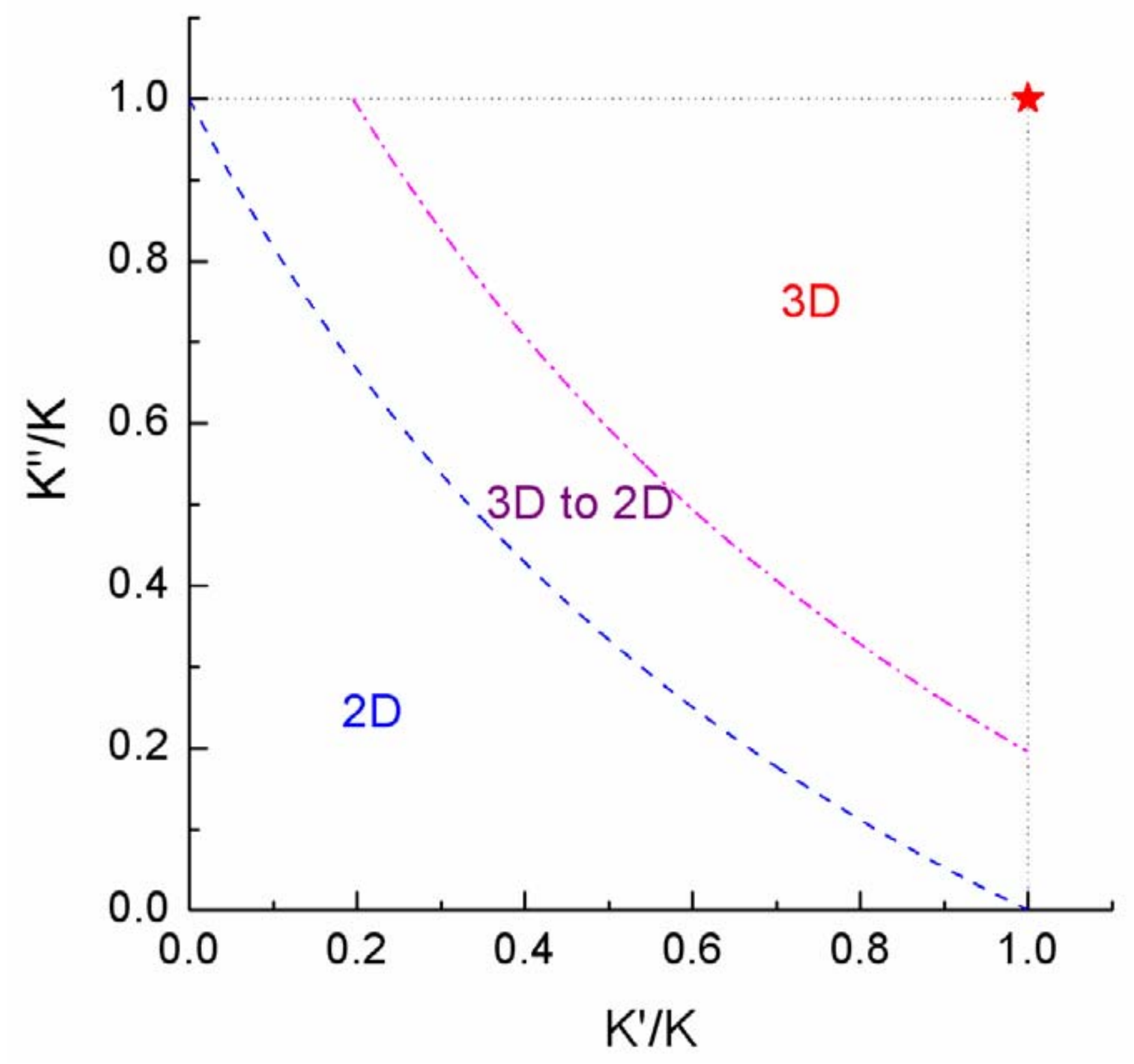

Fig. 4